\documentclass[a4paper,12pt]{article}

\usepackage{latexsym}
\usepackage{amsmath}
\usepackage{amssymb}
\usepackage{amsfonts}
\usepackage{amsthm}
\usepackage{bbold}
\usepackage[latin1]{inputenc}
\usepackage{graphicx,subfigure}
\usepackage{float,times}
\usepackage[usenames]{color}
\usepackage{longtable}
\usepackage{rotating}
\usepackage[round]{natbib}
\usepackage{fancyhdr}
\usepackage{multirow}
\usepackage[margin=2cm]{geometry}
\usepackage{ragged2e}
\usepackage{titlesec}
\usepackage{todonotes}
\usepackage{hyperref} 

\newtheoremstyle{example}{\topsep}{\topsep}%
     {}
     {}
     {\bfseries}
     {}
     {\newline}
     {\thmname{#1}\thmnumber{ #2}\thmnote{ #3}}

\theoremstyle{example}
\theoremstyle{theorem}
\theoremstyle{theorem}
\theoremstyle{proposition}
\theoremstyle{corollary}

\def\d{\mbox{d}}
\def\given{\,|\,}

\titleformat*{\section}{\large\bfseries\sffamily}
\titleformat*{\subsection}{\bfseries\sffamily}

\title{Model Averaging and its Use in Economics}

\author{
Mark F.J. Steel \thanks{Department of Statistics, University of Warwick, Coventry, CV4 7AL, UK; email: m.steel@warwick.ac.uk. I am grateful to the Editor for giving me the opportunity to write this article and for many stimulating and constructive comments. Insightful comments from five anonymous referees and from Jes\'us Crespo Cuaresma, Frank Diebold, Anabel Forte, Gonzalo Garc\'{\i}a-Donato, Tomas Havranek, Anne Miloschewski, Chris Papageorgiou, Adrian Raftery, David Rossell and Hal Varian were very gratefully received and led to substantial improvements of the paper. In July 2013, Eduardo Ley, who has made key contributions in this area, tragically passed away. He was a very dear friend and a much valued coauthor and this paper is dedicated to his memory.}\\ {Department of Statistics, University of Warwick}}

\date{\small{\today}}

\begin{document}

\maketitle

\begin{abstract}
\linespread{0.9}\selectfont
\noindent \sffamily \footnotesize  The method of model
averaging has become an important tool to deal with model uncertainty, for example in situations where a large amount of different theories exist, as are common in economics. Model averaging is a natural and formal response to model uncertainty in a  Bayesian framework, and most of the paper deals with Bayesian model averaging.  The important role of the prior assumptions in these Bayesian
procedures is highlighted. In addition, frequentist model averaging methods are also discussed. Numerical methods to implement these methods are explained, and I point the reader to some freely available computational resources. The main focus is on uncertainty regarding the choice of covariates in normal linear regression models, but the paper also covers other, more challenging, settings, with particular emphasis
on sampling models commonly used in economics. Applications of model averaging in economics are reviewed and discussed in a wide range of areas, among which growth economics, production modelling, finance and forecasting macroeconomic quantities.
(JEL: C11, C15, C20, C52, O47).
\end{abstract}


\section{Introduction}


This paper is about model averaging, as a solution to the problem of model uncertainty and focuses mostly on the theoretical developments over the last two decades and its uses in applications in economics. This is a topic that has now gained  substantial maturity and is generating a rapidly growing literature. Thus, a survey seems timely. The discussion focuses mostly on uncertainty about covariate inclusion in regression models (normal linear regression and its extensions), which is arguably the most pervasive situation in economics. Advances in the context of models designed to deal with more challenging situations, such as data with dependency over time or in space or endogeneity (all quite relevant in economic applications) are also discussed. Two main strands of model averaging are distinguished: Bayesian model averaging (BMA), based on probability calculus and naturally emanating from the Bayesian paradigm by treating the model index as an unknown, just like the model parameters and specifying a prior on
both; and frequentist model averaging (FMA), where the chosen weights are often determined so as to obtain desirable properties of the resulting estimators under repeated sampling and asymptotic optimality.

In particular, the aims of this paper are:
\begin{itemize}

\item To provide a survey of the most important methodological contributions in model averaging, especially aimed at economists. The presentation is formal, yet accessible, and uses a consistent notation. This review takes into account the latest developments, which is important in such a rapidly developing literature. Technicalities are not avoided, but some are dealt with by providing the interested reader with the relevant references. Even though the list of references is quite extensive, this is not claimed to be an exhaustive survey. Rather, it attempts to identify the most important developments that the applied economist needs to know about for an informed use of these methods. This review complements and extends other reviews and discussions; for example by \cite{Hoeting_etal_99} on BMA, \cite{ClydeGeorge04} on model uncertainty, \cite{Moral-Benito15} on model averaging in economics and  \cite{Wangetal09} on FMA. \cite{Dorman_etal_18} present an elaborate survey of model averaging methods used in ecology. A recent book on model averaging is \cite{Fletcher_18}, which is aimed at applied statisticians and has a mostly frequentist focus.
    Further, a review of weighted average least squares is provided in \cite{MagnusSurvey} while \cite{FragosoNeto_15} develop a conceptual classification scheme to better describe the literature in BMA. \cite{Koop_17} discusses the use of BMA or prior shrinkage as responses to the challenges posed by big data in empirical macroeconomics. This paper differs from the earlier surveys mainly through the combination of a more ambitious scope and depth and the focus on economics.

\item By connecting various strands of the literature, to enhance the insight of the reader into the way these methods work and why we would use them. In particular, this paper attempts to tie together disparate literatures with roots in econometrics and statistics, such as the literature on forecasting, often in the context of time series and linked with information criteria, fundamental methodology to deal with model uncertainty and shrinkage in statistics\footnote{Choosing covariates can be interpreted as a search for parsimony, which has two main approaches  in Bayesian statistics: through the use of shrinkage priors, which are absolutely continuous priors that shrink coefficients to zero but where all covariates are always included in the model, and through allocating prior point mass at zero for each of the regression coefficients, which allows for formal exclusion of covariates and implies that we need to deal with many different models, which is the approach recommended here.}, as well as more ad-hoc ways of dealing with variable selection. I also discuss some of the theoretical properties of model averaging methods.

\item To discuss, in some detail, key operational aspects of the use of model averaging. In particular, the paper covers the various commonly used numerical methods to implement model averaging (both Bayesian and frequentist) in practical situations, which are often characterized by very large model spaces. For BMA, it is important to understand that the weights (based on posterior model probabilities) are typically quite sensitive to the prior assumptions, in contrast to the usually much more robust results for the model parameters given a specific model. In addition, this sensitivity does not vanish as the sample size grows \citep{kassraftery1995,BergerPericchi_01}. Thus, a good understanding of the effect of (seemingly arbitrary) prior choices is critical.

\item To review and discuss how model averaging has already made a difference in economics. The paper lists a number of, mostly recent, applications of model averaging methods in economics, and presents some detail on a number of areas where model averaging has furthered our understanding of economic phenomena. For example, I  highlight the contributions to growth theory, where BMA has been used to shed light on the relative importance of the three main growth theories (geography, integration and institutions) for development, as well as on the existence of the so-called natural resource curse for growth; the use of BMA in combining inference on impulse responses from models with very different memory characteristics; the qualification of the importance of established early warning signals for economic crises; the combination of inference on production or cost efficiencies through different models, etc.
    Model averaging provides a natural common framework in which to interpret the results of different empirical  analyses and as such should be an important tool for economists to resolve differences.

\item To provide sensible recommendations for empirical researchers about which modelling framework to adopt and how to implement these methods in their own research. In the case of BMA, I recommend the use of prior structures that are easy to elicit and are naturally robust. I include a separate section on freely available computational resources that will allow applied researchers to try out these methods on their own data, without having to incur a prohibitively large investment in implementation. In making  recommendations, it is inevitable that one draws upon personal experiences and preferences, to some extent. Thus, I present the reader with a somewhat subjective point of view, which I believe, however, is well-supported by both theoretical and empirical results.

\end{itemize}



Given the large literature, and in order to preserve a clear focus, it is important to set some limits to the coverage of the paper. As already explained above, the paper deals mostly with covariate uncertainty in regression models, and does not address issues like the use of BMA in classification trees \citep{Hernandez_etal_15} or in clustering and density estimation \citep{Russell_etal_15}. The large literature in machine learning related to nonparametric approaches to covariate uncertainty \citep{Hastie_etal_09} will also largely be ignored. The present paper focuses on averaging over (mostly nontrivial) models as a principled and formal statistical response to model uncertainty and does not deal with data mining or machine learning approaches, as further briefly discussed in Subsection \ref{Sec:NLM}.
In addition, this paper considers situations where the number of observations exceeds the number of potential covariates as this is most common in economics (some brief comments on the opposite case can be found in footnote \ref{fn:kgtn}).

As mentioned above, I discuss Bayesian and frequentist approaches to model averaging. This paper is mostly concerned with the Bayesian approach for the following reasons: \begin{itemize}
\item BMA benefits from a number of appealing statistical properties, such as point estimators and predictors that  minimize prior-weighted Mean Squared Error (MSE), and the calibration of the associated intervals \citep{RafteryZheng_03}. In addition, probabilistic prediction is optimal in the log score sense. Furthermore, BMA is typically consistent and is shown to display optimal shrinkage in high-dimensional problems. More details on these properties and the conditions under which they hold can be found in Subsection \ref{sec:propertiesBMA}.
\item Computationally, BMA is much easier to implement in large model spaces than FMA, since efficient MCMC algorithms are readily available.
\item In contrast to FMA methods, BMA immediately leads to readily interpretable posterior model probabilities and probabilities of inclusion of possible determinants in the model.
\item I personally find the finite-sample and probability-based nature of the Bayesian approach very appealing. 
    I do realize this is, to some extent, a personal choice, but I prefer to operate within a 
    methodological framework that immediately links to prediction and decision theory. 
\item There is a large amount of recent literature using the Bayesian approach to resolve model uncertainty, both in statistics and in many areas of application, among which economics features rather prominently. Thus, this focus on Bayesian methods is in line with the majority of the literature 
    and seems to reflect the perceived preference of many researchers in economics.

\end{itemize}

Of course, as \cite{Wright_08} states: ``One does not have to be a subjectivist Bayesian to believe
in the usefulness of BMA, or of Bayesian shrinkage techniques
more generally. A frequentist econometrician can interpret these
methods as pragmatic devices that may be useful for out-of-sample
forecasting in the face of model and parameter uncertainty.'' A comprehensive overview of model averaging from a mostly frequentist perspective (but also discussing BMA) can be found in \cite{Fletcher_18}.

This paper is organised as follows: in Section \ref{Sec:model uncertainty} I discuss the issue of model uncertainty and the way it can naturally be addressed through BMA. This section also comments on the construction of the model space and introduces the specific context of covariate uncertainty in the normal linear model. Section \ref{BMA} provides a detailed account of BMA, focusing on the prior specification, its properties and its implementation in practice. This section also provides a discussion of various generalizations of the sampling model and of a number of more challenging models, such as dynamic models and models with endogenous covariates. Section \ref{FMA} describes FMA, its computational implementation, and its links with forecast combinations. Section \ref{Sec:Appl} mentions some of the literature where model averaging methods have been applied in economics and discusses how model averaging methods have contributed to our understanding of a number of economic issues. In Section \ref{Sec:software} some freely available computational resources are briefly discussed, and the final section concludes.  

\section{Model uncertainty} \label{Sec:model uncertainty}

It is hard to overstate the importance of model uncertainty for economic modelling. Almost invariably, empirical work in economics will be subject to a large amount of  uncertainty about model specifications.  
This may be the consequence of the existence of many different theories\footnote{Or perhaps more precisely, the lack of a universally accepted theory, which has been empirically verified as a (near) perfect explanation of reality, clearly a chimera in the social sciences.}  or of many different ways in which theories can be implemented in empirical models (for example, by using various possible measures of theoretical concepts or various functional forms) or of other aspects such as assumptions about heterogeneity or independence of the observables. It is important to realize that this uncertainty is an inherent part of economic modelling, whether we acknowledge it or not. Putting on blinkers and narrowly focusing on a limited set of possible models implies that we may fail to capture important aspects of economic reality.
Thus, model uncertainty affects virtually all modelling in economics and its consequences need to be taken into account. There are two main strategies that have been employed in the literature:
\begin{itemize}
\item Model selection: such methods attempt to choose the best of all models considered, according to some criterion. Examples of this abound and some of the main model selection strategies used in the context of a linear regression model are briefly described in Subsection \ref{Sec:NLM}. The most important common characteristic of model selection methods is that they choose a model and then conduct inference conditionally upon the assumption that this model actually generated the data. So these methods only deal with the uncertainty in a limited sense: they try to select the ``best'' model, and their inference can only be relied upon if that model happens to be (a really good approximation to) the data generating process. In the much more likely case where the best model captures some aspects of reality, but there are other models that capture other aspects, model selection implies that our inference is almost always misleading, either in the sense of being systematically wrong or overly precise. Model selection methods simply condition on the chosen model and ignore all the evidence contained in the alternative models, thus typically leading to underestimation of the uncertainty.
\item Model averaging: here we take into account all the models contained in the model space we consider (see Subsection \ref{sec:modelspace}) and our inference is averaged over all these models, using weights that are either derived from Bayes' theorem (BMA) or from sampling-theoretic optimality considerations (FMA). This means our inference takes into account a possible variation across models and its precision is adjusted for model uncertainty. Averaging over models is a very natural response to model uncertainty, especially in a Bayesian setting, as explained in some detail later in this section.
\end{itemize}

As it is unlikely that reality (certainly in the social sciences) can be adequately captured by any single model, it is often quite risky to rely on a single selected model for inference, forecasts and (policy) conclusions. It is much more likely that an averaging method gives a better approximation to reality and it will almost certainly improve our estimate of the uncertainty associated with our conclusions.

One could argue that the choice between model selection and model averaging methods boils down to the underlying question that one is interested in answering. If that question relates to identifying the ``true'' model within a model space that is known to contain the data generating process, then model selection might be the appropriate strategy. However, if the question relates to, for example, the effect of primary education on GDP growth, then there is no reason at all to artificially condition the inference on choosing a single model. More precisely, the choice between model averaging and model selection is related to the decision problem that we aim to solve. In most typical situations, however, the implicit loss function we specify will lead to model averaging. Examples are where we are interested in maximizing accuracy of prediction or in the estimation of covariate effects. So it makes sense to use model averaging, 
whenever we are (as usual) interested in quantities that are not model-specific. 
Within economics, we can immediately identify three broad and important categories of questions that are not related to specific models:
\begin{itemize}
\item Prediction. Here we are interested in predicting an observable quantity (for example, a country's GDP,  growth or inflation, a company's sales or a person's wages) and we clearly do not wish to condition our predictive inference on any particular model. The latter would not be a natural question to ask and would, almost invariably, lead to biased or overconfident predictions. The discussion in Subsection \ref{sec:averaging} shows that a Bayesian predictive distribution naturally leads to model averaging. There is a long history in economics of using averaging for forecasting, some of which is discussed in Section \ref{FMA} (particularly Subsection \ref{sec:CombForecasts}). Subsection \ref{sec:output} lists some examples of the use of model averaging in forecasting output or inflation.
\item Identifying the factors or determinants driving economic processes. An example, which is discussed in more detail in Subsection \ref{sec:institutions}, concerns the empirical evidence for the three main types of economic growth determinants traditionally mentioned in the literature: geography, integration (trade) and institutions (often linked to property rights and rule of law). Earlier influential papers in growth theory have tended to consider only a limited number of possible models, focusing on a particular theory but without adequately covering possible alternative theories. This led \cite{Acemoglu_etal_01} and \cite{Rodrik_etal_04} to conclude that the quality of institutions is the only robust driver of development, while \cite{FrankelRomer_99} find that trade is the dominating determinant. Analyses using BMA in \cite{Lenkoski_etal_14} and \cite{EicherNewiak_13} lead to much more balanced conclusions, where all three main theories are seen to be important for growth. This highlights the  importance of accounting for a large enough class of possible models and dealing with model uncertainty in a principled and statistically sound manner.
\item Policy evaluation, where the focus is on assessing the consequences of certain policies. In the context of the evaluation of macroeconomic policy, \cite{Brock_etal} describe and analyse some approaches to dealing with the presence of uncertainty about the structure of the economic environment under study. Starting from a decision-theoretic framework, they recommend model averaging as a key tool in tackling uncertainty. \cite{BrockDurlauf_15} specifically focus on policy evaluation and provide an overview of different approaches, distinguishing between cases in which the analyst can and cannot provide conditional probabilities for the effects of policies. As an example, \cite{Durlauf_etal_12b} examine the effect of different substantive assumptions about the homicide process
on estimates of the deterrence effect of capital punishment\footnote{A systematic investigation of this issue goes back to \cite{Leamer83}.}. Considering four different types of model uncertainty, they find a very large spread of effects, with the estimate of net lives saved per execution ranging from -63.6 (so no deterrence effect at all) to 20.9. The latter evidence was a critical part of the National Academy of Sciences report that concluded there is no evidence in favour of or against a deterrent effect of capital punishment.
This clearly illustrates that the issue of model uncertainty needs to be addressed before we can answer questions such as this and many others of immediate relevance to society.

\end{itemize}

As already mentioned, one important and potentially dangerous consequence of neglecting model uncertainty, either by only considering one model from the start or by choosing a single model through model selection, is that we assign more precision to our inference than is warranted by the data, and this leads to overly confident decisions and predictions. In addition, our inference can be severely biased. See \cite{Chatfield} and \cite{Draper} for extensive discussions of model uncertainty.  

Over the last decade, there has been a rapidly growing awareness of the importance of dealing with model uncertainty for economics. As examples, the {\it European Economic Review} has recently  published a special issue on ``Model Uncertainty in Economics'' which was also the subject of the 2014 Schumpeter lecture in \cite{Marinacci_15}, providing a decision-theory perspective. In addition, a book written by two Nobel laureates in economics \citep{HansenSargent_14}, focuses specifically on the effects of model uncertainty on rational expectations equilibrium concepts.  

\subsection{Why averaging?}\label{sec:averaging}

In line with probability theory, the formal Bayesian response to dealing with uncertainty is to average. When dealing with parameter uncertainty, this involves averaging over parameter values with the posterior distribution of that parameter in order to get the predictive distribution. Analogously, model uncertainty is also resolved through averaging, but this time averaging over models with the (discrete) posterior model distribution. The latter procedure is usually called BMA and was already described in \cite{leamer78} and later used in \cite{MinZellner}, \cite{OsiewalskiSteel_93}, 
\cite{Koop_etal_97} and \cite{Raftery_etal_97}. BMA thus appears as a direct consequence of Bayes' theorem (and hence probability laws) in a model uncertainty setting and is perhaps best introduced by considering the concept of a predictive distribution, often of interest in its own right. In particular, assume we are interested in predicting the unobserved quantity $y_f$ on the basis of the observations $y$. Let us denote the sampling model\footnote{For ease of notation, I will assume continuous sampling models with real-valued parameters throughout, but this can immediately be extended to other cases.} for $y_f$ and $y$ jointly by $p(y_f|y,\theta_j,M_j)p(y|\theta_j,M_j)$, where $M_j$ is the model selected from a set of $K$ possible models, and $\theta_j\in\Theta_j$ groups the (unknown) parameters of $M_j$. In a Bayesian framework, any uncertainty is reflected by a probability distribution\footnote{Or, more generally, a measure.} so we assign a (typically continuous) prior $p(\theta_j|M_j)$ for the parameters and a discrete prior $P(M_j)$ defined on the model space. We then have all the building blocks to compute the predictive distribution as
\begin{equation}\label{pred}
p(y_f|y)=\sum_{j=1}^K \left[\int_{\Theta_j} p(y_f|y,\theta_j,M_j) p(\theta_j|y,M_j) \d\theta_j \right] P(M_j|y),
\end{equation}
where the quantity in square brackets is the predictive distribution given $M_j$ obtained using the posterior of $\theta_j$ given $M_j$, which is computed as
\begin{equation}\label{post}
p(\theta_j|y,M_j)=\frac{p(y|\theta_j,M_j) p(\theta_j|M_j)}{\int_{\Theta_j}p(y|\theta_j,M_j) p(\theta_j|M_j) \d\theta_j} \equiv \frac{p(y|\theta_j,M_j) p(\theta_j|M_j)}{p(y|M_j)},
\end{equation}
with the second equality defining $p(y|M_j)$, which is used in computing the posterior probability assigned to $M_j$ as follows:
\begin{equation}\label{postmodel}
P(M_j|y)=\frac{p(y|M_j) P(M_j)}{\sum_{i=1}^K p(y|M_i) P(M_i)}\equiv \frac{p(y|M_j) P(M_j)}{p(y)}.
\end{equation}
Clearly, the predictive in (\ref{pred}) indeed involves averaging at two levels: over (continuous) parameter values, given each possible model, and discrete averaging over all possible models. The denominators of  both averaging operations are not immediately obvious from (\ref{pred}), but are made explicit in (\ref{post}) and (\ref{postmodel}). The denominator (or integrating constant) $p(y|M_j)$ in (\ref{post}) is the so-called marginal likelihood of $M_j$ and is a key quantity for model comparison. In particular, the Bayes factor between two models is the ratio of their marginal likelihoods and the posterior odds are directly obtained as the product of the Bayes factor and the prior odds. The denominator in (\ref{postmodel}), $p(y)$, is defined as a sum and the challenge in its calculation often lies in the sheer number of possible models, i.e. $K$.

BMA as described above is thus the formal probabilistic way of obtaining predictive inference, and is, more generally, the approach to any inference problem involving quantities of interest that are not model-specific. So it is also the Bayesian solution to conducting posterior inference on {\it e.g.}~the effects of covariates or of a certain policy decision. Formally, the posterior distribution of any quantity of interest, say $\Delta$, which has a common interpretation across models  is a mixture of the model-specific posteriors with the posterior model probabilities as weights, {\it i.e.}
\begin{equation}
P_{\Delta|y}= \sum_{j=1}^K P_{\Delta \given y, M_j} P(M_j\given y).\label{BMAeq}
\end{equation}

\begin{figure}[h!]
\begin{center}
\subfigure{\includegraphics[width=1.0\textwidth]{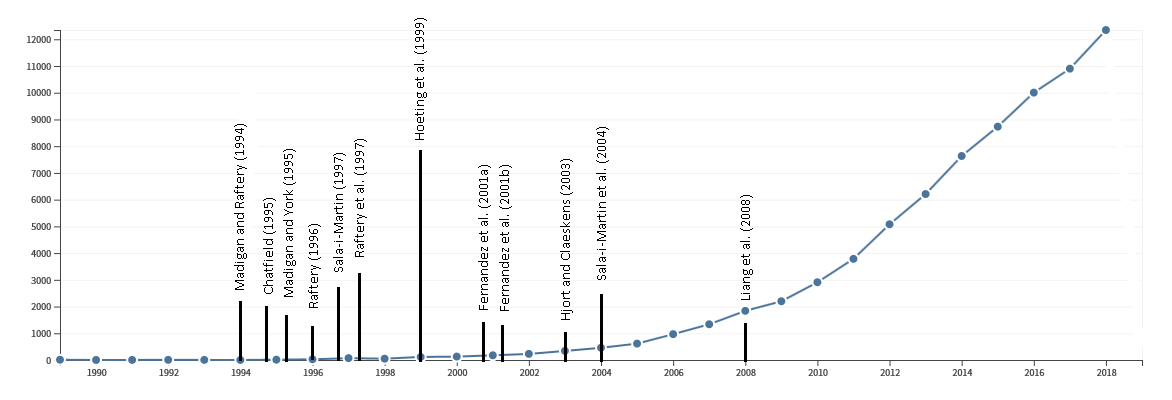}}
\caption{\footnotesize Total number of citations to papers with topic ``model averaging'' over years 1989-2018. 
Papers in economics or statistics journals with at least 250 citations are indicated by vertical lines proportional to the number of citations received. Source: Web of Science, January 29, 2019.} 
\label{citations}
\end{center}
\end{figure}

The rapidly growing importance of model averaging as a solution to model uncertainty is illustrated by Figure \ref{citations}, which plots the citation profile over time of papers with the topic ``model averaging'' in the literature. The figure also indicates influential papers (with 250 citations or more) published  in either economics or statistics journals\footnote{There are also some heavily cited papers on model averaging in a number of other application areas, in particular biology, ecology, sociology, meteorology, psychology and hydrology. The number of citations is, of course, an imperfect measure of influence and the cutoff at 250 leaves out a number of key papers, such as \cite{BrockDur01} and \cite{ClydeGeorge04} with both over 200 citations.}. A large part of the literature uses BMA methods, reflected in the fact that citations to papers with the topic ``Bayesian'' and ``model averaging'' account for more than 70\% of the citations in Figure \ref{citations}. 
The sheer number of recent papers in this area is evidenced by the fact that Google Scholar  returns over 52,000 papers in a search for ``model averaging'' and over 40,000 papers when searching for ``Bayesian'' and ``model averaging'', over half of which date from the last decade (data from January 29, 2019).

\subsection{Construction of the model space}\label{sec:modelspace}
An important aspect of dealing with model uncertainty is the precise definition of the space of all models that are being considered. The idea of model averaging naturally  assumes a well-defined space of possible models, over which the averaging takes place. This is normally a finite (but potentially very large) space of models, denoted by $\cal M$. There are also situations where we might consider an infinite space of models, for example when we 
consider data transformations of the response variable within a single family, such as the Box-Cox family\footnote{\cite{Hoeting_etal_02} use a number of specific values for the Box-Cox parameter, to aid interpretation, which gets us back to a finite model space}.  In these cases where models are indexed by continuous parameters, BMA is done by integration over these parameters and is thus perhaps less obvious. In other words, it is essentially a part of the standard Bayesian treatment of unknown parameters. Another example is given in \cite{Brock_etal}, who mention ``hierarchical models in which the parameters of a
model are themselves functions of various observables and unobservables. If these relationships
are continuous, one can trace out a continuum of models.'' Again, Bayesian analysis of hierarchical models is quite well-established.

In economics arguably the most common case of model uncertainty is where we are unsure about which covariates should be included in a linear regression model, and the associated model space is that constructed by including all possible subsets of covariates. This case is discussed in detail in the next subsection. A minor variation is where some covariates are always included and it is inclusion or exclusion of the ``doubtful'' ones that defines the
model space.
In order to carefully construct an appropriate model space, it is useful to distinguish various common types of uncertainty. \cite{Brock_etal} identify three main types of uncertainty that typically need to be considered:
\begin{itemize}
\item Theory uncertainty. This reflects the situation where economists disagree over fundamental aspects of the economy and is, for example, illustrated by the ongoing debates over which are important drivers for economic growth (see the discussion in Subsection \ref{sec:growth}) or what are useful early warning signals for economic crises (see Subsection \ref{sec:crises}).
\item Specification uncertainty. This type of uncertainty is about how the various theories that are considered will be implemented, in terms of how they are translated into specific models. Examples are the choice of available variables as a measure of theoretical constructs, the choice of lag lengths\footnote{Particularly relevant in {\it e.g.}~forecasting and VAR modelling in Subsections \ref{sec:output} and \ref{sec:VAR}.}, parametric versus semi- or nonparametric specifications, transformations of variables, functional forms (for example, do we use linear or non-linear models) and distributional assumptions (which also include assumptions about dependence of observables).
\item Heterogeneity uncertainty. This relates to model assumptions regarding different observations. Is the same model appropriate for all, or should the models include differences that are designed to accommodate observational  heterogeneity?  A very simple example would be to include dummies for certain classes of observations. Another example is given by \cite{Doppelhofer_etal_16} who introduce heterogeneous measurement error variance in growth regressions.

\end{itemize}

The definition of the model space is intricately linked with the model uncertainty that is being addressed. For example, if the researcher is unsure about the functional form of the models and about covariate inclusion, both aspects should be considered in building $\cal M$. Clearly, models that are not entertained in $\cal M$ will not contribute to the model-averaged inference and the researcher will thus be blind to any insights provided by these models. Common sense should be used in choosing the model space: if one wants to shed light on the competing claims of various papers that use different functional forms and/or different covariates, it would make sense to construct a model space that combines all functional forms considered (and perhaps more variations if they are reasonable) with a wide set of possibly relevant and available covariates. The fact that such spaces can be quite large should not be an impediment.\footnote{Certainly not  for a Bayesian analysis, where novel numerical methods have proven to be very efficient.} In practice, not all relevant model spaces used in model averaging analyses are large.  For example, to investigate the effect of capital punishment on the murder rate (see the discussion earlier in this section), \cite{Durlauf_etal_12b} build a bespoke model space by considering the following four model features: the probability model (linear or logistic regression), the specification of the covariates (relating to the probabilities of sentencing and execution), the presence of state-level heterogeneity, and the treatment of zero observations for the murder rate. In all, the model space they specify only contains 20 models, yet leads to a large range of deterrence effects. Another example of BMA with a small model space is the analysis of impulse response functions in \cite{Koop_etal_97}, who use two different popular types of univariate time series models with varying lag lengths, leading to averaging over only 32 models (see Subsection \ref{sec:output}). Here the 
model space only reflects specification uncertainty. An example of theory uncertainty leading to a model space with a limited number of models can be found in  \cite{Liu}, who compares WALS and various FMA methods on cross-country growth regressions. Following \cite{Magnus_etal_10}, \cite{Liu} always includes  a number of core regressors and allows for a relatively small number of auxiliary regressors. Models differ in the inclusion of the auxiliary regressors, leading to model spaces with sizes of 16 and 256.

It is important to distinguish between the case where  the model space contains the ``true'' data-generating model and the case where it does not. These situations are respectively referred to as $\cal M$-closed and $\cal M$-open in the statistical literature \citep{BernardoSmith_94}. Most theoretical results (such as consistency of BMA in Subsection \ref{cons}) are obtained in the simpler $\cal M$-closed case, but it is clear that in economic modelling the $\cal M$-open framework is a more realistic setting. Fortunately, model selection consistency results\footnote{As explained in Subsection \ref{cons}, model selection consistency is the property that the posterior probability of the data-generating model tends to unity with sample size in an $\cal M$-closed setting.} can often be shown to extend to $\cal M$-open settings in an intuitive manner \citep{Mukh_etal_15,Mukh_etal_17} and \cite{George_99b} states that ``BMA is well
suited to yield predictive improvements over single
selected models when the entire model class is
misspecified. In a sense, the mixture model elaboration
is an expansion of the model space to include
adaptive convex combinations of models. By incorporating
a richer class of models, BMA can better
approximate models outside the model class.'' A decision-theoretic  approach to implementing BMA in an $\cal M$-open environment is provided in \cite{ClydeIversen_13}, who treat models not as an extension of the parameter space, but as part of the action space. The main objection to using BMA in the $\cal M$-open framework is the perceived logical tension between knowing the ``true'' model is not in $\cal M$ and assigning a prior on the models in $\cal M$. However, in keeping with most of the literature, we will assume that the user is comfortable with assigning a prior on $\cal M$, even in  $\cal M$-open situations.\footnote{Personally, I prefer to think of the prior over models as a reflection of prior beliefs about which models would be ``useful proxies for'' (rather than ``equal to'') the data-generating process, so I do not feel the $\cal M$-open setting leads to a significant  additional challenge for BMA.}

\subsection{Covariate uncertainty in the normal linear regression model}\label{Sec:NLM}

Most of the relevant literature assumes the simple case of the normal linear sampling model. This helps tractability, and it is fortunately also a model that is often used in empirical work. In addition, it is a canonical version for nonparametric regression\footnote{A typical nonparametric regression approach is to approximate the unknown regression function for the mean of $y$ given $x$ as a linear combination of a finite number of basis functions of $x$.}, which is gaining in popularity. I shall follow this tradition, and will assume for most of the paper\footnote{Section \ref{Extension} explores some important extensions, {\it e.g.}~to the wider class of Generalized Linear Models (GLMs) and some other modelling environments that deal with specific challenges in economics.} that the sampling model is normal with a mean which is a linear function of some covariates\footnote{This is not as restrictive as it may seem. It certainly does not mean that the effects of determinants on the modelled phenomenon are linear; we can simply include regressors that are nonlinear transformations of determinants, interactions etc.}.
I shall further assume, again in line with the vast majority of the literature (and many real-world applications) that the model uncertainty relates to the choice of which covariates should be included in the model, {\it i.e.}~under model $j$ the $n$ observations in $y$ are generated from
\begin{equation}
y|\theta_j,M_j \sim N(\alpha\iota+Z_j\beta_j, \sigma^2). \label{NLM}
\end{equation}
Here $\iota$ represents a $n\times 1$-dimensional vector
of ones, $Z_j$ groups $k_j$ of the possible $k$ regressors ({\it i.e.}~it selects $k_j$ columns from an $n\times k$ matrix $Z$, corresponding to the full model) and $\beta_j\in\Re^{k_j}$ are its corresponding regression coefficients. Furthermore, all considered models contain
an intercept $\alpha\in\Re$ and the scale $\sigma>0$ has a common interpretation across all models. I standardize the regressors by subtracting their means, which makes them orthogonal to the intercept and renders the interpretation of the intercept common to all models. The model space is then formed by all possible subsets of the covariates and thus contains $K=2^k$ models in total\footnote{This can straightforwardly be changed to a (smaller) model space where some of the regressors are always included in the models.}. Therefore, the model space includes the null model (the model with only the intercept and $k_j=0$) and the full model (the model where $Z_j=Z$ and $k_j=k$). This definition of the model space is consistent with the typical situation in economics, where theories regarding variable inclusion do not necessarily contradict each other. \cite{BrockDur01} refer to this as the ``open-endedness'' of the theory\footnote{In the context of growth theory, \cite{BrockDur01} define this concept as ``the idea that the validity of one causal theory
of growth does not imply the falsity of another. So, for example, a causal relationship
between inequality and growth has no implications for whether a causal relationship
exists between trade policy and growth.''\label{fn:openended}}. Throughout, the matrix formed by adding a column of ones to $Z$ is assumed to have full column rank\footnote{For economic applications this is generally a reasonable assumption, as typically $n>k$, although they may be of similar orders of magnitude. In other areas such as genetics this is usually not an assumption we can make. However, it generally is enough that for each model we consider to be a serious contender the matrix formed by adding a column of ones to $Z_j$ is of full column rank, and that is much easier to ensure. Implicitly, in such situations we would assign zero prior and posterior probability to models for which $k_j\ge n$. Formal approaches to use $g$-priors in situations where $k>n$ include \cite{MaruyamaGeorge_11} and \cite{Berger_etal_16}, based on different ways of generalizing the notion of inverse matrices.\label{fn:kgtn}}.

This model uncertainty problem is very relevant for empirical work, especially in the social sciences where typically competing theories abound on which are the important determinants of a phenomenon.
Thus, the issue has received quite a lot of attention both in statistics and economics, and various approaches have been suggested. We can mention:

\begin{itemize}

\item[1.] Stepwise regression: this is a sequential procedure for entering and deleting variables in a regression model based on some measure of ``importance'', such as the $t$-statistics of their estimated coefficients (typically in ``backwards'' selection where covariates are considered for deletion) or (adjusted) $R^2$ (typically in ``forward'' selection when candidates for inclusion are evaluated).

\item[2.] Shrinkage methods: these methods aim to find a set of sparse solutions (i.e. models with a reduced set of covariates) by shrinking coefficient estimates toward zero. Bayesian shrinkage methods rely on the use of shrinkage priors, which are such that some of the estimated regression coefficients in the full model will be close to zero. 
    A common classical method is    penalized least squares, such as LASSO (least absolute shrinkage and selection operator), introduced by \cite{Tibshirani}, where the regression ``fit'' is maximized subject to a complexity penalty. Choosing a different penalty function, \cite{FanLi01} propose the smoothly clipped absolute deviation (SCAD) penalized regression estimator.

\item[3.] Information criteria: these criteria can be viewed as the use of the classical likelihood ratio principle combined with penalized likelihood (where the penalty function depends on the model complexity). A common example is the Akaike information criterion (AIC). The Bayesian information criterion (BIC) implies a stronger complexity penalty and was originally motivated through asymptotic equivalence with a Bayes factor \citep{Schwarz78}. Asymptotically, AIC selects a single model that minimizes the mean squared error of prediction. BIC, on the other hand, chooses the ``correct'' model with probability tending to one as the sample size grows to infinity if the model space contains a true model of finite dimension. So BIC is consistent in this setting, while AIC has better asymptotic behaviour if the true model is of infinite dimension\footnote{A careful classification of the asymptotic behaviour of BIC, AIC and similar model selection criteria can be found in \cite{Shao_97} and its discussion.}. \cite{Spiegelhalter_etal02} propose the Deviance information criterion (DIC) which can be interpreted as a Bayesian generalization of AIC.\footnote{DIC is quite easy to compute in practice, but has been criticized for its dependence on the parameterization and its lack of consistency.}


\item[4.] Cross-validation: the idea here is to use only part of the data for inference and to assess how well the remaining observations are predicted by the fitted model. This can be done repeatedly for random splits of the data and models can be chosen on the basis of their predictive performance.

\item[5.] Extreme Bounds Analysis (EBA): this procedure was proposed in \cite{Leamer83,Leamer85} and is based on distinguishing between ``core'' and ``doubtful'' variables. Rather than a discrete search over models that include or exclude subsets of the variables, this sensitivity analysis answers the question: how extreme can the estimates be if any linear homogenous restrictions on a selected subset of the coefficients (corresponding to doubtful covariates) are allowed? An extreme bounds analysis chooses the linear combinations of doubtful variables that, when included along with the core variables, produce the most extreme estimates for the coefficient on a selected core variable. If the extreme bounds interval is small enough to be useful, the coefficient of the core variable is reported to be ``sturdy''. A useful discussion of EBA and its context in economics can be found in \cite{ChristensenMiguel_18}.

\item[6.] $s$-values: proposed by \cite{Leamer_16_EER,Leamer_16_JE} as a measure of ``model ambiguity''. 
    Here $\sigma$ is replaced by the ordinary least squares (OLS) estimate and no prior mass points at zero are assumed for the regression coefficients. 
    For each coefficient, this approach finds the interval bounded by the extreme estimates (based on different prior variances, elicited through $R^2$); the $s$-value ($s$ for sturdy) then summarizes this interval of estimates in the same way that a $t$-statistic summarizes a confidence interval (it simply reports the centre of the interval divided by half its width). A small $s$-value then indicates fragility of the effect of the associated covariate, by measuring the extent to which the sign of the estimate of a regression coefficient depends on the choice of model. 

\item[7.] General-to-specific modelling: this approach starts from a general unrestricted model and uses a pre-selected set of misspecification tests as well as individual $t$-statistics to reduce the model  to a parsimonious representation. I refer the reader to \cite{HooverPerez_99} and \cite{HendryKrolzig_05} for background and details. \cite{HendryKrolzig_04} present an application of this technique to the cross-country growth dataset of \cite{FLS01b} (``the FLS data'', which record average per capita GDP growth over 1960-1992 for $n=72$ countries with $k=41$ potential regressors).

\item[8.] The Model Confidence Set (MCS): this approach to model uncertainty consists in constructing a set of models such that it will contain the
best model with a given level of confidence. This was introduced by \cite{Hansen_etal_11} and only requires the specification of a collection of competing objects (model space) and a
criterion for evaluating these objects empirically.
The MCS is constructed through a sequential testing procedure, where an equivalence test determines whether all objects in the current set are equally good. If not, then an elimination rule is used to delete an underperforming object. The same significance level is used in all tests, which allows one to control the $p$-value of the resulting set and each of its elements. The appropriate critical values of the tests are determined by bootstrap procedures. \cite{Hansen_etal_11} apply their procedure to e.g.~US inflation forecasting, and \cite{WeiCao_17} use it for modelling Chinese house prices, using predictive elimination criteria.

\item [9.] Best subset regression of \cite{Hastie_etal_09}, called full subset regression in \cite{Hanck_16}. This method considers all possible models: for a given model size $k_j$ it selects the best in terms of fit (the lowest sum of squared residuals).
As all these models have $k_j$ parameters, none has an unfair advantage over the others using
this criterion. Of the resulting set of optimal models of a given dimension, the procedure then
chooses the one with the smallest value of some criterion such as Mallows' $C_p$ \footnote{Mallows' $C_p$ was developed for selecting a subset of regressors in linear regression problems. For model $M_j$ with $k_j$ parameters $C_p = \frac{SSE_j}{\hat{\sigma}^2} -n+ 2k_j$ where $SSE_j$ is the error sum of squares from $M_j$ and $\hat{\sigma}^2$ the estimated error variance. $E(C_p)=k_j$ (approximately) and regressions with low $C_p$ are favoured.\label{fn:Cp}}. \cite{Hanck_16} does a small simulation exercise to conclude that log runtime for complete enumeration methods is roughly linear in $k$, as expected. Using the FLS data and a best subset regression approach which uses a leaps and bounds algorithm (see Section \ref{sec:numerical}) to avoid complete enumeration of all models, he finds that the best model for the FLS data has 22 (using $C_p$) or 23 (using BIC) variables. These are larger model sizes than indicated by typical BMA results on these data\footnote{For example, using the prior setup later described in (\ref{FLSprior}) with fixed $g$, \cite{LS6} find the models with highest posterior probability to have between 5 and 10 regressors for most prior choices. Using random $g$, the results in \cite{leysteel2012} indicate that a typical average model size is between 10 and 20.}.

\item[10.] Bayesian variable selection methods based on decision-theory. Often such methods avoid specifying a prior on model space and employ a utility or loss function defined on an all-encompassing model, {\it i.e.}~a model that nests all models being considered. An early contribution is \cite{Lindley_68}, who proposes to include costs in the utility function for adding covariates, while \cite{Brown_etal_99} extend this idea to multivariate regression. Other Bayesian model selection procedures that are based on optimising some loss or utility function can be found in {\it e.g.}~\cite{GelfandGhosh_98}, \cite{DraperFouskakis_00} and \cite{DupuisRobert_03}. Note that decision-based approaches do need the specification of a utility function, which is arguably at least as hard  to formulate as a model space prior.

\item[11.] BMA, discussed here in detail in Section \ref{BMA}.

\item[12.] FMA, discussed in Section \ref{FMA}.

\end{itemize}

In this list, methods 5-8 were specifically motivated by and introduced in economics.
Note that all but the last two methods do not involve model averaging and essentially aim at uncovering a single ``best'' model (or a set of models for MCS). In other words, they are model selection methods, as opposed to methods for model averaging, which is the focus here. As discussed before, model selection strategies condition the inference on the chosen model and ignore all the evidence contained in the alternative models, thus typically leading to an underestimating of the uncertainty.
BMA methods can also be used for model selection, by {\it e.g.}~simply selecting the model with the highest posterior probability\footnote{Another possibly interesting model is the median probability model of \cite{BarbieriBerger}, which is the model including those covariates which have marginal posterior inclusion probabilities of 0.5 or more. This is the best single model
for prediction in orthogonal and nested correlated designs under commonly used priors.}. Typically, the opposite is not true as most model selection methods do not specify prior probabilities on the model space and thus can not provide posterior model probabilities.

Some model averaging methods in the literature combine aspects of both frequentist and Bayesian reasoning. Such hybrid methods will be discussed along with BMA if they employ a prior over models (thus leading to posterior model probabilities and inclusion probabilities of covariates), and in the FMA section if they do not. Thus, for example BACE (Bayesian averaging of classical estimates) of \cite{SDM} will be discussed in Section \ref{BMA} (Subsection
\ref{Sec:BIC}) and weighted average least squares (WALS) of \cite{Magnus_etal_10} is explained in Section \ref{FMA}. As a consequence, all methods discussed in Section \ref{BMA} can be used for model selection, if desired, while the model averaging methods in Section \ref{FMA} can not lead to model selection.

Comparisons of some methods (including the method by \cite{Benjamini_H_95} aimed at controlling the false discovery rate) can be found in \cite{DeckersHanck_14}  in the context of cross-sectional growth regression. \cite{Blazejowski_etal_18} replicate the long-term UK inflation model (annual data for 1865-1991) obtained through general-to-specific principles in \cite{Hendry_01} and compare this with the outcomes of BACE (using $k=20$). They find that the single model selected in \cite{Hendry_01} contains all variables that were assigned very high posterior inclusion probabilities in BACE. However, by necessity, the model selection procedure of \cite{Hendry_01} conditions the inference on a single model, which has a posterior probability of less than 0.1 in the BACE analysis (it is the second most probable model, with the top model obtaining 20\% of the posterior mass).

\cite{Wangetal09} claim that there are model selection methods that automatically incorporate model uncertainty by selecting variables
and estimating parameters simultaneously. Such  approaches are {\it e.g.}~the SCAD penalized regression of \cite{FanLi01} and adaptive LASSO methods as in \cite{Zou06}. These methods sometimes possess the
so-called oracle property\footnote{The oracle property implies that an estimating procedure identifies the ``true'' model asymptotically if the latter is part of the model space and has the optimal square root convergence rate. See \cite{FanLi01}.}.
However, the oracle property is asymptotic and assumes that the ``true'' model is one of the models considered (the ${\cal M}$-closed setting). So in the practically much more relevant context of finite samples and with true models (if they can even be formulated) outside the model space these procedures will very likely still underestimate uncertainty.


Originating in machine learning,  a number of algorithms aim to construct a prediction model by combining
the strengths of a collection of simpler base models, like random forests, boosting
or bagging \citep{Hastie_etal_09}. As these methods typically exchange the neat, possibly structural, interpretability of a
simple linear specification for the flexibility of nonlinear and nonparametric models and cannot provide probability-based uncertainty intervals, I do not consider them in this article. Various machine learning algorithms use model averaging ideas, but they are quite different from the model averaging methods discussed in this paper in that they tend to focus on combining ``poor'' models, since weak base learners can be boosted to lower predictive errors than strong learners \citep{Hastie_etal_09}, they work by always combining a large number of models and their focus is purely predictive, rather than on parameter estimation or the identification of structure.
In line with their main objective, they do often provide good predictive performance, especially in classification problems\footnote{\cite{Domingos_00} finds that BMA can fail to beat the machine learning methods in classification problems, and conjectures that this is a consequence of BMA ``overfitting'', in the sense that the sensitivity of the likelihood to small changes in the data carries over to the weights in (\ref{BMAeq}).}. An intermediate method was proposed in \cite{Hernandez_etal_15}, who combine elements of both Bayesian additive regression trees and random forests, to offer a model-based algorithm which can deal with high-dimensional data.
For discussions on the use of machine learning methods in economics, see \cite{Varian_14}, \cite{KapetaniosPap_18} and \cite{Korobilis_18}.




\section{Bayesian model averaging} \label{BMA}

The formal Bayesian response to model uncertainty is BMA, as already explained in Section \ref{sec:averaging}. Here, BMA methods are defined as those model averaging procedures for which the weights used in the averaging are based on exact or approximate posterior model probabilities and the parameters are integrated out for prediction, so there is a 
prior for both models and model-specific parameters.

\subsection{Prior Structures}\label{sec:prior}

As we will see, prior assumptions can be quite important for the final outcomes, especially for the posterior model  probabilities used in BMA. Thus, a reasonable question is whether one can assess the quality of priors or limit the array of possible choices. Of course, the Bayesian paradigm prescribes a strict separation between the information in the data being analysed and that used for the prior\footnote{This is essentially implicit in the fact that the prior times the likelihood should define a joint distribution on the observables and the model parameters, so that {\it e.g.}~the numerator in the last expression in (\ref{post}) is really $p(y,\theta_j|M_j)$ and we can use the tools of probability calculus.}. 
In principle, any coherent\footnote{This means the prior is in agreement with the usual rules of probability, and prevents ``Dutch book'' scenarios, which would guarantee a profit in a betting setting, irrespective of the outcome.} prior which does not use the data can be seen as ``valid''.
Nevertheless, there are a number of legitimate questions one could (and, in my view, should) ask about the prior:
\begin{itemize}
\item Does it adequately capture the prior beliefs of the user? Is the prior a ``sensible'' reflection of prior ideas, based on aspects of the model that can be interpreted? This could, for example, be assessed through (transformations of) parameters or predictive quantities implied by the prior. At the price of making the prior data-dependent, priors can even be judged on the basis of posterior results. \cite{Leeper_etal_96} introduce the use of priors in providing appropriate structure for Bayesian VAR modelling and propose the criterion ``reasonableness of results'' as a general desirable property of priors. They state that ``Our procedure differs from the standard practice of empirical researchers in economics only in being less apologetic. Economists adjust their models until they both fit the data and give `reasonable' results. There is nothing unscientific or dishonest about this. It would be unscientific or dishonest to hide results for models that fit much better than the one presented (even if the hidden model seems unreasonable), or for models that fit about as well as the one reported and support other interpretations of the data that some readers might regard as reasonable.''

\item Does it matter for the results? If inference and decisions regarding the question of interest are not much affected over a wide range of ``sensible'' prior assumptions, it indicates that you need not spend a lot of time and attention to finesse these particular prior assumptions. This desirable characteristic is called ``robustness'' in \cite{Brock_etal}. Unfortunately, when it comes to model averaging, the prior is often surprisingly important, and then it is important to find structures that enhance the robustness, such as the hierarchical structures in Subsections \ref{Sec:PriorModel} and \ref{EB_hier}.

\item What is the predictive ability (as measured by {\it e.g.}~scoring rules)? The immediate availability of probabilistic forecasts that formally incorporate both parameter and model uncertainty provides us with a very useful tool for checking the quality of the model. If a Bayesian model predicts unobserved data well, it reflects well upon both the likelihood and the prior components of this model. Subsection \ref{Pred} provides more details in the context of model averaging.

\item Are the desiderata of \cite{Bayarri_etal_12} for ``objective'' priors satisfied? These key theoretical principles, such as consistency and invariance, can be used to motivate the main prior setup in this paper. Here I focus on the most commonly used prior choices, based on (\ref{FLSprior}) introduced in the next subsection. These prior structures have been shown \citep{Bayarri_etal_12} to  possess very useful properties. 
    For example, they are measurement and group invariant and satisfy exact predictive matching.\footnote{See \cite{Bayarri_etal_12} for the precise definition of these criteria.}

\item What are the frequentist properties of the resulting Bayesian procedure? Even though frequentist arguments are, strictly speaking, not part of the underlying rationale for Bayesian inference, these procedures often perform well in repeated sampling experiments, and BMA is not an exception\footnote{However, frequentist performance necessarily depends on the assumptions made about the ``true'' data generating model, so there is no guarantee that BMA will do well in all situations and, for example, there is anecdotal evidence that it can perform worse in terms of, say, mean squared error than simple least squares procedures for situations with small $k$.}. This is discussed in Subsection \ref{sec:freqpropBMA}.

\item Can it serve as a benchmark? This is mentioned in \cite{Brock_etal}, who argue that priors ``should be flexible enough to allow for their use across similar studies and thereby facilitate comparability of results.'' \cite{Leeper_etal_96} use the terminology ``reference prior''\footnote{In the statistical literature, this name is typically given to a prior which is somewhat similar in spirit but derived from a set of precise rules designed to minimize the information in the prior; see \cite{BernardoSmith_94}.\label{fn:refprior}} as a prior which ``only reflects a simple summary of beliefs that are likely to be uncontroversial across a wide range of users of the analysis.''

\end{itemize}

\subsubsection{Priors on model parameters}

When deciding on the priors for the model parameters, i.e. $p(\theta_j|M_j)$ in (\ref{post}), it is important to realize that the prior needs to be proper on model-specific parameters. Indeed, any arbitrary constant in $p(\theta_j|M_j)$ will similarly affect the marginal likelihood $p(y|M_j)$ defined in (\ref{post}). Thus, if this constant emanating from an improper prior multiplies $p(y|M_j)$ and not the marginal likelihoods for all other models, it clearly follows from (\ref{postmodel}) that posterior model probabilities are not determined. If the arbitrary constant relates to a parameter that is common to all models, it will simply cancel in the ratio (\ref{postmodel}), and for such parameters we can thus employ improper priors \citep{FLS01a,BergerPericchi_01}. 
In our normal linear model in (\ref{NLM}), the common parameters are the intercept $\alpha$ and the variance $\sigma^2$, and the model-specific parameters are the $\beta_j$s.

This paper will primarily focus on the prior structure proposed by \cite{FLS01a}, which is in line with the majority of the current literature\footnote{Textbook treatments of this approach can be found in Chapter 11 of \cite{Koop_03} and Chapter 2 of \cite{Fletcher_18}.}. \cite{FLS01a} start from a proper conjugate prior specification, but then   adopt Jeffreys-style non-informative priors for the common parameters $\alpha$ and $\sigma^2$. For the model-specific regression coefficients $\beta_j$, they propose a $g$-prior specification \citep{Zellner86} for the covariance structure\footnote{In line with most of the literature, in this paper $g$ denotes a variance factor rather than a precision factor as in \cite{FLS01a}. Interestingly, the $g$-prior appears earlier in the context of combining forecasts by \cite{DieboldPauly_90}, who use a regression-based forecast combination framework as a means to introduce shrinkage in the weights and adopt an empirical Bayes (see Subsection \ref{EB_hier}) approach to selecting $g$.}.  The prior density\footnote{For the null model, the prior is simply $p(\alpha,\sigma)\propto \sigma^{-1}$.} is then as follows:
\begin{equation}
p(\alpha,\beta_j,\sigma \,|\, M_j)\propto
\sigma^{-1}f_N^{k_j}(\beta_j\vert 0, \sigma^2 g
(Z_j'Z_j)^{-1}), \label{FLSprior}
\end{equation}
where $f_N^q(\cdot | m, V)$ denotes the density function of a
$q$-dimensional Normal distribution with mean $m$ and
covariance matrix $V$.  It is worth pointing out that the dependence of the $g$-prior on the design matrix is not in conflict with the usual Bayesian precept that the prior should not involve the data, since the model in (\ref{NLM}) is a model for $y$ given $Z_j$, so we simply condition on the regressors throughout the analysis.
The regression coefficients not appearing in
$M_j$ are exactly zero, represented by a prior point mass at zero. The amount of prior information requested from the user is limited to a single scalar $g>0$, which can either be fixed or assigned a hyper-prior distribution.
In addition, the marginal likelihood for each model (and thus the Bayes factor between each pair of models) can be calculated in closed form   \citep{FLS01a}.
In particular, the posterior distribution for the model parameters has an analytically known form as follows:
\begin{eqnarray}
p(\beta_j\given \alpha, \sigma, M_j) &=& f_N^{k_j}(\beta_j \given \delta (Z_j'Z_j)^{-1} Z_j'y, \sigma^2\delta(Z_j'Z_j)^{-1})\label{Eq:postbeta}\\
p(\alpha\given \sigma, M_j)&=& f_N^1(\alpha \given \bar y, \sigma^2/n)\\
p(\sigma^{-2}\given M_j) &=& f_{Ga}\left(\sigma^{-2} \given \frac{n-1}{2}, \frac{s_\delta}{2}\right),
\end{eqnarray}
where $\delta=g/(1+g)$, $\bar y=\frac{1}{n}\sum_{i=1}^n y_i$, $s_\delta=\left[\delta y'Q_{X_j}y + (1-\delta) (y-\bar y \iota)'(y- \bar y \iota)\right]$ with $Q_W=I_n-W(W'W)^{-1}W'$ for a full column rank matrix $W$ and $X_j=(\iota : Z_j)$ (assumed of full column rank, see footnote \ref{fn:kgtn}). Furthermore,  $f_{Ga}(\cdot\given a, b)$ is the density function of a Gamma distribution with mean $a/b$. The conditional independence between $\beta_j$ and $\alpha$ (given $\sigma$) is a consequence of demeaning the regressors.
After integrating out the model parameters as above, we can write the marginal likelihood as
\begin{equation}
p(y|M_j)\propto  (1+g)^{\frac{n-1-k_j}{2}}
[1+g(1-R_j^2)]^{-{\frac{n-1}{2}}},
\label{MargLikelihood}
\end{equation}
where $R^2_j$ is the usual coefficient of determination for model
$M_j$, defined through $1-R_j^2=y'Q_{X_j}y/[(y-\bar y \iota)'(y-\bar y \iota)]$, and the proportionality constant is the same for all models,
including the null model. In addition, for each model $M_j$, the marginal posterior distribution of the regression coefficients $\beta_j$ is a $k_j$-variate Student-$t$ distribution with $n-1$ degrees of freedom, location $\delta (Z_j'Z_j)^{-1} Z_j'y$ (which is the mean if $n>2$) and scale matrix $\delta s_\delta (Z_j'Z_j)^{-1}$ (and variance $\frac{\delta s_\delta}{n-3} (Z_j'Z_j)^{-1}$ if $n>3$). The out-of-sample predictive distribution for each given model (which in a regression model will of course also depend on the covariate values associated with the observations we want to predict) is also a Student-$t$ distribution with $n-1$ degrees of freedom. Details can be found in equation (3.6) of \cite{FLS01a}. Following (\ref{BMAeq}), we can then conduct posterior or predictive inference by simply averaging these model-specific distributions using the posterior model weights computed (as in (\ref{postmodel})) from (\ref{MargLikelihood}) and the prior model distributions described in the next subsection.

There are a number of suggestions in the literature for the choice of fixed values for $g$, among which the most popular ones are:
\begin{itemize}
\item The unit information prior of \cite{KassWasserman_95} corresponds to the amount of information
contained in one observation. For regular parametric
families, the ``amount of information'' is defined through
Fisher information. This gives us $g = n$, and leads to
log Bayes factors that behave asymptotically like the BIC \citep{FLS01a}.

\item The risk inflation criterion prior, proposed by \cite{FosterGeorge_94}, is based on the Risk inflation criterion (RIC) which leads to $g = k^2$ using a minimax perspective.

\item The benchmark prior of \cite{FLS01a}. They examine various choices of $g$ depending on the
sample size $n$ or the model dimension $k$ and recommend $g = \max(n,k^2)$.

\end{itemize}

When faced with a variety of possible prior choices for $g$, a natural Bayesian response is to formulate a hyperprior on $g$. This was already implicit in \cite{ZellnerSiow80} who use a Cauchy prior on the regression coefficients, corresponding to an inverse gamma prior on $g$. This idea was investigated further in \cite{Liang_etal_08}, where hyperpriors on $g$ are shown to alleviate certain paradoxes that appear with fixed choices for $g$. Sections \ref{EB_hier} and \ref{cons} will provide more detail.

The $g$-prior is a relatively well-understood and convenient prior with nice properties, such as invariance under rescaling and translation of the covariates (and more generally, invariant to reparameterization under affine transformations), and automatic adaptation to situations with near-collinearity between
different covariates \citep[p.~193]{Robert2007}. It can also be interpreted as the conditional posterior of the regression coefficients
 given a locally uniform prior and an imaginary sample of zeros with design
matrix $Z_j$ and a scaled error variance.

This idea of imaginary data is also related to the power
prior approach \citep{IbrahimChen_00}, initially developed on the basis of the availability of historical data ({\it i.e.}~data arising from previous similar studies).
In addition, the device of imaginary training samples forms the basis of
the expected-posterior prior \citep{PerezBerger_02}.  In \cite{FouskakisNtzoufras_16a} the power-conditional-expected-posterior prior is developed by
combining the power prior and the expected-posterior prior approaches for the regression parameters conditional on the error
variance.

\cite{Som_etal_15} introduce the block hyper-$g/n$ prior for so-called ``poly-shrinkage'', which is a
collection of ordinary mixtures of $g$-priors applied separately to groups of predictors. Their motivation is to avoid certain paradoxes, related to different asymptotic behaviour for different subsets of predictors.
\cite{MinSun_16} consider the situation of grouped covariates (occurring, for example, in ANOVA models where each factor has various levels) and propose separate $g$-priors for the associated groups of regression coefficients. This also circumvents the fact that in ANOVA models the full design matrix is often not of full rank.

A similar idea is used in \cite{Zhang_etal_16BA} where a two-component extension of the $g$-prior is proposed, with each regressor being assigned one of two possible values for $g$. Their prior is proper by treating the intercept as part of the regression vector in the $g$-prior and by using a ``vague'' proper prior\footnote{Note that this implies the necessity to choose the associated  hyperparameters in a sensible manner, which is nontrivial as what is sensible depends on the scaling of the data.} on $\sigma^2$. They focus mostly on variable selection.

A somewhat different approach was advocated by \cite{GeorgeMcCulloch_93,GeorgeMcCulloch_97}, who use a prior on the regression coefficient which does not include point masses at zero. In particular, they propose
a normal prior with mean zero on the entire $k$-dimensional vector of regression coefficients $\beta$ given the model $M_j$ which assigns a small prior variance to the coefficients of the variables that are ``inactive''\footnote{Formally, all variables appear in all models, but the coefficients of some variables will be shrunk to zero by the prior, indicating that their role in the model is negligible.} in $M_j$ and a larger variance to the remaining coefficients. In addition, their overall prior is proper and does not assume a common intercept.

\cite{Raftery_etal_97} propose yet another approach and use a proper conjugate\footnote{Conjugate prior distributions combine analytically with the likelihood to give a posterior in the same class of distributions as the prior.} prior with a diagonal covariance structure for the regression coefficients (except for categorical predictors where a $g$-prior structure is used).

\subsubsection{Priors over models}\label{Sec:PriorModel}
\begin{figure}[ht!]
\begin{center}
\subfigure{\includegraphics[width=8cm]{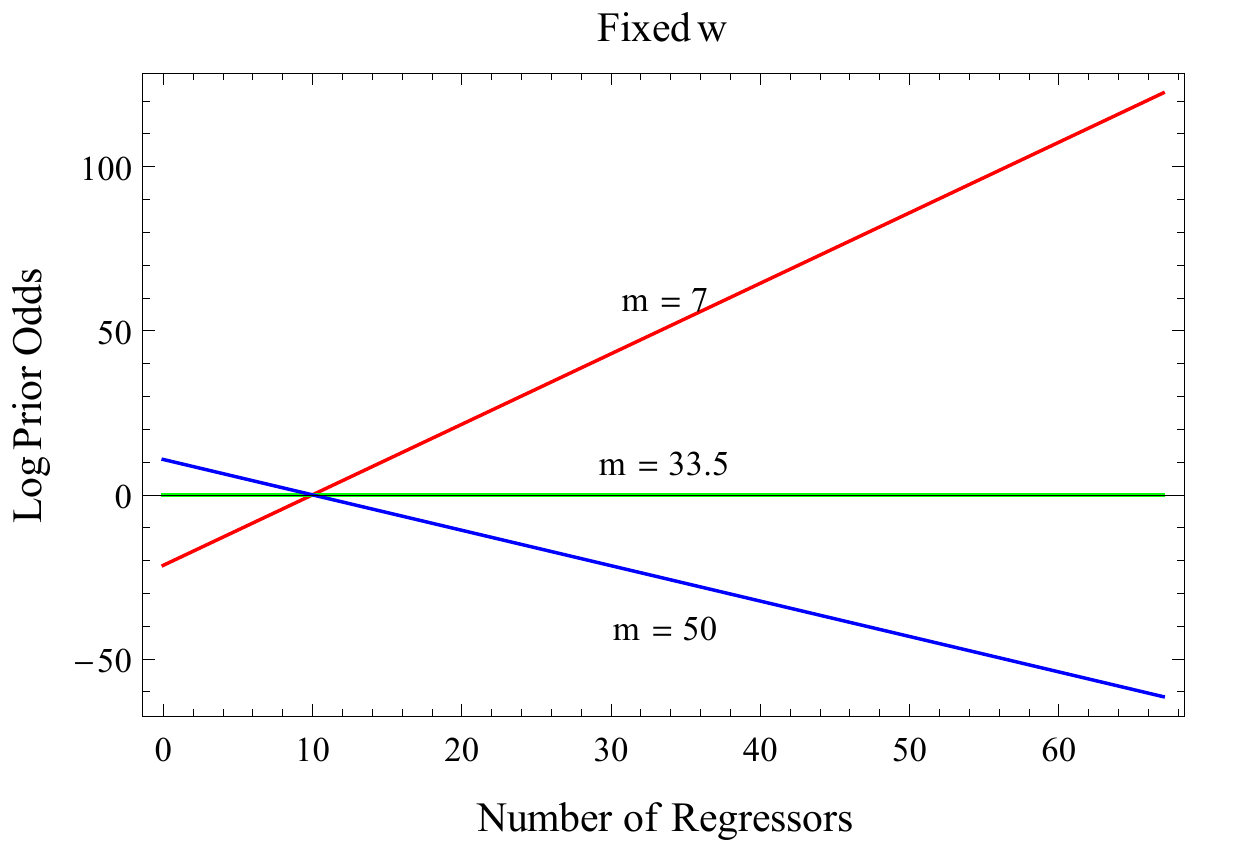}}
\subfigure{\includegraphics[width=8cm]{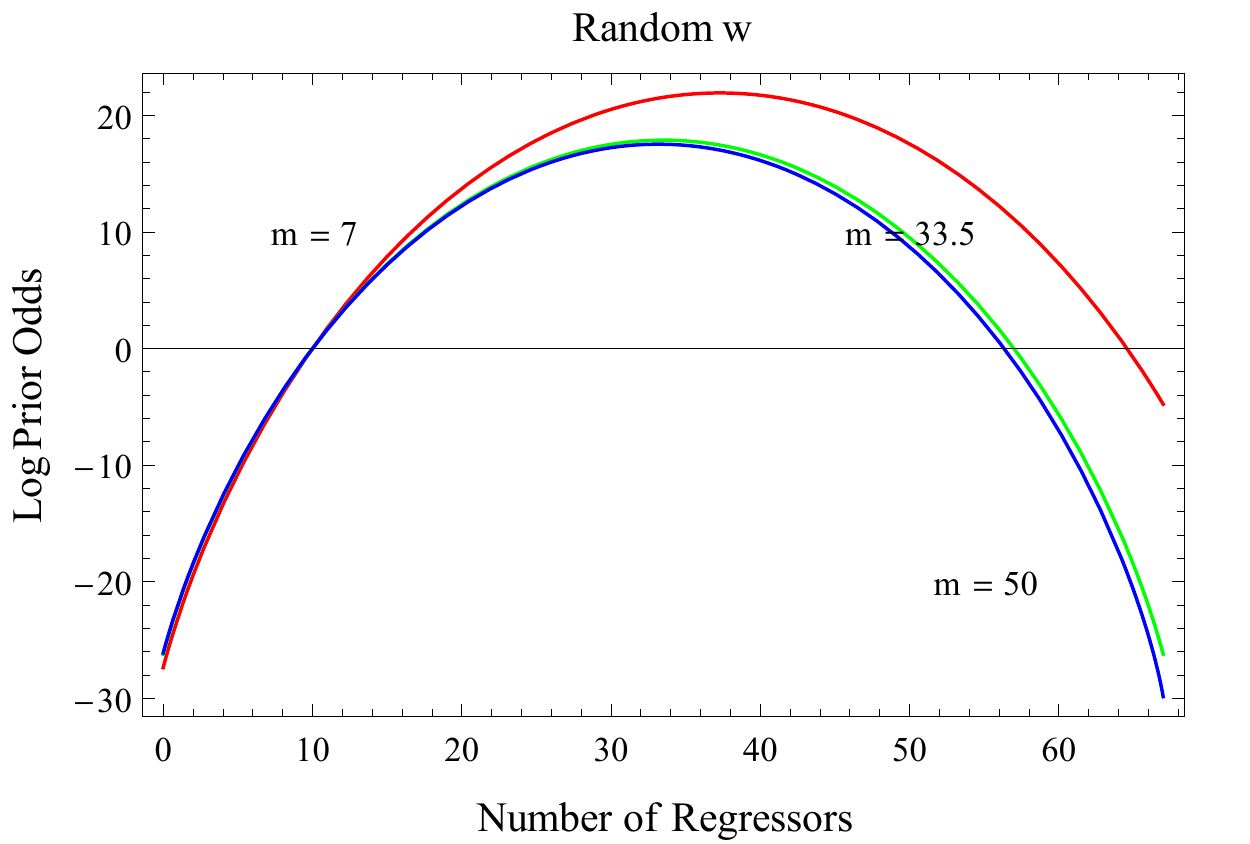}}
\caption{\footnotesize Log of Prior Odds: $k_i=10$ vs varying $k_j$. From \cite{LS6}.}
\label{PO1}
\end{center}
\end{figure}

The prior $P(M_j)$ on model space is typically constructed by considering the probability of inclusion of each covariate. If the latter is the same for each variable, say $w$, and we assume inclusions are prior independent, then
\begin{equation}\label{PM_fixed}
P(M_j)=w^{k_j}(1-w)^{k-k_j}.
\end{equation}
This implies that prior odds will favour larger models if $w>0.5$ and the opposite if $w<0.5$. For $w=0.5$ all model have equal prior probability $1/k$. Defining model size as the number of included regressors in a model, a simple way to elicit $w$ is through the prior mean model size, which is $wk$.\footnote{So, if our prior belief about mean model size is $m$, then we simply choose $w=m/k$.} As the choice of $w$ can have a substantial effect on the results, various authors \citep{Brown_etal_98,ClydeGeorge04,LS6,ScottBerger} have suggested to put a Beta$(a,b)$ hyperprior on $w$. This results in
\begin{equation}\label{PM_hyper}
P(M_j)=\frac{\Gamma(a+b)}{\Gamma(a)\Gamma(b)} \frac{\Gamma(a+k_j)\Gamma(b+k-k_j)}{\Gamma(a+b+k)},
\end{equation}
which leads to much less informative priors in terms of model size. \cite{LS6} compare both approaches and suggest choosing $a=1$ and $b=(k-m)/m$, where $m$ is the chosen prior mean model size. This means that the user only needs to specify a value for $m$. The large differences between the priors in (\ref{PM_hyper}) and (\ref{PM_fixed}) can be illustrated by the prior odds they imply.
Figure \ref{PO1} compares the log prior odds induced by the fixed and random
$w$ priors, in the situation where $k=67$ (corresponding to the growth dataset first used in \cite{SDM}) and using
$m=7, 33.5$ and $50$. For fixed $w$, this corresponds to $w=7/67, w=1/2$ and $w=50/67$ while for random $w$, I have used the specification of \cite{LS6}. The figure displays the prior odds in favour
of a model with $k_i=10$ versus models with varying $k_j$.

Note that
the random $w$ case always leads to down-weighting of models
with $k_j$ around $k/2$, irrespectively of $m$. This counteracts the
fact that there are many more models with $k_j$ around $k/2$ in the
model space than of size nearer to $0$ or
$k$.\footnote{This reflects the multiplicity issue analysed more generally in \cite{ScottBerger} who propose to use (\ref{PM_hyper}) with $a=b=1$ implying a prior mean model size of $k/2$. The number of models with $k_j$ regressors in $\cal M$ is
given by ${k}\choose{k_j}$. For example, with $k=67$, we have 1
model with $k_j=0$ and $k_j=k$, $8.7\times 10^8$ models with $k_j=7$
and $k_j=60$ and a massive $1.4\times 10^{19}$ models with $k_j=33$
and $34$.\label{fn:mult}} In contrast, the prior with fixed $w$ does not take
the number of models at each $k_j$ into account and simply always
favours larger models when $m>k/2$ and smaller ones when $m<k/2$.
Note also the much wider range of values that the log prior odds take in the case of fixed $w$.
Thus, the choice of $m$ is critical for the prior with fixed $w$, but much
less so for the hierarchical prior structure, which is naturally
adaptive to the data observed.

It is often useful to elicit prior ideas by focusing on model size, as it is an easily understood concept. In addition, there will often be a preference for somewhat smaller models due to their interpretability and simplicity. The particular choice of $m=7$ (mentioned above) was used in \cite{SDM} in the context of growth regression and has become a rather popular choice in a variety of applied contexts. \cite{SDM} sensibly argue that the prior mean model size should not be linearly increasing with $k$, but provide little motivation for specifically choosing $m=7$. The origins of this choice may be related to computational restrictions faced by earlier empirical work ({\it e.g.}~the EBA analysis of \cite{LevineRenelt_92} was conducted on a restricted set of models that never had more than 8 regressors). I think that any particular prior choice should be considered within the appropriate context and I would encourage the use of sensitivity analyses and robust prior structures (such as the hierarchical prior leading to (\ref{PM_hyper})). \cite{Giannone_etal_18} investigate whether sparse modelling is a good approach to predictive problems in economics on the basis of a number of datasets from macro, micro and finance. They find 
that artificially tight model priors\footnote{They use a very tight prior indeed, which corresponds to a prior mean model size $m=k/(k+1)$ which is less than one!} focused on small models induce sparsity at the expense of predictive performance and model fit. They conclude that ``predictive model uncertainty seems too
pervasive to be treated as statistically negligible. The right approach to scientific reporting
is thus to assess and fully convey this uncertainty, rather than understating it through the
use of dogmatic (prior) assumptions favoring low dimensional models.'' 

\cite{George_99} raises the issue of  ``dilution'', which occurs when posterior probabilities are spread among many similar models, and suggest that prior model probabilities could have a built-in adjustment to compensate for dilution by down-weighting prior probabilities on sets of similar models. \cite{George_10} suggests three distinct approaches for the construction of these so-called ``dilution priors'', based on tessellation determined neighbourhoods, collinearity adjustments, and pairwise distances between models. Dilution priors were implemented in economics by \cite{Durlauf_etal_08} to represent priors that are uniform on theories ({\it i.e.}~neighbourhoods of similar models) rather than on individual models, using a collinearity adjustment factor.
A form of dilution prior in the context of models with interactions of covariates is the heredity prior of \cite{Chipman_etal_97} where interaction are only allowed to be included if both main effects are included (strong heredity) or at least one of the main effects (weak heredity).  In the context of examining the sources of growth in Africa, \cite{Cuaresma_11} comments that the use of a strong heredity prior leads to different conclusions than the use of a uniform prior in the original paper by \cite{MasanPapa_08}.\footnote{See also \cite{Papa_11}, which is a reply to the comment by Crespo Cuaresma.} Either prior is, of course, perfectly acceptable, but it is clear that the user needs to reflect which one best captures the user's own prior ideas and intended interpretation of the results. Using the same data, \cite{MoserHofmarcher_14} compare a uniform prior with a strong heredity prior and a tesselation dilution prior and find quite similar predictive performance (as measured by LPS and CRPS, explained in Section \ref{Pred}) but large differences in posterior inclusion probabilities (probably related to the fact that both types of dilution priors are likely to have quite different responses to multicollinearity).

\cite{Womack_etal_15} propose viewing the model space as a partially ordered set. When the
number of covariates increases, an isometry argument leads to the Poisson distribution as the unique,
natural limiting prior over model dimension. This limiting prior is derived using two constructions that
view an individual model as though it is a ``local'' null hypothesis and compares its prior probability to
the probability of the alternatives that nest it. They show that this prior induces a posterior that concentrates
on a finite true model asymptotically.

Another interesting recent development is the use of a loss function to assign a model prior. Equating information loss as measured by the expected minimum Kullback-Leibler divergence between any model and its nearest model and by the ``self-information loss''\footnote{This is a loss function (also known
as the log-loss function) for probability statements, which is given by the negative logarithm of the probability.} while adding an adjustment for complexity, \cite{VillaLee_16} propose the prior
$P(M_j)=\exp(-ck_j)$
for some $c>0$. This builds on an idea of \cite{VillaWalker_15}.

\subsubsection{Empirical Bayes versus Hierarchical Priors}\label{EB_hier}

The prior in (\ref{FLSprior}) and (\ref{PM_fixed}) only depends on two scalar quantities, $g$ and $w$. Nevertheless, these quantities can have quite a large influence on the posterior model probabilities and it is very challenging to find a single default choice for $g$ and $w$ that performs well in all cases, as explained in {\it e.g.}~\cite{FLS01a}, \cite{BergerPericchi_01} and \cite{LS6}. One way of reducing the impact of such prior choices on the outcome is to use hyperpriors on $w$ and $g$, which fits seamlessly with the Bayesian paradigm. Hierarchical priors on $w$ are relatively easy to deal with and were already discussed in the previous section.

\cite{ZellnerSiow80} used a multivariate Cauchy prior on the regression coefficients rather than the normal prior in (\ref{FLSprior}). This was inspired by the argument in \cite{jeffreys1961} in favour of heavy-tailed priors\footnote{The reason for this was the limiting behaviour of the resulting Bayes factors as we consider models with better and better fit. In this case, you would want these Bayes factors, with respect to the null model, to tend to infinity. This criterion is called ``information consistency'' in \cite{Bayarri_etal_12} and its absence is termed ``information paradox'' in \cite{Liang_etal_08}.\label{fn:InfCons}}. Since a Cauchy is a scale mixture of normals, this means that implicitly the Zellner-Siow prior uses an Inverse-Gamma$(1/2,n/2)$ prior on $g$.

\cite{Liang_etal_08} introduce the hyper-$g$ priors, which correspond to the following family of priors:
\begin{equation}
p(g)=\frac{a-2}{2}(1+g)^{-a/2}\label{hyper-g}
\end{equation}
where $a>2$ in order to have a proper distribution for $g>0$. This includes the priors proposed in \cite{Strawderman} in the context of the normal means problem. A value of $a=4$ was suggested by \cite{CuiGeorge_08} for model selection with known $\sigma$, while \cite{Liang_etal_08} recommend values $2<a\le 4$. \cite{FZ09} propose to use a hyper-$g$ prior with a value of $a$ that leads to the same mean of the, so-called, shrinkage factor\footnote{The name ``shrinkage factor'' derives from the fact that the posterior mean of the regression coefficients for a given model is the OLS estimator times this shrinkage factor, as clearly shown in (\ref{Eq:postbeta}) and the ensuing discussion.} $\delta=g/(1+g)$ as for the unit information or the RIC prior. \cite{leysteel2012} consider the more general class of beta priors on the shrinkage factor where a Beta($b,c$) prior on $\delta$ induces the following prior on $g$:
\begin{equation}
p(g)=\frac{\Gamma(b+c)}{\Gamma(b)\Gamma(c)}g^{b-1}(1+g)^{-(b+c)}\label{beta-g}.
\end{equation}
This is an inverted beta distribution\footnote{Also known as a gamma-gamma distribution (Bernardo
and Smith, 1994, p.~120)\nocite{BernardoSmith_94}.} (Zellner, 1971,
p.~375)\nocite{zellner1971} which clearly reduces to the hyper-$g$ prior in (\ref{hyper-g}) for  $b=1$ and $c=(a/2)-1$. Generally, the hierarchical prior on $g$ implies that the marginal likelihood of a given model is not analytically known, but is the integral of (\ref{MargLikelihood}) with respect to the prior of $g$. \cite{Liang_etal_08} propose the use of a Laplace approximation for this integral, while \cite{leysteel2012} use a Gibbs sampler approach to include $g$ in the Markov chain Monte Carlo procedure (see footnote \ref{fn:MCMC}). Some authors have proposed Beta shrinkage priors as in (\ref{beta-g}) that lead to analytical marginal likelihoods by making the prior depend on the model: the robust prior of \cite{Bayarri_etal_12} truncate the prior domain to $g>[(n+1)/(k_j+1)]-1$ and \cite{MaruyamaGeorge_11} adopt the choice $b+c=(n-k_j-1)/2$ with $c<1/2$. However, the truncation of the robust prior is potentially problematic for cases where $n$ is much larger than a typical model size (as is often the case in economic applications). \cite{leysteel2012} propose to use the beta shrinkage prior in (\ref{beta-g}) with mean shrinkage equal to the one corresponding to the benchmark prior of \cite{FLS01a} and the second parameter chosen to ensure a reasonable prior variance. They term this the benchmark beta prior and recommend using $b=c\max\{n,k^2\}$ and $c=0.01$.


An alternative way of dealing with the problem of selecting $g$ and $w$ is to adopt so-called ``empirical Bayes'' (EB) procedures, which use the data to suggest appropriate values to choose for  $w$ and $g$. Of course, this amounts to using data information in selecting the prior, so is not formally in line with the Bayesian paradigm. Often, such EB methods are adopted for reasons of convenience and because they are sometimes shown to have good properties. In particular, they provide ``automatic'' calibration of the prior and avoid the (relatively small) computational complications that typically arise when we adopt a hyperprior on $g$.
Motivated by
information theory, \cite{HansenYu_01} proposed a local EB method which uses a different $g$ for each model estimated by maximizing the marginal likelihood. \cite{GeorgeFoster_00} develop a global EB approach, which assumes one common $g$ and $w$ for all models and borrows strength
from all models by estimating $g$ and $w$ through maximizing the marginal likelihood, averaged over all models. \cite{Liang_etal_08} propose specific ways of estimating $g$ in this context.



There is some evidence in the literature regarding comparisons between fully Bayes and EB procedures: \cite{CuiGeorge_08} largely favour (global) EB in the context of known $\sigma$ and $k=n$, whereas \cite{Liang_etal_08} find that there is little difference between EB and fully Bayes procedures (with unknown $\sigma$ and $k<n$).
\cite{ScottBerger} focus on EB and fully Bayesian ways of dealing with $w$, which, respectively, use maximum likelihood\footnote{This is the value of $w$ that maximizes  the marginal likelihood of $w$ summed over model space, or $\arg \max_w p(y)$ in (\ref{postmodel}), which can be referred to as type-II maximum likelihood.} and a Beta(1,1) or uniform hyperprior on $w$. They remark that both fully Bayesian and EB procedures exhibit clear multiplicity
adjustment: as the number of noise variables increases, the posterior inclusion
probabilities of variables decrease (the analysis with fixed $w$ shows no
such adjustment; see also footnote \ref{fn:mult}). However, they highlight some theoretical differences, for example the fact that EB will assign probability one to either the full
model or the null model whenever one of these models has the
largest marginal likelihood. They also show rather important differences in various applications, one of which uses data on GDP growth. Overall, they recommend the use of fully Bayesian procedures.
\cite{LiClyde_17} compare EB and fully Bayes procedures in the more general GLM context (see Section \ref{GLM}), and find that local EB does badly in simulations from the null model in that it almost always selects the full model.




\subsection{Properties of BMA}\label{sec:propertiesBMA}

\subsubsection{Consistency and paradoxes}\label{cons}


One of the desiderata in \cite{Bayarri_etal_12} for objective model selection priors is model selection consistency (introduced by \cite{FLS01a}), which implies that if data  have been generated
by $M_j\in {\cal M}$, then the posterior probability of $M_j$ should converge to unity with sample size\footnote{So this is a concept defined in the $\cal M$-closed framework, mentioned in Subsection \ref{sec:modelspace}.}. \cite{FLS01a} present general conditions for the case with non-random $g$ and show that consistency holds for {\it e.g.}~the unit information and benchmark priors (but not for the RIC prior). When we consider hierarchical priors on $g$, model selection consistency is achieved by the Zellner-Siow prior in \cite{ZellnerSiow80} but not by local and global EB priors nor by the hyper-$g$ prior in \cite{Liang_etal_08}, who therefore introduce a consistent modification, the hyper-$g/n$ prior, which corresponds to a beta distribution on $g/(n+g)$.
Consistency is shown to hold for the priors of \cite{MaruyamaGeorge_11}, \cite{FZ09} (based on the unit information prior) and the benchmark beta prior of \cite{leysteel2012}.

\cite{Moreno_etal_15} consider model selection consistency when the number of potential regressors $k$ grows with sample size. Consistency is found to depend not only on the priors for the model parameters, but also on the priors in model space. They conclude that if $k=O(n^b)$, the unit information prior, the Zellner-Siow prior and the intrinsic prior\footnote{Intrinsic priors were introduced to justify the intrinsic
Bayes factors \citep{BergerPericchi1996}. In principle, these are often based on improper reference (see footnote \ref{fn:refprior}) or Jeffreys priors and the use of a so-called minimal training sample to convert the improper prior to a proper posterior. The latter is then used as a prior for the remaining data, so that Bayes factors can be computed. As the outcome depends on the arbitrary choice of the minimal training sample, such Bayes factors are typically ``averaged'' over all possible training samples. Intrinsic priors are priors that (at least asymptotically) mimic these intrinsic Bayes factors.} lead to
consistency for $0\le b <1/2$ under the uniform prior over model space, while consistency holds for $0\le b \le 1$ if we use a Beta(1,1) hyperprior on $w$ in (\ref{PM_hyper}). \cite{WangMaruyama_16} investigate Bayes factor consistency associated with the prior structure in (\ref{FLSprior}) for the
problem of comparing nonnested models under a variety of scenarios where model dimension grows with sample size. They show that
in some cases, the Bayes factor is consistent whichever the true model is, and that in
others, the consistency depends on the pseudo-distance between the
models. In addition,
they find that the asymptotic behaviour of Bayes factors and intrinsic
Bayes factors are quite similar.


\cite{Mukh_etal_15}  show that in situations where the true model is not
one of the candidate models (the $\cal M$-open setting), the use of $g$-priors leads to selecting a model that is in an intuitive sense closest to the true model. In addition, the loss incurred in estimating the unknown regression function under the selected model tends to that under the true model. These results have been shown under appropriate conditions on the rate of growth of $g$ as $n$ grows and for both the cases when the number of potential
predictors remains fixed  and when $k = O(n^b)$ for some $0 < b < 1$.\footnote{Unlike \cite{Moreno_etal_15}, they do not explicitly find different results for  different priors on the model space, which looks like an apparent contradiction. However, their results are derived under an assumption (their (A3)) bounding the ratio of prior model probabilities. Note from our Figure \ref{PO1} that ratio tends to be much smaller when we use a hyperprior on $w$.} \cite{Mukh_etal_17} extend this to the situation of mixtures of $g$-priors and derive consistency properties for growing $k$ under a modification of the Zellner-Siow prior, that continue to hold for more general error distributions.

Using Laplace approximations, \cite{Xiang_etal_16} prove that in the case of hyper-$g$ priors with growing model sizes, the
Bayes factor is consistent when $k = O(n^b)$ for some $0 < b \le 1$, even when the true model is the null model. For the case when the true model is not the null model, they show that Bayes factors are always consistent when the true model is nested within the model under consideration, and they give conditions for the non-nested case.
In the specific context of analysis-of-variance (ANOVA) models, \cite{Wang_17} shows that the Zellner-Siow
prior and the beta shrinkage prior of \cite{MaruyamaGeorge_11} yield inconsistent Bayes factors when $k$ is proportional to $n$ due to the presence of an inconsistency region around the null model. To solve the latter inconsistency, \cite{Wang_17}
propose a variation on the hyper-$g/n$ prior, which generalizes the prior arising from a Beta distribution on  $g/(\frac{n}{k}+g)$.
Consistency for the power-expected-posterior approach using independent Jeffreys baseline priors is shown by \cite{FouskakisNtzoufras_16}.

\cite{Sparks_etal_15} consider posterior consistency for parameter estimation, rather than model selection. They consider posterior consistency
under the sup vector norm (weaker than the usual $l_2$ norm) in situations where $k$ grows with sample size and derive necessary and sufficient conditions for consistency under the standard $g$-prior, the empirical Bayes specification of \cite{GeorgeFoster_00} and the hyper-$g$ and Zellner-Siow mixture priors.

In addition, BMA also has important optimality properties in terms of shrinkage in high-dimensional problems. In particular, \cite{Castillo_etal_15} prove that BMA in linear regression leads to an optimal rate of contraction of the posterior on the regression coefficients to a sparse ``true'' data-generating model (a model where many of the coefficients are zero), provided the prior sufficiently penalizes model complexity. \cite{RossellTelesca_17} show that BMA leads to fast shrinkage for spurious coefficients (and explore so-called nonlocal priors introduced by \cite{JohnsonRossell_10} that provide even faster shrinkage in the BMA context).

A related issue is that Bayes factors can asymptotically behave in the same way as information criteria. \cite{KassWasserman_95} investigate the relationship between BIC (see Section \ref{Sec:NLM}) 
and Bayes factors using unit information priors for testing non-nested hypotheses and \cite{FLS01a} show that log Bayes factors with $g_j=n/f(k_j)$ (with $f(\cdot)$ some function which is finite for finite arguments) tend to BIC. When $k$ is fixed, this asymptotic equivalence to BIC extends to the Zellner-Siow and \cite{MaruyamaGeorge_11} priors \citep{Wang_17} and also the intrinsic prior \citep{Moreno_etal_15}.

\cite{Liang_etal_08} remark that analyses with fixed $g$ tend to lead to a number of paradoxical results. They mention the Bartlett (or Lindley) paradox, which is induced by the fact that very large values of $g$ will induce support for the null model, irrespective of the data\footnote{This can be seen immediately by considering (\ref{PostOdds}) in Subsection \ref{Sec:RolePrior}, which behaves like a constant times $(1+g)^{(k_j-k_i)/2}$ as $g\to\infty$.}. Another paradox they explore is the information paradox, where as $R_i^2$ tends to one, the Bayes factor in favour of $M_i$ versus, say, the null model does not tend to $\infty$ but to a constant depending on $g$ (see also footnote \ref{fn:InfCons}). From  (\ref{PostOdds}) in Subsection \ref{Sec:RolePrior} this latter limit is $(\frac{w}{1-w})^{k_i} (1+g)^{(n-k_i-1)/2}$. \cite{Liang_etal_08} show that this information paradox is resolved by local or global EB methods, but also by using hyperpriors $p(g)$ that satisfy $\int (1+g)^{(n-k_i-1)/2} p(g) \d g=\infty$ for all $k_i\le k$, which is the case for the Zellner-Siow prior, the hyper-$g$ prior and the benchmark beta priors (the latter two subject to a condition, which is satisfied in most practical cases).




\subsubsection{Predictive performance}\label{Pred}

Since any statistical  model will typically not eliminate uncertainty and it is important to capture this uncertainty in forecasting, it is sensible to consider probabilistic forecasts, which have become quite popular in many fields. In economics, important forecasts such as the quarterly Bank of England inflation report are presented in terms of predictive distributions, and in the field of finance the area of risk management focuses on  probabilistic forecasts of portfolio values. Rather than having to condition on estimated parameters, the Bayesian framework has the advantage that predictive inference can be conducted on the basis on the predictive distribution, as in (\ref{pred}) where all uncertainty regarding the parameters and the model is properly incorporated. This can be used to address a genuine interest in predictive questions, but also as a model-evaluation exercise. In particular, if a model estimated on a subset of the data manages to more or less accurately predict data that were not used in the estimation of the model, that intuitively suggests satisfactory performance.

In order to make this intuition a bit more precise, scoring rules provide useful summary measures for the evaluation
of probabilistic forecasts. Suppose the forecaster wishes to optimize the scoring rule. If the scoring rule is proper, the forecaster has no incentive to predict any other distribution than his or her true belief for the forecast distribution. Details can be found in \cite{GneitingRaft_07}.

Two important aspects of probabilistic forecasts are calibration and sharpness. Calibration refers to the compatibility between the forecasts and the observations and is a joint property of the predictions
and the events that materialize. Sharpness refers to the concentration of the predictive
distributions and is a property of the forecasts only. Proper scoring rules address both of these issues simultaneously.
Popular scoring rules, used in assessing predictive performance in the context of BMA are
\begin{itemize}
\item The logarithmic predictive score (LPS), which is the negative of the logarithm of the predictive density evaluated
at the observation. This was introduced in \cite{Good_52} and used in the BMA context in \cite{Madigan_etal_95},  \cite{FLS01a,FLS01b} and \cite{LS6}.
\item The continuous ranked probability score (CRPS). The CRPS measures
the difference between the predicted and the observed cumulative distributions as follows\footnote{An alternative expression is given by \cite{GneitingRaft_07} as
$CRPS(Q,x)=E_Q|X-x| -{\frac{1}{2}} E_Q|X-Z|$,
where $X$ and $Z$ are independent copies of a random variable
with distribution function $Q$ and finite first moment.  This shows that CRPS generalizes the absolute error, to which it reduces if $Q$ is a point forecast.}:
\begin{equation}
CRPS(Q, x)=\int_{-\infty}^\infty \left[Q(y)-{\mathbb{1}}(y\ge x)\right]^2 {\d}y,
\end{equation}
where $Q$ is the predictive distribution, $x$ is the observed outcome and $\mathbb{1}(\cdot)$ is the indicator function. CRPS was found in \cite{GneitingRaft_07} to be less sensitive to outliers than LPS and was introduced in the context of growth regressions by
\cite{Eicheretal11}.

\end{itemize}

Simple point forecasts do not allow us to take into account the uncertainty associated with the prediction, but are popular in view of their simplicity, especially in more complicated models incorporating {\it e.g.} dynamic aspects or endogenous regressors. Such models are often evaluated in terms of the mean squared forecast
error or the mean absolute forecast error calculated with respect to a point forecast.

There is a well-established literature indicating the predictive advantages of BMA. For example, \cite{MadiganRaftery_94} state that  BMA predicts at least as well\footnote{This optimality holds under the assumption that the data is generated by the predictive in (\ref{pred}) rather than a single ``true'' model. \cite{George_99b} comments that ``It is tempting to criticize BMA because it does
not offer better average predictive performance than
a correctly specified single model. However, this fact
is irrelevant when model uncertainty is present because
specification of the correct model with certainty
is then an unavailable procedure. In most
practical applications, the probability of selecting
the correct model is less than 1, and a mixture
model elaboration seems appropriate.''\label{fn:freq}} as any single model in terms of LPS and \cite{MinZellner} show that expected squared error loss of point (predictive mean) forecasts is always minimized by BMA provided the model space includes the model that generated the data\footnote{The importance of this latter assumption is discussed in \cite{Diebold_91}, who points out that violation of this assumption could negatively affect large sample performance of BMA in point forecasting.}.
On the basis of empirical results, \cite{Raftery_etal_97} report that predictive coverage is improved by BMA with respect to prediction based on a single model. Similar results were obtained by \cite{FLS01a}, who also use LPS as a model evaluation criterion in order to compare various choices of $g$ in the prior (\ref{FLSprior}). \cite{FLS01b} find, on the basis of LPS that BMA predicts substantially better than single models (such as the model with highest posterior probability) in growth data. \cite{LS6} corroborate these findings, especially with a hyperprior on $w$, as used in (\ref{PM_hyper}).
\cite{PiirVeht_17} focus on model selection methods, but state that ``From the predictive point of view, best results are generally obtained by accounting for the model uncertainty and forming the full BMA solution over the candidate models, and one should not expect to do better by selection.''
In the context of volatility forecasting of non-ferrous metal futures, \cite{Lyocsa_etal_17} show that averaging of forecasts substantially improves the results, especially where the averaging is conducted through BMA.




\subsubsection{Frequentist performance of point estimates}\label{sec:freqpropBMA}

Point estimation is not a particularly natural Bayesian concept, and the Bayesian paradigm does not rely on frequentist considerations (unlike FMA, where often the model weights are chosen to optimize some frequentist criterion). Nevertheless, it is interesting to know the properties of point estimates resulting from BMA under repeated sampling. Whenever we consider sampling-theoretic properties, we require an assumption about the sampling model under which the properties will be studied.  Using the assumption that the data is generated by the predictive in (\ref{pred}), \cite{RafteryZheng_03} state a number of important results about Bayesian model selection and BMA. Firstly, they state a result by \cite{jeffreys1961}, which is that for two nested models, model choice based on Bayes factors minimizes the sum of type I and type II errors.
Secondly, they mention properties of BMA point estimates (of parameters and observables). In particular, BMA leads to point estimates that minimize MSE and BMA estimation intervals are calibrated in the sense that the average coverage probability of a BMA interval with posterior probability $\alpha$ is at least equal to $\alpha$. Of course, this does not provide any guarantee for the frequentist properties when the actual observations come from a different distribution, but there exists no compelling argument for any particular choice of such a ``true'' distribution. I believe it is highly unlikely in economics (as in most other areas of application) that the data are truly generated by a single model within the model space (assumed in {\it e.g.}~\cite{Castillo_etal_15}) or a local neighbourhood around such a model that shrinks with sample size\footnote{Insightful comments concerning the practical relevance of this particular assumption can be found in \cite{RafteryZheng_03} and other discussions of \cite{ClaeskensHjort_03} and \cite{HjortClaeskens_03}.\label{fn:localmis}} (as in {\it e.g}~\cite{ClaeskensHjort_03} and \cite{HjortClaeskens_03}) (see also footnote \ref{fn:freq}). 

\subsection{BMA in practice: Numerical methods}\label{sec:numerical}

One advantage of the prior structure in (\ref{FLSprior}) is that integration of the model parameters can be conducted analytically, and the Bayes factor between any two given models can be computed quite easily, given $g$, through (\ref{MargLikelihood}). The main computational challenge is then constituted by the often very large model space, which makes complete enumeration impossible. In other words, we simply can not try all possible models, as there are far too many of them\footnote{In areas such as growth economics, we may have up to $k=100$ potential covariates. This implies a model space consisting of $K=2^{100}=1.26\times 10^{30}$ models. Even with fast processors, this dramatically exceeds the number of models that can be dealt with exhaustively. In other fields, the model space can even be much larger: for example, in genetics $k$ is the number of genes and can well be of the order of tens of thousands.}.

A first possible approach is to (drastically) reduce the number of models under consideration. One way to do this is the Occam's window algorithm, which was proposed by \cite{MadiganRaftery_94} for graphical models and extended to linear regression in \cite{Raftery_etal_97}. It uses a search strategy to weed out the models that are clearly dominated by others in terms of posterior model probabilities and models that have more likely submodels nested within them.
An algorithm for finding the best models is the so-called leaps and bounds method used by \cite{Raftery1995} for BMA, based on the all-subsets regression algorithm of \cite{FurnivalWilson}. The resulting set of best models can then still be reduced further through Occam's window if required. The Occam's window and the leaps and bounds algorithms are among the methods implemented in the BMA {\tt R} package of \cite{Raftery_etal_soft} and the leaps and bounds algorithm was used in {\it e.g.}~\cite{MasanPapa_08} and \cite{Eicheretal11}.

However, this tricky issue of exploring very large model spaces is now mostly dealt with through so-called Markov chain Monte Carlo (MCMC) methods\footnote{Suppose we have a distribution, say $\pi$, of which we do not know the properties analytically and which is difficult to simulate from directly. MCMC methods construct a Markov chain that has $\pi$ as its invariant distribution and conduct inference from the generated chain. The draws in the chain are of course correlated, but ergodic theory still forms a valid basis for inference. Various algorithms can be used to generate such a Markov chain. An important one is the Metropolis-Hastings algorithm, which takes an arbitrary Markov
chain and adjusts it using a simple accept-reject mechanism to ensure the stationarity of $\pi$ for the
resulting Markov chain. Fairly mild conditions then ensure that the values in the realized chain actually converge to draws from $\pi$. Another well-known algorithm is the Gibbs sampler, which partitions the vector of random variables which have distribution $\pi$ into components and replaces
each component by a draw from its conditional distribution given the current values of all other
components. Various combinations of these algorithms are also popular ({\it e.g.}~a Gibbs sampler where one or more conditionals are not easy to draw from directly and are treated through a Metropolis-Hastings algorithm). More details can be found in {\it e.g.}~\cite{RobertCasella_04} and \cite{Chib_11}\label{fn:MCMC}.}.
In particular, a popular strategy is to run an MCMC algorithm in model space, sampling the models that are most promising: the one that is most commonly used is a random-walk Metropolis sampler usually referred to as MC$^3$, introduced in  \cite{MadiganYork} and used in {\it e.g.}~\cite{Raftery_etal_97} and \cite{FLS01a}. On the basis of the application in \cite{MasanPapa_08}, \cite{Cuaresma_11} finds that MC$^3$ leads to rather different results from the leaps and bounds method, which does not seem to explore the model space sufficiently well.

The original prior in \cite{GeorgeMcCulloch_93} is not conjugate in that the prior variance of $\beta$ does not involve $\sigma^2$ (unlike (\ref{FLSprior})); this means that marginal likelihoods are not available analytically, but an MCMC algorithm can easily be implemented by a Gibbs sampler on the joint space of the parameters and the models. This procedure is usually denoted as Stochastic Search Variable Selection (SSVS). \cite{GeorgeMcCulloch_97} also introduce an alternative prior which is conjugate, leading to an analytical expression for the marginal likelihoods and inference can then be conducted using an MCMC sampler over only the model space (like MC$^3$).

\cite{Clydeetal11} remark that while the standard algorithms MC$^3$ and SSVS are easy to implement, they may mix poorly when covariates are highly correlated. More
advanced algorithms that utilize other proposals can then be considered, such as adaptive MCMC\footnote{MCMC methods often require certain parameters (of the proposal distribution) to be appropriately tuned for the algorithm to perform well. Adaptive MCMC algorithms achieve such
tuning automatically. See \cite{AtchadeRosenthal_05} for an introduction.} \citep{NottKohn_05} or evolutionary
Monte Carlo \citep{LiangWong_00}.  \cite{Clydeetal11} propose a Bayesian
adaptive sampling algorithm (BAS), that samples models without replacement from the
space of models. In particular, the probability of a model being sampled is proportional to some
probability mass function  with known normalizing constant. Every time a new model is sampled, one needs to account
for its mass by subtracting off its probability from the probability mass function to ensure that there is no duplication
and then draw a new model from the renormalized distribution. The model space is represented by a binary tree structure indicating inclusion or exclusion of each variable, and marginal posterior inclusion probabilities are set at an initial estimate and then adaptively updated using the marginal likelihoods from the sampled models.

Generic numerical methods were compared in \cite{GarciaDonato}, who identify two different strategies:
\begin{itemize}
\item[i)] MCMC methods to sample from the posterior (\ref{postmodel}) in combination with estimation based on model visit frequencies and
\item[ii)] searching methods looking for ``good'' models with estimation based on renormalization ({\it i.e.}~with weights defined by the analytic expression of posterior probabilities, such as in (\ref{PostOdds})).
\end{itemize}
Despite the fact that it may, at first sight, appear that ii) should be a more efficient strategy,
they show that i) is potentially more precise than ii) which could be biased by the searching procedure. Nevertheless, implementations of ii) have lead to fruitful contributions, and a lot of the most frequently used software (see Section \ref{Sec:software}) uses this method. Of course, if the algorithm simply generates a chain through model space in line with the posterior model probabilities (such as MC$^3$ using the prior in (\ref{FLSprior})) then both strategies can be used to conduct inference on quantities of interest, {\it i.e.}~to compute the model probabilities to be used in (\ref{BMAeq}). Indeed, \cite{FLS01a} suggest the use of the correlation between posterior model probabilities based on i) and ii) as an indicator of convergence of the chain.
However, some other methods only lend themselves to one of the strategies above. For example, the prior of \cite{GeorgeMcCulloch_93} does  not lead to closed form expressions for the marginal likelihood, so SVSS based on this prior necessarily follows the empirical strategy i). Examples of methods that can only use strategy ii) are BAS in \cite{Clydeetal11}, which only samples each model once, and the implementation in \cite{Raftery1995}  based on a leaps and bound algorithm which is used only to identify the top models. These methods need to use the renormalization strategy, as model visit frequencies are not an approximation to posterior model probabilities in their case.

MC$^3$ uses a Metropolis sampler which proposes models from a small neighbourhood of the current model, say $M_j$, namely all models with one covariate less or more. Whereas this works well for moderate values of $k$, it is not efficient in situations with large $k$ where we expect parsimonious
models to fit the data well. This is because the standard MC$^3$ algorithm (using a uniform distribution on the model neighbourhood) will propose to add a covariate with probability $(k-k_j)/k$, which is close to 1 if $k>>k_j$. Therefore, the algorithm will much more frequently propose to add a variable than to delete one. However, the acceptance rate of adding a new
variable is equal to the acceptance rate of deleting a variable if the chain is in equilibrium.
Thus, a large number of adding moves are rejected and this leads to a low between-model
acceptance rate.
\cite{Brown_etal_98} extend the MC$^3$ proposal by adding a ``swap'' move where one included and one excluded covariate are selected at random and the proposed model is the one where they are swapped. They suggest to generate a candidate model by either using the MC$^3$ move or the swap move. \cite{Lamnisos_etal_09} extend this further by decoupling the MC$^3$ move into an ``add'' and a ``delete'' move (to avoid proposing many more additions than deletions) and uniformly at random choosing whether the candidate model is generated from an ``add'', a ``delete'' or a ``swap'' move. In addition, they allow for less local moves by adding, deleting or swapping more than one covariate at a time. The size of the blocks of variables used for these moves is drawn from a binomial distribution. This allows for faster exploration of the model space. In \cite{Lamnisos_etal_13} an adaptive MCMC sampler is introduced where the success probability  of the binomial distribution (used for block size) is tuned adaptively to generate a target acceptance probability of the proposed models. They successfully manage to deal with problems like finding genetic links to colon tumours with $n = 62$ and $k = 1224$ genes in a (more challenging) probit model context (see Section \ref{sec:Probit}), where their algorithm is almost 30 times more efficient\footnote{The efficiency is here standardized by CPU time. Generally, the efficiency of a Monte Carlo method is proportional to the reciprocal of the variance of the
sample mean estimator normalized by the size of the
generated sample.} than MC$^3$ and the adaptive Gibbs sampler of \cite{NottKohn_05}. Problems with even larger $k$ can be dealt with through more sophisticated adaptive MCMC algorithms. \cite{Griffin_etal_17} propose such algorithms  which exploit the observation that in these settings the vast majority of the inclusion indicators of the variables will be virtually uncorrelated
a posteriori. They are shown to lead to orders of magnitude improvements
in efficiency compared to the standard Metropolis-Hastings algorithm, and  are successfully applied to extremely challenging  problems ({\it e.g.}~with $k=22,576$
possible covariates and $n=60$ observations).

In situations with more complicated models, it may not be practical to run an MCMC chain over model space. One option would be to use an easily computed approximation to the marginal likelihoods  and to base the MCMC model moves on that approximation (discussed in Subsection \ref{Sec:BIC}), but this may not always work well. If the number of models is not too large, another option is to treat all $K$ models separately and to compute their marginal likelihoods directly so that BMA inference can be implemented. This is not a straightforward calculation on the basis of MCMC output for each model, but there are a number of standard methods in the statistics literature. \cite{Chib_11} discusses the method of \cite{chib1995} which is essentially based on Bayes' formula evaluated at a particular value of the parameters. Another popular method is bridge sampling \citep{mengwong1996}, which is a generalization of importance sampling with a high degree of robustness with respect to the relative
tail behaviour of the importance function.
For nested models, we can often compute Bayes factors through the Savage-Dickey density ratio \citep{verdinelliwasserman1995}, which is typically both easy to calculate and accurate.
An in-depth discussion of all of these methods can be found in \cite{DiCiccio_etal_97}.

\subsection{Role of the prior}\label{Sec:RolePrior}

It has long been understood that the effect of the prior distribution on posterior model probabilities can be much more pronounced than its effect on posterior inference given a model \citep{kassraftery1995,FLS01a}. Thus, it is important to better understand the role of the prior assumptions in BMA. While \cite{FLS01a} examined the effects of choosing fixed values for $g$, a more systematic investigation of the interplay between $g$ and $w$ was conducted in \cite{LS6} and \cite{Eicheretal11}.

From combining the marginal likelihood in (\ref{MargLikelihood}) and the model space prior in (\ref{PM_fixed}), we obtain the posterior odds between models, given $g$ and $w$:
\begin{equation}
{\frac{P(M_i\,|\,y,w,g)}{P(M_j\,|\, y,w,g)}}=\left({\frac{w}{1-w}}\right)^{k_i-k_j}
(1+g)^{\frac{k_j-k_i}{2}}\left(\frac{1+g(1-R^2_i)}{1+g(1-R^2_j)}\right)^{-{\frac{n-1}{2}}}.\label{PostOdds}
\end{equation}
The three factors on the right-hand side of (\ref{PostOdds}) correspond to, respectively, a model size (or complexity) penalty induced by the prior odds on the model space, a model size penalty resulting from the marginal likelihood (Bayes factor) and a lack-of-fit penalty from the marginal likelihood. It is clear that for fixed $g$ and $w$, the complexity penalty increases with $g$ and decreases with $w$ (see also the discussion in Section \ref{sec:illusion} and in \cite{Eicheretal11}).
\cite{leysteel2012} consider each of the three factors separately, and 
define penalties as minus the logarithm of the corresponding odds factor, which ties in well with classical information criteria, which, in some cases, correspond to the limits of log posterior odds \citep{FLS01a}.
The complexity penalty induced by the prior odds can be in favour of the smaller or the larger model, whereas the penalties induced by the Bayes factor are always in favour of the smaller and the better fitting models. 

\begin{figure}[ht!]
\begin{center}
\subfigure{\includegraphics[width=10cm]{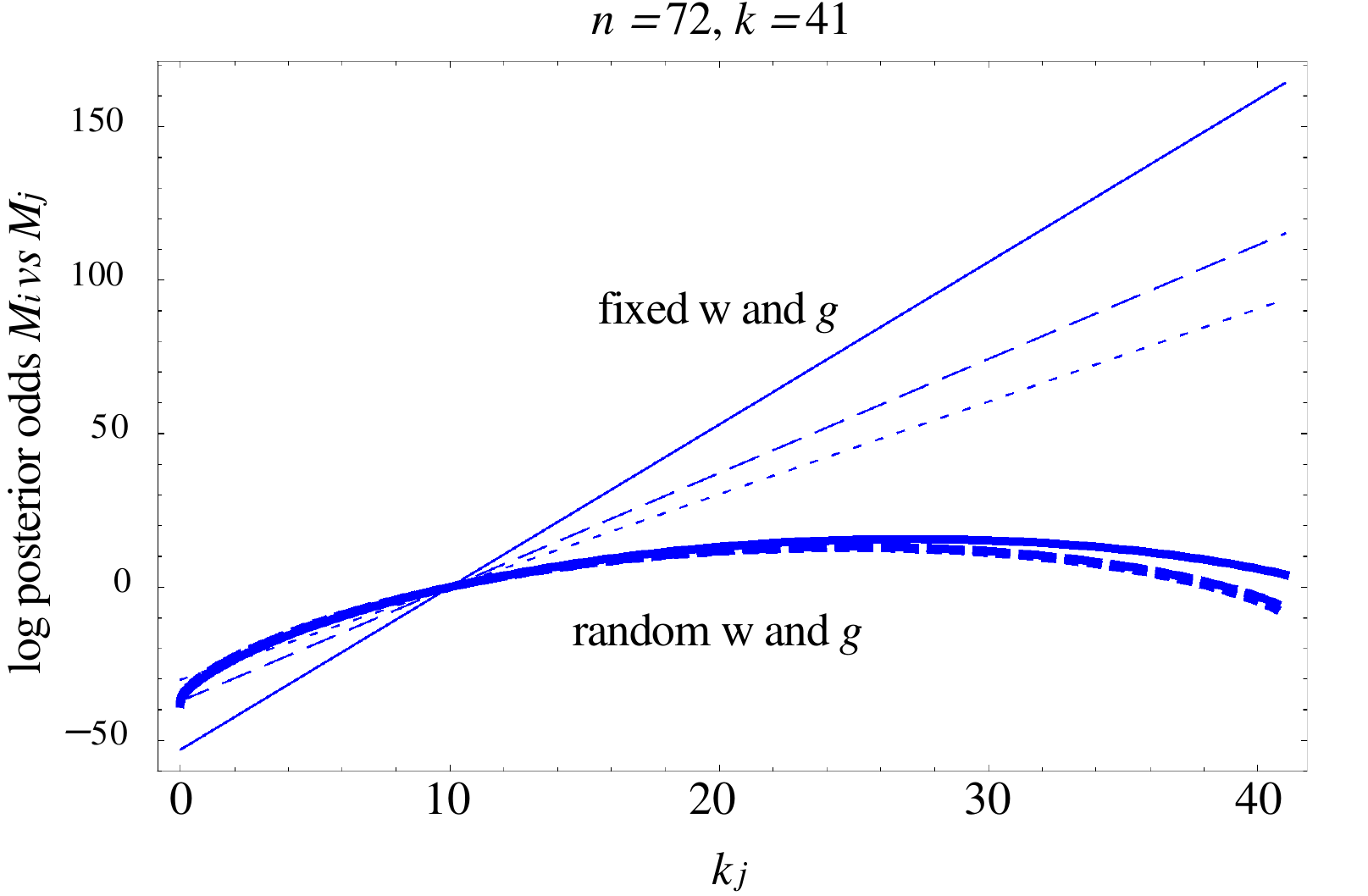}}
\caption{\footnotesize Posterior odds as a function of $k_j$ when $k_i=10$ with equal fit, using $m=7$ (solid), $m=k/2$ (dashed), and $m=2k/3$ (dotted).
 Bold lines correspond to random $w$ and $g$. From \cite{leysteel2012}.}
\label{Complex}
\end{center}
\end{figure}

\cite{leysteel2012} find that the choice of the hyperpriors on $g$ and $w$ can have a large effect on the induced penalties for model complexity but does not materially affect the impact of the relative fit of the models. They also investigate how the overall complexity penalty behaves if we integrate over $g$ and $w$. Figure \ref{Complex} plots the logarithm of the approximate posterior odds for $M_i$ versus $M_j$ as a function of $k_j$ when fixing $k_i=10$, for different values of the prior mean model size, $m$, using a beta hyperprior on $w$ as in \cite{LS6} and the benchmark beta prior on $g$ in (\ref{beta-g}) with $c=0.01$. It uses $n=72$ and $k=41$ (as in the FLS growth data) and assumes $R_i^2=R_j^2$. The figure contrasts these graphs with those for fixed values of $w$ and $g$ (corresponding to the values over which the priors are centered) as derived from (\ref{PostOdds}). For fixed values of $w$ and $g$, the log posterior odds are linear in $(k_i-k_j)$, but they are much less extreme for the case with random $w$ and $g$, and always penalize models of size around $k/2$. This reflects the multiplicity penalty (see Section  \ref{Sec:PriorModel}) which is implicit in the prior and analyzed in \cite{ScottBerger} in a more general context, and in \cite{LS6} in this same setting. No fixed $w$ can induce a multiplicity correction. The behaviour is qualitatively similar to that of the prior odds in Figure \ref{PO1}. The difference with Figure \ref{PO1} is that we now consider the complexity penalty in the posterior, which also includes an (always positive) size penalty resulting from the Bayes factor.  As in Figure \ref{PO1}, the (relatively arbitrary) choice of $m$ matters very little for the case with random $w$ (and $g$), whereas it makes a substantial difference if we keep $w$ (and $g$) fixed. 

Thus, marginalising out the posterior model probabilities with the hyperpriors on $w$ and $g$ induces a much flatter model size penalty over the entire range of model sizes. In addition, it makes the analysis less dependent on (usually somewhat arbitrary) prior assumptions and increases the relative importance of the data contribution (the model fit) to the posterior odds. 

\subsection{Data Robustness}

Generally, in economics, the quality of the data may be problematic.  An important issue is whether policy conclusions and key insights change when data are revised to eliminate errors, incorporate improved data or account for new price benchmarks. For example, the Penn World Table (PWT) income data, a
dataset frequently used in cross-country empirical work in economics, have undergone periodic revisions. \cite{CicconeJaros_10} applied the methodologies of \cite{FLS01b} and \cite{SDM} for investigating the determinants of cross-country growth to data generated as in \cite{SDM} from three different versions of the PWT, versions 6.0-6.2. Both methods led to substantial variations in posterior inclusion probabilities of certain covariates between the different datasets.

It is, of course, not surprising that solving a really complicated problem (assessing the posterior distribution on a model space that contains huge quantities of models) on the basis of a relatively small number of observations is challenging, and if we modify the data in the absence of strong prior information, we can expect some (perhaps even dramatic) changes in our inference. Clearly, if we add prior information such changes would normally be mitigated. Perhaps the most relevant question in practice is whether we can conduct meaningful inference using BMA with the kinds of prior structures that are discussed in this paper, such as (\ref{FLSprior}), on the basis of available data.

Using  BMA, \cite{FeldkircherZeugner_12} examine more in detail what causes the lack of robustness found in \cite{CicconeJaros_10}. One first conclusion is that the changes are roughly halved if the analyses with the different PWT data use the same set of countries. They also stress that the use of the fixed value of $g$ as in the benchmark prior leads to a very large $g$ and it is clear from (\ref{PostOdds}) that this amplifies the effect of differences in the fit on posterior odds. Thus, small differences in the data can have substantial impact on the results. They propose to use a hyper-$g$ prior which allows the model to adjust $g$ to the data, and this dramatically reduces the instability. Interestingly, this is not a stronger prior, but a less informative one. The important thing is that fixing $g$ at a value which is not warranted by the data quality leads to an exaggerated impact of small difference in model fit. They find that the analysis with stochastic $g$ leads to much smaller values of $g$. The same behaviour was also found in \cite{leysteel2012} where three datasets were analysed: two cross-country growth datasets as in \cite{FLS01b} (with $n=72$ and $k=41$) and \cite{SDM} (with $n=88$ and $k=67$) and the returns-to-schooling data of \cite{TobiasLi} ($n=1190$ and $k=26$). In all these examples, the data favour\footnote{This can be inferred from the likelihood which is marginalised with respect to all parameters but $g$ and averaged over models; see expression (9) in \cite{leysteel2012} which is plotted in their Figure 9.} values of $g$ in the range 15-50, which contrasts rather sharply with the fixed values of $g$ that the benchmark prior would suggest, namely 1681, 4489 and 1190, respectively. As a consequence of the smaller $g$, differences between models will be less pronounced and this can be seen as a quite reasonable reaction to relatively low-quality data.

\cite{RockeyTemple_16} consider restricting the model space by imposing the presence of initial GDP per capita and
regional dummies, {\it i.e.}~effectively using a more informative prior on the model space. They conclude that this enhances robustness even when the analysis is extended to more recent vintages of the Penn World Table (they also consider PWT 6.3-8.0).

\subsection{Collinearity and Jointness}

One of the primary outputs of a BMA analysis is the posterior distribution of the regression coefficients, which is a mixed distribution (for each coefficient a continuous distribution with mass point at zero, reflecting exclusion of the associated regressor) of dimension $k$, which is often large. Thus, this is a particularly hard object to describe adequately. Summarizing this posterior distribution merely by its $k$ marginals is obviously a gross simplification and fails to capture the truly multivariate nature of this distribution. Thus, efforts have been made to define measures that more adequately reflect the posterior distribution.
Such measures should be well suited for
extracting relevant pieces of information. It is important that they provide additional
insight into properties of the posterior that are of particular interest, and that they are easy to interpret.
\cite{LeySteel_07} and \cite{DW_09}  propose various measures of ``jointness'', or the tendency of variables to appear together in a regression model. \cite{LeySteel_07} formulate four desirable criteria for such measures to possess:
\begin{itemize}
\item Interpretability: any jointness measure should have either a formal statistical or a
clear intuitive meaning in terms of jointness.
\item Calibration: values of the jointness measure should be calibrated against some
clearly defined scale, derived from either formal statistical or intuitive arguments.
\item Extreme jointness: the situation where two variables always appear together should
lead to the jointness measure reaching its value reflecting maximum jointness.
\item Definition: the jointness measure should always be defined whenever at least one of
the variables considered is included with positive probability.
\end{itemize}
The jointness measure proposed in \cite{LeySteel_07}  satisfies all of these criteria and is defined as the posterior odds ratio
between those models that include a set of variables and the models that only include proper subsets. If we consider the simple case of bivariate jointness between variables $i$ and $j$, and we define the events $\tilde{\imath}$ and $\tilde{\jmath}$ as the exclusion of $i$ and $j$, respectively, this measure is
the  probability of joint inclusion relative to the
probability of including either regressor, but not both:
$$
{\cal J}^\circ_{ij} =\frac{P(i\cap j\given y)}{ P(i \cap \tilde \jmath\given y)+P( \tilde
\imath\cap j\given y)}. 
$$

An alternative measure, suggested by \cite{DW_09}, takes the form
$$
{\cal J}_{ij} =\ln\frac{P(i\cap j\given y) P(\tilde \imath \cap \tilde \jmath\given y)}{P(i \cap \tilde \jmath\given y) P( \tilde
\imath\cap j\given y)}, 
$$
which has the interpretation of the log of the posterior odds of including $i$ given that $j$ is included
divided by the posterior odds of including $i$ given that $j$ is not included.
\cite{LeySteel_09b} discuss how these and another jointness measure proposed by \cite{Strachan_09} compare on the basis of the criteria above. They highlight that ${\cal J}_{ij}$ is undefined when a variable is always included or excluded, and for such cases \cite{DW_09} propose to use the log posterior odds of inclusion of the variable with inclusion probability in $(0,1)$. This, however, implies that pairwise jointness involving covariates that are almost always in the model will depend entirely on the low-probability models not involving this variable.
As ${\cal J}^\circ$ is a posterior odds ratio, its values can be immediately interpreted as evidence in favour of jointness (values above one) or disjointness (values below one, suggesting that variables are more likely to appear on their own than jointly). Disjointness can occur, {\it e.g.}, when variables are highly collinear and are proxies or substitutes for each-other. In the context of two growth data sets, \cite{LeySteel_07} find evidence of jointness only between important variables, which are complements in that each of them has a
separate role to play in explaining growth. They find many more occurrences of disjointness, where regressors are substitutes and really should not appear together. However, these latter regressors tend to be fairly unimportant drivers of growth.

\cite{Man_17} compares different jointness measures 
using data from a variety of disciplines and finds that results differ substantially between the measures of \cite{DW_09} on the one hand and \cite{LeySteel_07} on the other hand. In contrast, results appear quite robust across different prior choices. \cite{Man_17} suggests the use of composite indicators, which
combine the information contained in the different concepts, often by simply averaging over different indicators.
Given the large differences in the definitions and the properties of the jointness measures considered, I would actually expect to find considerable differences. I would recommend selecting a measure that makes sense to the user while making sure the interpretation of the results is warranted by the properties of the specific measure chosen. The use of composite indicators, however interesting it may be from the perspective of combining information, seems to me to make interpretation much harder.

\cite{GhoshGattas_16} investigate the consequences of strong collinearity for Bayesian variable selection. They find that strong
collinearity may lead to a multimodal posterior distribution
over models, in which joint summaries are more appropriate
than marginal summaries. They recommend a routine calculation of the joint
inclusion probabilities for correlated covariates, in addition to
marginal inclusion probabilities, for assessing the importance
of regressors in Bayesian variable selection.



\cite{Crespo_etal_16} propose a different approach to deal with patterns of inclusion such as jointness among covariates.
They use a two-step approach starting from the posterior model distribution obtained from BMA methods, and then use clustering methods based on latent class analysis  to unveil clusters of model profiles. Inference in the second step is based on Dirichlet process clustering methods. They also indicate that the jointness measures proposed in the literature (and mentioned earlier in this subsection) relate closely to measures used in data mining (see their footnote 1). These links are further explored in \cite{Crespo_etal_17_joint}, who propose a new measure of bivariate jointness which is a regularised version of the so-called Yule's $Q$ association coefficient, used in the machine learning literature on association rules. They use insights from the latter to extend the set of desirable criteria outlined above, and show they are satisfied by the measure they propose.

\subsection{Approximations and hybrids}\label{Sec:BIC}
The use of the prior structure in (\ref{FLSprior}) for the linear normal model in (\ref{NLM}) immediately leads to a closed-form marginal likelihood, but for other Bayesian models this may not be the case. In particular, more complex models (such as described in Section \ref{Extension}) often do not allow for an analytic expression. One approach to addressing this problem is to use an approximation to the marginal likelihood, which is based on the ideas underlying the development of BIC (or the Schwarz criterion). In normal (or, more generally, regular\footnote{Regular models are such that the sampling distribution of the maximum likelihood estimator is asymptotically normal around the true value with covariance matrix equal to the inverse expected Fisher information matrix.}) models, BIC can be shown (see Schwarz, 1978\nocite{Schwarz78} and Raftery, 1995\nocite{Raftery1995}) to provide an asymptotic approximation to the log Bayes factor. In the specific context of the normal linear model with prior  (\ref{FLSprior}), \cite{FLS01a} provide a direct link between the BIC approximation and the choice of $g=n$ (the unit information prior, which essentially leaves the asymptotics unaffected). Thus, in situations where closed-form expressions for the Bayes factors are not available (or very costly to compute), BIC has been used to approximate the actual Bayes factor. For example, some  available procedures for models with endogenous regressors and models assuming Student-$t$ sampling are based on BIC approximations to the marginal likelihood.

An alternative approximation of posterior model probabilities is through the (smoothed) AIC. \cite{BurnhamAnderson_02}  provide a Bayesian justification
for AIC (with a different prior over the models than the BIC approximation) and suggest the use of AIC-based weights as posterior model probabilities. The smoothed AIC approximation is used in the context of assessing the pricing determinants of credit default swaps in \cite{PelsterVilsmeier_16}\footnote{\cite{PelsterVilsmeier_16} comment that ``Burnham and Anderson (2002) strongly suggest replacing BIC by the (smoothed)
Akaike Information Criterion (S-AIC) since BIC aims to identify the models with the
highest probability of being the true model for the data assuming a true model exists.
Since a true model in reality does not exist, BIC tends to assign too much weight to the
best model. AIC, by the contrary, tries to select the model that most adequately fits the
unknown model, and can be interpreted as the probability that a model is the expected
best model in repeated samples. Hansen (2007) reports rather poor performance of
BIC weights compared to S-AIC weights, particularly if the sample size is large.''}.

\cite{SDM} use asymptotic reasoning in the specific setting of the linear model with a $g$-prior to avoid specifying a prior on the model parameters and arrive at a BIC approximation in this manner. They call the resulting procedure BACE (Bayesian averaging of classical estimates). 
This approach was generalized to panel data by \cite{Moral-Benito12}, who proposes Bayesian averaging of maximum likelihood estimates
(BAMLE). In a similar vein, hybrids of frequentist and Bayesian methods were used in
\cite{Durlauf_etal_08} and \cite{Durlauf_etal_11} to deal with the difficult issue of endogenous regressors (later discussed in Subsection \ref{sec:endogeneity}). In particular, they propose to use BIC approximations to posterior model probabilities for averaging over classical two-stage least squares (2SLS) estimates. 
\cite{Durlauf_etal_08} comment that: ``Hybrids of this type are controversial from the perspective of the philosophical foundations of statistics and we do not pursue such issues here. Our concern is exclusively with communicating the evidentiary support across regressions; our use of averaging is simply
a way of combining cross-model information
and our posterior probabilities are simply relative weights.''

\subsection{Prior robustness: illusion or not?}\label{sec:illusion}
\begin{figure}[ht!]
\begin{center}
\subfigure{\includegraphics[width=0.45\textwidth]{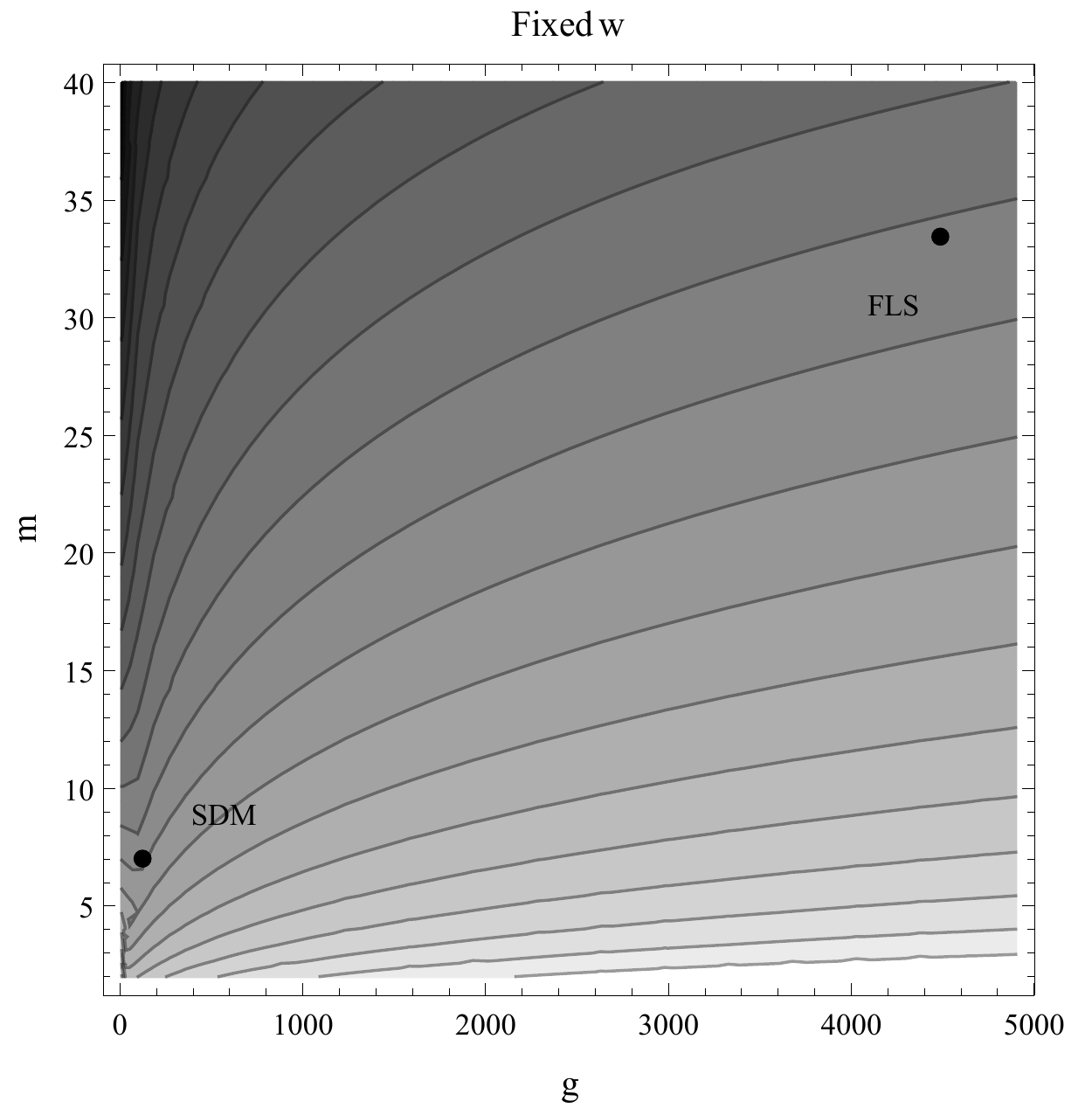}}
\subfigure{\includegraphics[width=0.45\textwidth]{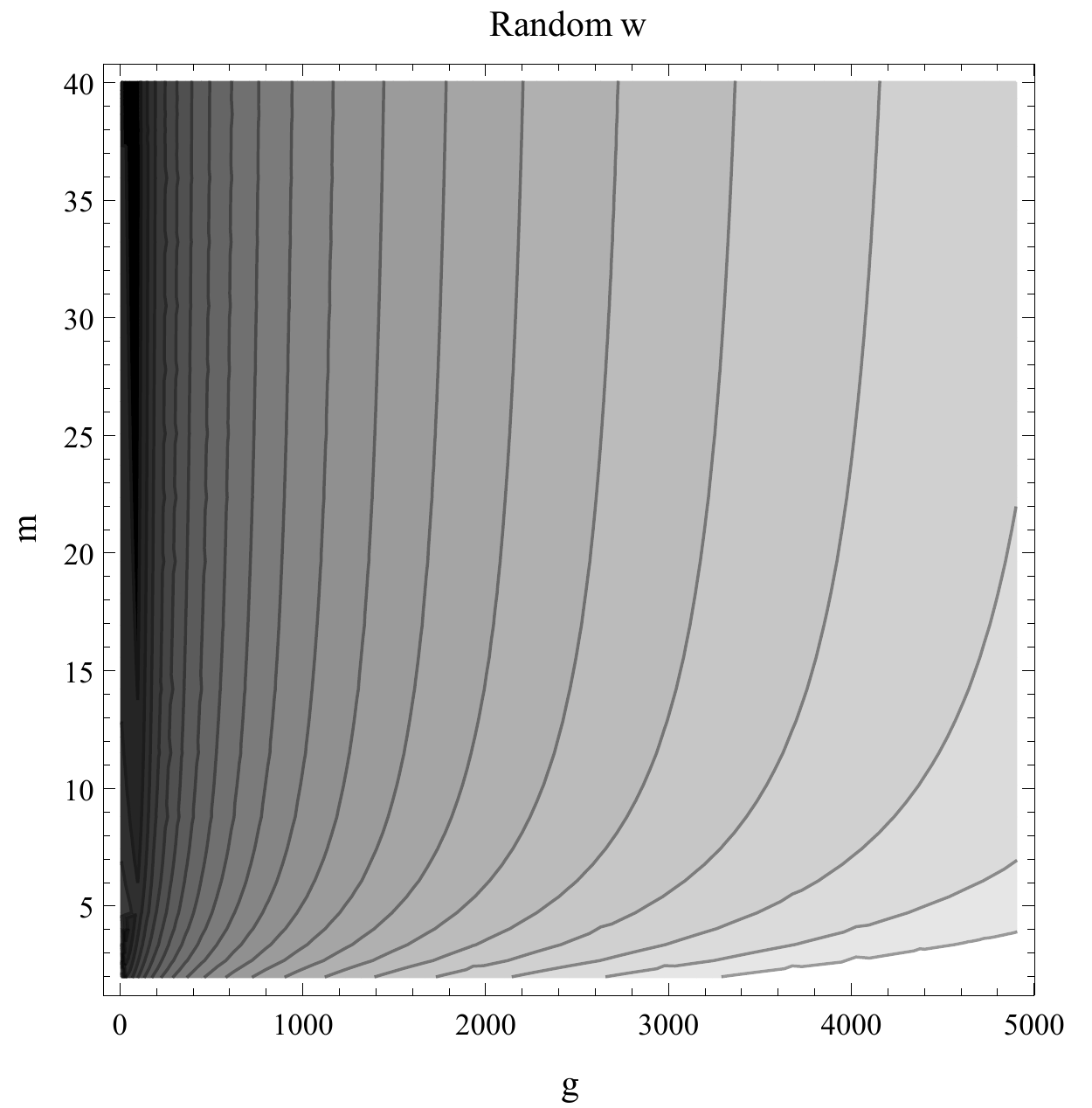}}
\caption{\footnotesize Equal Posterior Probability Contours
for different values of $R_i^2$, using $n=88, k=67, k_i=8, k_j=7$, and $R_j^2=0.75$.  The left panel is for fixed $w$ and also indicates the choices of $(g,m)$ in \cite{FLS01b} (FLS) and \cite{SDM} (SDM). The right panel corresponds to random $w$. Adapted from \cite{LS6}.}
\label{LS6Fig4}
\end{center}
\end{figure}

Previous sections have already stressed the importance of the choices of the hyperparameters and have made the point that settings for  $g$ and $w$ are both very important for the results. However, there are examples in the literature where rather different choices of these hyperparameters led to relatively similar conclusions, which might create the impression that these choices are not that critical. For example, if we do not put hyperpriors on $g$ and $w$, we note that the settings used in \cite{FLS01b} and in \cite{SDM} lead to rather similar results in the analysis of the growth data of \cite{SDM}, which have $n=88$ and $k=67$. The choices made in \cite{FLS01b} are $w=0.5$ and $g=k^2$, whereas the BACE analysis in \cite{SDM} is based on $w=7/k$ (giving a prior mean model size $m=7$) and $g=n$. As BACE attempts to avoid specifying a prior on the model parameters, the latter is not explicit, but follows from the expression used in \cite{SDM} for the Bayes factors, which is a close approximation to Bayes factors for the model with prior (\ref{FLSprior}) using $g=n$. There is, however, an important tradeoff between values for $g$ and $w$, which was visually clarified in \cite{LS6} and was also mentioned in \cite{Eicheretal11}. In particular, \cite{LS6} present a version of Figure \ref{LS6Fig4} which shows the contours in $(g, m)$ space of the values of fit ($R_i^2$) of Model $M_i$ that would give it equal posterior probability to $M_j$, when $n=88, k=67, k_i=8, k_j=7$, and $R_j^2=0.75$.

From the left panel in Figure \ref{LS6Fig4} the particular combinations of $(g, w)$ values underlying the analyses in \cite{FLS01b} and in \cite{SDM} are on contours that are quite close, and thus require a very similar increase in $R^2$ to compensate for an extra regressor (in fact, the exact values are $R_i^2=0.7731$ for FLS and $R_i^2=0.7751$ for SDM). Remember from Section \ref{Sec:RolePrior} that the model complexity penalty increases with $g$ and decreases with $w$ (or $m=wk$), so the effects of increasing both $g$ and $w$ (as in \cite{FLS01b} with respect to \cite{SDM}) can  cancel each-other out, as they do here.

In conclusion, it turns out that certain (often used) combinations happen to give quite similar results. However, this does not mean that results are generally robust with respect to these choices, and there is ample evidence in the literature \citep{LS6,Eicheretal11} that these choices matter quite crucially. Also, it is important to point out that making certain prior assumptions implicit (as is done in BACE) does not mean they no longer matter.
Rather, it seems to me more useful to be transparent about prior choices and to attempt to robustify the analysis by using prior structures that are less susceptible to subjectively chosen quantities. This is illustrated in the right panel of Figure \ref{LS6Fig4}, where the equal probability contours are drawn for the case with a Beta$(1,(k-m)/m)$ hyperprior on $w$. As discussed in Section \ref{Sec:PriorModel}, this prior is much less informative, which means that the actual choice of $m$ matters much less and the trade-off between $g$ and $w$ has almost disappeared. A hyperprior can also be adopted for $g$, as in Section \ref{EB_hier}, which would further robustify the analysis (see also Section \ref{Sec:RolePrior}).


\subsection{Other sampling models}\label{Extension}

This section describes the use of BMA in the context of other sampling models, which are sometimes fairly straightforward extensions of the normal linear regression model (for example, the \cite{Hoeting_etal_96} model for outliers in Section \ref{Sec:Outlier} or the Student-$t$ model mentioned in Section \ref{sec:otherExt}) and sometimes imply substantial challenges in terms of prior elicitation and numerical implementation. Many of the models below are inspired by issues arising in economics, such as dynamic models, spatial models, models for panel data and models with endogenous covariates.

\subsubsection{Generalized linear models}\label{GLM}

Generalized Linear Models (GLMs) describe a more general class of models \citep{McCullaghNelder_89} that covers the normal linear regression model but also regression models where the response variable is non-normal, such as binomial (e.g.~logistic or logit regression models, probit models), Poisson, multinomial (e.g.~ordered response models, proportional odds models) or gamma distributed.

An early contribution to BMA with GLMs is \cite{Raftery_96}, who proposes to use approximations for the Bayes factors, based on the Laplace method for integrals. He also suggests a way to elicit reasonable (but data-dependent) proper priors.

 \cite{BoveHeld_11} consider the interpretation of the $g$-prior in linear models as the conditional posterior of the regression coefficients
 given a locally uniform
prior and an imaginary sample of zeros with design
matrix $Z_j$
 and a scaled error variance, and extend this to the GLM context. Asymptotically, this leads to a prior which is very similar to the standard $g$-prior, except that it has an extra scale factor $c$ and a weighting matrix $W$ in the covariance structure. In many cases, $c=1$ and $W=I$, which leads to exactly the same structure as (\ref{FLSprior}). This idea was already used in the conjugate prior proposed by \cite{ChenIbrahim_03},
although they only considered the case with $W=I$ and do not treat the intercept separately. For priors on $g$, \cite{BoveHeld_11} consider a Zellner-Siow prior and a hyper-$g/n$ prior. Both choices are shown to lead to consistent model selection in \cite{Wu_etal_15}.

The priors on the model parameters designed for GLMs in \cite{LiClyde_17} employ a different type of ``centering'' of the covariates (induced by the observed information matrix at the maximum likelihood estimator (MLE) of the coefficients), leading to a $g$-prior that displays local orthogonality properties at the MLE. In addition, they use a wider class of (potentially truncated) hyper-priors for $g$\footnote{In particular, they use the class of compound confluent hypergeometric distributions, which contains most hyperpriors used in the literature as special cases.}. Their results rely on approximations, and, more importantly, their prior structures are data-dependent (depending on $y$, not just the design matrix). Interestingly, on the basis of theoretical and empirical findings in the GLM context, they recommend similar hyper-priors\footnote{Namely, the hyper-$g/n$ prior and the benchmark beta prior.} as recommended by \cite{leysteel2012} in a linear regression setting.

The power-conditional-expected-posterior prior of \cite{FouskakisNtzoufras_16a} has also been extended to the GLM setting in \cite{Perrakis_etal_15}. \cite{Wu_etal_18} propose and investigate the use of nonlocal priors (as in \cite{JohnsonRossell_10}, but with a hyperprior on scale parameters) designed for GLMs.

\subsubsection{Probit models}\label{sec:Probit}

A popular approach for modelling dichotomous responses uses the probit model, which is an example of a GLM. If we observe $y_1,\dots,y_n$ taking the values either zero or one, this model assumes that the probability that $y_i=1$ is
modeled by
$
y_i|\eta_i\sim \mbox{Bernoulli}(\Phi\left(\eta_i)\right)
$
where $\Phi$ is the cumulative distribution function of a
standard normal random variable and ${\eta}=(\eta_1,\eta_2,\dots,\eta_n)'$ is a vector of linear
predictors modelled as $
{\eta} =\alpha\iota+{Z_j}{\beta_j},
$
where
  $\alpha$, ${\beta_j}$ and $Z_j$ are as in (\ref{NLM}).

Typical priors have a product structure with a normal prior on $\beta_j$ (for example a $g$-prior) and an improper uniform on $\alpha$.  Generally, posterior inference for the probit model can be facilitated by using
the data augmentation approach of \cite{AlbertChib_93}.

When dealing with model uncertainty, this model is often analysed through a Markov chain Monte Carlo method on the joint space of models and model parameters, since the marginal likelihood is no longer analytically available.  This complicates matters with respect to the linear regression model as this space is larger than model space and the dimension of the model parameters varies with the model. Thus, reversible jump Metropolis-Hastings methods (Green, 1995) are typically used here.
Details and comparison of popular algorithms can be found in \cite{Lamnisos_etal_09}.

\subsubsection{Generalized additive models}

Generalized additive models are generalized linear models in which the linear predictor depends linearly on unknown smooth functions of the covariates, so these models can account for nonlinear effects; see e.g.~\cite{Hastie_etal_09}. In the context of the additive model\footnote{This is where the link function is the identity link, so we have a normally distributed response variable.}, \cite{SabanesHeld_11} consider using fractional polynomials for these smooth functions in combination with a hyper-$g$ prior. They combine covariate  uncertainty with flexible modelling of additive effects by expanding the model space to include different powers of each potential regressor. To explore this large model space, they propose an MCMC algorithm which adapts the Occam's window strategy of \cite{Raftery_etal_97}. Using splines for the smooth functions, \cite{Sabanes_etal_15} propose hyper-$g$ priors based on an iterative weighted least squares approximation to the nonnormal likelihood. They conduct inference using an algorithm which is quite similar to that in \cite{BoveHeld_11}.

\subsubsection{Outliers}\label{Sec:Outlier}
The occurrence of outliers (atypical observations) is a general problem that may affect both parameter estimation and model selection, and the issue is especially relevant if the modelling assumptions are restrictive, for example by imposing normality ({\it i.e.}~thin tails). In the context of normal linear regression, \cite{Hoeting_etal_96} propose a Bayesian method for simultaneous variable selection and outlier identification, using variable inflation to model outliers.  They use a proper prior and recommend the use of a pre-screening procedure to generate a list of potential outliers, which are then used to define the model space to consider. \cite{Ho_15} applies this methodology to explore the cross-country variation in the output impact of the global financial crisis in
2008-9.

Outliers are also accommodated in \cite{Doppelhofer_etal_16}. In the context of growth data, they also introduce heteroscedastic measurement error, with variance potentially differing with country and data vintage. The model also accounts for vintage fixed effects and outliers. They use data from eight vintages of the PWT (extending the data used in \cite{SDM}) to estimate the model, and conclude that 18 variables are relatively robustly  associated with GDP growth over the period 1960
to 1996, even when outliers are allowed for. The quality of the data seems to improve in later vintages and varies quite a bit among the different countries. They estimate the model using JAGS, a generic MCMC software package which determines the choice of sampling strategy, but this approach is very computer-intensive\footnote{They comment that a single MCMC run takes about a week to complete, even with the use of multiple computers and parallel chains.}.

Of course, the use of more flexible error distributions such as scale mixtures of normals (like, for example, the Student-$t$ regression model mentioned in the next section) can be viewed as a way to make the results more robust against outliers.

\subsubsection{Non-normal errors}\label{sec:otherExt}

\cite{DoppelhoferWeeks_11} use a Student-$t$ model as the sampling model, instead of the normal model in (\ref{NLM}) in order to make inference more robust with respect to outliers and unmodelled heterogeneity. They consider either fixing the degrees of freedom parameter of the Student-$t$ distribution or estimating it and they use the representation of a Student-$t$ as a continuous scale mixture of normals. Throughout, they approximate posterior model probabilities by the normality-based BIC, so the posterior model probabilities remain unaffected and only the estimates of the model parameters are affected\footnote{For each model they propose a simple Gibbs sampler setup after augmenting with the mixing variables.}. \cite{Oberdabernig_etal_16} use a Student-$t$ sampling model with fixed degrees of freedom in a spatial BMA framework  to investigate the drivers of differences in democracy levels across countries. A finite location-scale mixture of Normals for the sampling distribution is considered in \cite{Ranciati_etal_19}, who propose to use a Gibbs sampler with priors as in \cite{GeorgeMcCulloch_93}.

Non-normality can, of course, also be accommodated by transformations of the data. \cite{Hoeting_etal_02} combine selection of covariates with the simultaneous choice of a transformation of the dependent variable within the Box-Cox family of transformations. \cite{Charitidou_etal_18} consider four different families of transformations along with covariate uncertainty and use model averaging based on intrinsic and fractional Bayes factors.



\subsubsection{Dynamic models}\label{Sec:DMA}

In the context of simple AR(F)IMA time-series models, BMA was used in {\it e.g.}~
\cite{Koop_etal_97}, which will be discussed in more detail in Subsection \ref{sec:output}.

\cite{Raftery_etal_10} propose the idea of using state-space models in order to allow for the forecasting model to change
over time while also allowing for coefficients in each model to evolve over
time. Due to the use of approximations, the computations essentially boil down to
the Kalman filter.
In particular, they use the following dynamic linear model, where the subscript indicates time $t=1,\dots,T$:
\begin{eqnarray}
y_t&\sim& N(z_t^{(j)'}\theta_t^{(j)}, H^{(j)})\\
\theta_{t}^{(j)}&\sim& N(\theta_{t-1}^{(j)}, Q_t^{(j)}),\label{state}
\end{eqnarray}
and the superscript is the model index with models differing in the choice of covariates in the first equation. Choosing $Q_t^{(j)}$ sequences is not required as they propose to use a forgetting factor (discount factor) on the variance of the state equation (\ref{state}). Using another forgetting factor, \cite{Raftery_etal_10} approximate the model probabilities at each point in time, which greatly simplifies the calculations. Dynamic model averaging (DMA) is where these model weights are used to average in order to conduct inference, such as predictions, and dynamic model selection (DMS) uses a single model for such inference (typically the one with the highest posterior probability) at each point in time. \cite{KoopKorobilis12} apply DMA and DMS to inflation forecasting, and find that the best predictors change considerably over time and that DMA and DMS lead to improved forecasts with respect to the usual autoregressive and time-varying-parameter models.
\cite{Drachal_16} investigates the determinants of monthly spot oil prices between 1986 and 2015, using Dynamic Model Averaging (DMA) and Dynamic Model Selection (DMS). Although some interesting patterns over time were revealed, no significant evidence was found that DMA is superior in terms of forecast accuracy over, for example, a simple ARIMA model (although this seems to be based only on point forecasts, and not on predictive scores).
Finally, \cite{Onorante16} introduce a dynamic Occam's window to deal with larger model spaces.

\cite{vanderMaas_14} proposes a dynamic BMA framework that allows for time variation in the set of variables
that is included in the model, as well as structural breaks in the intercept and conditional variance.
This framework is then applied to real-time forecasting of
inflation.


Other time-varying Bayesian model weight schemes are considered in \cite{Hoogerheide_etal_10}, who find that they outperform
other combination forecasting schemes in terms of predictive and economic gains. They suggest forecast combinations based on a regression approach with the predictions of different models as regressors and with time-varying regression coefficients.

Another increasingly important area of dynamic model averaging is the use of BMA methods for ensembles of deterministic models, which was introduced for probabilistic weather forecasting in \cite{Raftery_etal_05}. BMA is used to post-process the forecasts provided by ensembles\footnote{An ensemble is a collection of runs of deterministic models which differ in initial conditions or physical model assumptions.}, where each forecast is associated with a predictive distribution and weights are computed by maximum likelihood (implemented through an expectation-maximization algorithm).


\subsubsection{Endogeneity}\label{sec:endogeneity}

Endogeneity occurs if one or more of the covariates is correlated with the error term in the equation corresponding to (\ref{NLM}).
In particular, consider the following extension of the model in (\ref{NLM}):
\begin{eqnarray}
y &=&\alpha \iota + x \gamma+ Z_j \beta_j +\varepsilon \\
x &=& W \delta + \nu,
\end{eqnarray}
where $x$ is an endogenous regressor\footnote{For simplicity, I focus the presentation on the case with one endogenous regressor, but this can immediately be extended.} and $W$ is a set of instruments, which are independent of $\varepsilon$.  
Finally, the error terms corresponding to observation $i$ are identically and independently distributed as follows:
\begin{equation}
(\varepsilon_i, \nu_i)' \sim N(0, \Sigma),
\end{equation}
with $\Sigma=(\sigma_{ij})$ a $2 \times 2$ covariance matrix. It is well-known that whenever $\sigma_{12}\ne 0$ this introduces a bias in the OLS estimator of $\gamma$ and a standard classical approach is the use of 2SLS estimators instead. For BMA it also leads to misleading posterior inference on coefficients and model probabilities, even as sample size grows, as shown in \cite{Miloschewski_16}.

\cite{Tsangarides04} and \cite{Durlauf_etal_08} consider the issue of endogenous regressors in a BMA
context. \cite{Durlauf_etal_08} focus on uncertainty surrounding the selection of the
endogenous and exogenous variables  and propose to average over 2SLS model-specific estimates for each single model.
\cite{Durlauf_etal_11} consider model averaging across just-identified models (with as many instruments as endogenous regressors). In this
case, model-specific 2SLS estimates coincide with LIML estimates, which means that
likelihood-based BIC weights have some formal justification.

\cite{Lenkoski_etal_14} extend BMA to formally account for model uncertainty not only
in the selection of endogenous and exogenous variables, but also in the selection of instruments. They propose a two-step procedure that first averages across the first-stage models ({\it i.e.}~linear regressions of the
endogenous variables on the instruments) and then, given the fitted endogenous regressors from the first
stage, it again takes averages in the second stage. Both steps use BIC weights. Their approach, named two-stage BMA (2SBMA), was used in \cite{EicherKuenzel_16} in the context of establishing the effect of trade on growth, where feedback (and thus endogeneity) can be expected.

\cite{Koop_et_al_12} use simulated tempering to design an MCMC method that can deal with BMA in the endogeneity context in one step. It is, however, quite a complicated and computationally costly algorithm and it is nontrivial to implement.

\cite{KarlLenkoski_12} propose IVBMA, which is based on the Gibbs sampler of \cite{Rossi_etal_06} for instrumental variables models and use conditional Bayes factors to account for model uncertainty in this Gibbs algorithm. Their algorithm hinges on certain restrictions ({\it e.g.}~joint Normality of the errors is important and the prior needs to be conditionally conjugate), but it is very efficient and is implemented in an {\tt R}-package (see Section \ref{Sec:software}).

\subsubsection{Panel data and individual effects}\label{Sec:Panel}

Panel (or longitudinal) data contain information on individuals ($i = 1,\dots,N$) over different time
periods ($t = 1,\dots, T$). Correlation between covariates and error term might arise through a
time-invariant individual effect, denoted by $\eta_i$ in the model
\begin{equation}\label{panelmodel}
y_{it} = z_{it}'\beta + \eta_i + \epsilon_{it}.
\end{equation}
\cite{Moral-Benito12} uses BMA in such a panel setting with strictly
exogenous regressors (uncorrelated with the $\epsilon_{it}$s but correlated with the individual effects).
In this framework, the vector of regressors can also include a lagged dependent variable ($y_{it-1}$) which is then correlated
with $\epsilon_{it-1}$.  \cite{Moral-Benito12} considers such a dynamic panel model within the BMA approach
by combining the likelihood function discussed in \cite{AlvarezArellano03} with the unit information $g$-prior.
Distinguishing between ``within'' and ``between'' estimators in a BMA context is advocated by \cite{Desbordes_etal_18}, who stress that cross-sectional and time-series relationships can be quite different. They use a BIC approximation for the log Bayes factors and average over classical estimators (both discussed in Subsection \ref{Sec:BIC}).

\cite{Tsangarides04} addresses the issues of endogenous and omitted
variables by incorporating a panel data system Generalized Method of Moments (GMM) estimator. This was extended to the limited information BMA (LIBMA) approach of \cite{MiresteanTsangarides_15} by \cite{Chen_etal_16}, in the context of short-$T$ panel models with endogenous covariates using a GMM approximation to the likelihood. They then employ a BIC approximation of the limited information marginal likelihood.
\cite{Moral-Benito_16} remarks on the controversial nature of combining frequentist GMM procedures with BMA, as it is not firmly rooted in formal statistical foundations and GMM methods may require mean stationarity. Thus, \cite{Moral-Benito_16} proposes the use of a suitable likelihood function (derived in \cite{Moral-Benito_13}) for dynamic panel data with fixed effects and weakly exogenous\footnote{This implies that past shocks to the dependent variable can be correlated with current covariates, so that there is feedback from the dependent variable to the covariates} regressors, which is argued to be the most relevant form of endogeneity in the growth regression context. Posterior model probabilities are based on the BIC approximation of the log Bayes factors with a unit-information $g$-prior (see Section \ref{Sec:BIC}) and on a uniform prior over model space.

\cite{Leon-GonzalezMontolio_15} develop BMA methods for models for panel data with individual effects and endogenous regressors, taking into account the uncertainty regarding the choice of instruments
and exogeneity restrictions. They use reversible jump MCMC methods (developed by \cite{Koop_et_al_12}) to deal with a model space that
includes models that differ in the set of regressors, instruments, and exogeneity
restrictions in a panel data context. 

\subsubsection{Spatial data}\label{sec:spatial}

If we wish to capture spatial interactions in the data, the model for panel data in (\ref{panelmodel}) can be extended to a Spatial Autoregressive (SAR) panel model as follows:
\begin{equation}\label{SARmodel}
y_{it} = \rho \sum_{j=1}^N w_{ij} y_{jt} + z_{it}'\beta + \eta_i + \xi_t+ \epsilon_{it},
\end{equation}
where $i=1,\dots,N$ denotes spatial location and $w_{ij}$ is the $(i,j)^{th}$ element of the spatial weight matrix reflecting spatial proximity
of the $N$ regions, with $w_{ii} = 0$ and the matrix is normalized to have row-sums of unity. Finally, there are regional effects $\eta_i$ and time effects $\xi_t, t=1\dots,T$.
BMA in this model was used in \cite{LeSage_14}, building on earlier work, such as \cite{LeSageParent_07}. \cite{Cuaresma_etal_17} use SAR models to jointly model income growth and human capital accumulation and mitigate the computational requirements by using an approximation based on spatial eigenvector filtering as in \cite{CuaresmaFeldkircher_13}.
\cite{HortasRios_16} investigate the drivers of urban income inequality using Spanish municipal data. They follow the framework of \cite{LeSageParent_07} to incorporate spatial effects in the BMA analysis.
\cite{PiribauerCuaresma_17} compare the relative performance of the BMA methods used in \cite{LeSageParent_07} with two different versions of the SVSS  method (see Section \ref{sec:numerical}) for spatial
autoregressive models. On simulation data the SVSS approaches tended to perform better in terms of both in-sample predictive performance and computational
efficiency.

An alternative approach was proposed by \cite{DearmonSmith_16}, who use the nonparametric technique of Gaussian process regression to accommodate spatial patterns and develop a BMA version of this approach. They apply it to the FLS growth data augmented with spatial information.

\subsubsection{Duration models}

BMA methods for duration models were first examined by \cite{Volinsky_etal_97} in the context of proportional hazard models and based on a BIC approximation. \cite{KourtellosTsangarides_16} set out to uncover the correlates of the duration of growth spells. In particular, they investigate the relationship between
inequality, redistribution, and the duration of growth spells in the presence of other possible determinants. They employ BMA for Cox hazards models and extend the
BMA method developed by \cite{Volinsky_etal_97} to allow for time-dependent
covariates in order to properly account for the time-varying feedback effect of the
variables on the duration of growth spells.
\cite{Traczynski_17} uses BMA for predicting firm
bankruptcies and defaults at a 12-month horizon using hazard models. The analysis is based on a Laplace approximations for the marginal likelihood, arising from the logistic likelihood and a $g$-prior. On model space, a collinearity-adjusted dilution prior is chosen.
Exact BMA methodology was used to identify risk factors associated with dropout and delayed graduation in higher education in \cite{VallejosSteel_17}, who employ a discrete time competing
risks survival model, dealing simultaneously with university outcomes and its associated
temporal component. For each choice of regressors, this amounts to a multinomial logistic regression model, which is a special case of a GLM. They use the prior as in \cite{BoveHeld_11} in combination with  the hyper-$g/n$ prior (see Subsection \ref{cons}). 


\section{Frequentist model averaging}\label{FMA}

Frequentist methods\footnote{These are based on the ``classical'' statistical methodology which still underlies most introductory textbooks in statistics and econometrics.}  are inherently different to Bayesian methods, as they tend to focus on estimators and their properties (often, but not always, in an asymptotic setting) and do not require the specification of a prior on the parameters. Instead, parameters are treated as fixed, yet unknown, and are not assigned any probabilistic interpretation associated with prior knowledge or learning from data. Whereas Bayesian inference on parameters typically centers around the uncertainty (captured by a full posterior distribution) that remains after observing the sample in question, frequentist methods usually focus on estimators that have desirable properties in the context of repeated sampling from a given experiment.


Early examples of FMA can be found in the forecasting literature, such as the forecast combinations of \cite{BatesGranger}. This literature on forecast combinations (discussed here more in detail in Subsection \ref{sec:CombForecasts}) has become quite voluminous, see {\it e.g.}~\cite{Granger_89} and \cite{StockWatson_06} for reviews, while useful surveys of  FMA
 can be found in  \cite{BurnhamAnderson_02}, \cite{Wangetal09}, \cite{UllahWang_13} and \cite{Dorman_etal_18}. 


In the context of the linear regression model in (\ref{NLM}), FMA estimators can be described as
\begin{equation}
{\hat\beta}_{FMA}=\sum^K_{j=1} \omega_j {\hat\beta}_j,
\label{eqFMA}
\end{equation}
where ${\hat\beta}_j$ is an estimator based on model $j$ and $\omega_j, j=1\dots,K$ are weights in the unit simplex within $\Re^K$. The critical choice is then how to choose the weights.

\cite{Buckland_etal_97} construct weights based on different
information criteria. They propose using
\begin{equation}
\omega_j = {\frac{\exp(-I_j/2)}{\sum_{i=1}^K\exp(-I_i/2)}},
\end{equation}
where $I_j$ is an information criterion for model $j$, which can be the AIC or the BIC. \cite{BurnhamAnderson_02} recommend the use of a modified AIC criterion, which has an additional small-sample second order bias correction term. They argue that this modified AIC should be used whenever $n/k<40$.

\cite{HjortClaeskens_03} build a general large-sample likelihood framework to describe limiting distributions and risk properties of estimators post-selection as well as of model averaged estimators. A crucial tool for this analysis is their assumption of local misspecification, which allows them to derive asymptotic results in a framework that mimicks the effects of model uncertainty (however, see footnote \ref{fn:localmis}). Their approach also explicitly takes modeling bias into account. Besides suggesting various FMA procedures (based on e.g.~AIC, the focused information criterion, FIC\footnote{The underlying idea of FIC is that it focuses the analysis on parameters of interest, by ignoring the risk or loss associated with parameters that are not of interest.}, of \cite{ClaeskensHjort_03} and empirical Bayes ideas), they provide a frequentist view of the performance of BMA schemes (in the sense of limiting distributions and large sample approximations to risks).

\cite{Hansen_07} proposed a least squares model averaging estimator with model
weights selected by minimizing the Mallows' criterion ($C_p$ as defined in footnote \ref{fn:Cp}). This
estimator, known as Mallows model averaging (MMA), is easily implementable for linear
regression models and has certain asymptotic optimality properties, since the Mallows' criterion is asymptotically equivalent to the squared predictive error. Therefore, the MMA estimator minimizes the
squared error in large samples. \cite{Hansen_07} shows that the weight vector chosen by
MMA achieves optimality in the sense conveyed by \cite{Li_87}. The assumed data generating process is an infinite linear regression model, but the model space is kept quite small by asking the user to order the regressors and by considering only the sequence of approximating models $M_1,\dots,M_K$ where the $j$th
model uses the first $k_j$ covariates, with $0\le k_1<k_2<\dots<k_K$ for some $K\le n$ such that 
the matrix of covariates for $M_K$ is of full column rank.

\cite{HansenRacine_12} introduced another
estimator within the FMA framework called jackknife model averaging
(JMA) that selects appropriate weights for averaging models by minimizing a cross-validation
(leave-one-out) criterion. JMA is asymptotically optimal in the sense of reaching
the lowest possible squared errors over the class of linear estimators. Unlike MMA, JMA has optimality properties under heteroscedastic errors and when the candidate models are non-nested.


\cite{Liu}  derives the limiting distributions of least squares averaging estimators for linear regression
models in a local asymptotic framework. The averaging estimators with fixed weights are shown to be
asymptotically normal and a plug-in averaging estimator is proposed that minimizes the sample analog
of the asymptotic mean squared error. This estimator is compared with the MMA and JMA estimators and the one based on FIC. The asymptotic distributions
of averaging estimators with data-dependent weights are shown to be nonstandard and a simple procedure to construct valid confidence
intervals is proposed.


\cite{Liu_etal_16} extend MMA to linear regression models with heteroscedastic errors, and propose a model averaging method that combines generalized least squares (GLS) estimators. They derive $C_p$-like criteria to determine the model weights and show they are optimal in the sense of asymptotically achieving the smallest possible MSE. They also consider feasible versions using both parametric and nonparametric estimates of the error variances. Their objective is to obtain an estimator that generates
a smaller MSE, which they achieve by choosing weights to minimize an estimate of the
MSE. They compare their methods with those of
\cite{Magnus_etal_11}, who also average feasible GLS estimators. 

Most asymptotically optimal FMA methods have been developed for linear models, but \cite{Zhang_etal_17} specifically consider GLMs (see Section \ref{GLM}) and generalized linear mixed-effects
models\footnote{These models are GLMs with so-called random effects, {\it e.g.}~effects that are subject-specific in a panel data context.} and propose weights based on a plug-in estimator
of the Kullback-Leibler loss plus a penalty term. They prove asymptotic optimality (in terms of  Kullback-Leibler loss) for fixed or growing numbers of covariates. Another extension of the linear model is the partial linear model, which includes both linear and nonparametric components in the regression function. FMA for such models is investigated in \cite{ZhangWang_18} who do not consider estimation of the parameters, but focus on finding weights that minimize the squared prediction error. They use kernel smoothing to estimate each individual model, which implies a serious computational burden when the number of candidate models is large. For larger model spaces, \cite{ZhangWang_18} recommend a model screening step (see Subsection \ref{Sec:compFMA}) prior to model averaging. This semiparametric setting is taken one step further in \cite{Zhu_etal_18}, who consider partial linear models with varying coefficients. They also allow for heteroscedastic errors and assume the ``true'' model is infinite-dimensional, as {\it e.g.}~in \cite{Hansen_07}. They propose Mallows-type weights based on minimizing expected predictive squared error, and prove asymptotic optimality of the resulting FMA estimator (in terms of squared error loss). \cite{Li_etal_18} also propose an FMA approach based on semiparametric varying coefficient models, using MMA weights. In a fully nonparametric regression context, \cite{HendersonParmeter_16}  propose a nonparametric regression estimator averaged
over the choices of kernel, bandwidth selection mechanism and local-polynomial
order, while \cite{Tu_18} uses local linear kernel estimation of individual models and combines them with weights derived through maximizing entropy.

FMA was used for forecasting with factor-augmented regression models in \cite{Cheng_Hansen_15}.
In the context of growth theory, \cite{SiM_97} uses (\ref{eqFMA}), and focuses on the ``level of confidence''\footnote{This is defined as the maximum probability mass one side of zero for a Normal distribution centred at the estimated regression coefficient with the corresponding estimated variance.}, using weights that are either uniform or based on the maximized likelihood.

Another model-averaging procedure that has been proposed in \cite{Magnus_etal_10} and
reviewed in \cite{MagnusSurvey} is weighted average
least squares (WALS), which can be viewed as being in between BMA and FMA. The weights it implies in (\ref{eqFMA}) can be given a Bayesian justification. However, it assumes no prior on the model space and thus can not produce inference on posterior model probabilities. WALS is easier to compute than BMA or FMA, but quite a bit harder to explain and inherently linked to a nested linear regression setting. \cite{MagnusSurvey} provide an in-depth description of WALS and its relation to BMA and FMA.  They state: ``The WALS procedure surveyed in this paper is a Bayesian combination of frequentist estimators. The
parameters of each model are estimated by constrained least squares, hence frequentist. However, after
implementing a semiorthogonal transformation to the auxiliary regressors, the weighting scheme is
developed on the basis of a Bayesian approach in order to obtain desirable theoretical properties such
as admissibility and a proper treatment of ignorance. The final result is a model-average estimator that
assumes an intermediate position between strict BMA and strict FMA estimators [....] Finally we emphasize (again) that WALS is a model-average procedure, not a model-selection
procedure. At the end we cannot and do not want to answer the question: which model is best? This
brings with it certain restrictions. For example, WALS cannot handle jointness \citep{LeySteel_07,DW_09}. The concept of jointness refers to the dependence between explanatory
variables in the posterior distribution, and available measures of jointness depend on posterior inclusion
probabilities of the explanatory variables, which WALS does not provide.''
The choice of the weight function in WALS is based on  risk (or MSE) considerations, and can be given a frequentist and a Bayesian flavour: in the latter framework admissibility is ensured, so it is favoured by \cite{MagnusSurvey}. In particular, for transformations of the regressions coefficients they consider robust\footnote{This is a specific implementation of the robustness idea in Subsection \ref{sec:prior}: a robust prior has fatter tails than a normal and is often defined as a prior for which $p'(x)/p(x)\to 0$ as $x\to \infty$.} priors in the class of reflected generalized gamma distributions (with a Laplace prior as a (non-robust) special case).
An extension called Hierarchical WALS was proposed in \cite{MagnusWang_14} to jointly deal with uncertainty in concepts and in measurements within each concept, in the spirit of dilution priors (see Section \ref{Sec:PriorModel}). Generally, however, WALS is rather closely linked to the context of covariate uncertainty in a normal linear regression framework and the possibility of extensions to other settings seems limited. One important extension is to the general GLM framework, developed in \cite{DeLuca_etal_18}. They base WALS on a combination of one-step MLEs and consider its asymptotic frequentist properties in the $\cal M$-closed local misspecification framework of \cite{HjortClaeskens_03}.


\cite{WagnerHlouskova_15} consider FMA for principal components augmented
regressions illustrated with the FLS data set on economic growth determinants. In addition, they compare and contrast their method and findings with BMA and with the WALS approach, finding some differences but also some variables that are important in all analyses. Another comparison of different methods on growth data can be found in \cite{Amini_Parmeter_11}. They consider BMA, MMA and WALS and find that results (in as far as they can be compared: for example, MMA and WALS do not provide posterior inclusion probabilities) for three growth data sets are roughly similar.
In the context of GLMs, a comparison of BMA (as in \cite{ChenIbrahim_03}, various FMA methods ({\it e.g.}~based on smoothed versions of AIC, BIC and FIC), a number of model selection procedures ({\it e.g.}~LASSO, SCAD and stepwise regression) and WALS is conducted by \cite{DeLuca_etal_18}. They find that WALS performs similarly to BMA and FMA based on FIC, and that these model averaging procedures outperform the model selection methods.


\subsection{Estimating the sampling variance of FMA estimators}

The construction of confidence intervals associated with FMA estimators has to take into account the model uncertainty, in order to achieve the expected coverage probability. The traditional classical confidence interval is derived conditional upon the chosen model, and thus underestimates uncertainty and leads to confidence intervals that are too narrow. \cite{Buckland_etal_97} propose an adjustment to the standard error to be used in this traditional formula, but \cite{HjortClaeskens_03} show that the
asymptotic distribution of the FMA estimator in (\ref{eqFMA}) is not normal in their local misspecification framework. They derive the limiting distribution and suggest an expression for a confidence interval with correct coverage probability. \cite{WangZhou_13} prove that this confidence interval is asymptotically equivalent to that constructed based only on the full model estimator (which is very easy to compute). In fact, they also show that the same equivalence result holds for the varying-coefficient partially linear model. This means that if the interest of the user is only in interval estimation (for linear combinations of regression
coefficients), there appears to be no need to consider FMA and the full model should be able to provide the appropriate intervals. However, in simulations FMA does beat model selection procedures and the full model in terms of finite sample MSE performance, which suggests that for point estimation there is a clear advantage to model averaging.

\subsection{Computational aspects of FMA}\label{Sec:compFMA}

An important difference between BMA and FMA is that the latter does not lead to estimated model probabilities\footnote{Nevertheless, \cite{BurnhamAnderson_02} have suggested interpreting the estimated weights with AIC as model probabilities.}. A consequence of this for computation in large model spaces is that simple MCMC algorithms where models are visited in line with their posterior probabilities are not readily available. This restricts the ability of FMA methods to deal with large model spaces, as it would need to enumerate the large number of models in (\ref{eqFMA}). In addition, some approaches, like JMA or other cross-validation schemes require non-trivial computational effort for each model weight computation.

Thus, researches with an interest in applying FMA often rely on ways of reducing the model space. In the literature, this has been implemented in a number of ways:
\begin{itemize}
\item Conduct a preliminary model-screening step to remove the least interesting models before applying FMA. This was done in {\it e,g,}~\cite{Claeskens_etal_06} (for logistic models), \cite{Zhang_etal_13} (for forecasting discrete response time series) and \cite{Zhang_etal_17} (for GLMs). Inference is then conducted conditionally upon the outcome of the screening, so that the uncertainty involved in this first step is not addressed by the model averaging. 
    This approach is not just used for FMA, as some BMA methods also include a preliminary screening step: in particular, the use of Occam's window (see Subsection \ref{sec:numerical}) proposed for graphical models in \cite{MadiganRaftery_94} and applied to linear regression in \cite{Raftery_etal_97}.
\item In the context of covariate uncertainty in the linear model, apply an orthogonal transformation of the regressors. WALS critically relies on a semiorthogonal transformation which effectively reduces the computational burden of WALS to the order $k$, in a model space of dimension $2^k$. Of course, this does not allow us to conduct inference on inclusion probabilities of the covariates and the inference is about estimating the effects of covariates. For growth regressions with large model spaces, \cite{AminiParmeter} introduce an operational version of MMA by using the same semiorthogonal transformations as adopted in WALS.
\item Always include a subset of the covariates in each model. This was already used as a device to simplify computations in early work in growth regressions, such as the EBA in \cite{LevineRenelt_92}, and is also  found in WALS and {\it e.g.}~the analysis of \cite{WagnerHlouskova_15}.
\end{itemize}



\subsection{Combining forecasts}\label{sec:CombForecasts}

\subsubsection{The forecast combination puzzle}

As mentioned earlier, there is a large literature in forecasting which combines point forecasts from different models in an equation such as (\ref{eqFMA}) to provide more stable and better-performing forecasts. Of course, the choice of weights in combination forecasting is important. For example, we could consider weighting better forecasts more heavily. In addition, time-varying weights have been suggested.  \cite{StockWatson_04} examine a number of weighting schemes in terms of the accuracy of point forecasts and find that forecast combinations can perform well in comparison with single models, but that the best weighting schemes are often the ones that incorporate little or no data adaptivity. This empirical fact that ``optimally'' estimated weights often perform worse than equal weights in terms of mean squared forecast error is known as the ``forecast combination puzzle''.

\cite{SmithWallis_09} provide an explanation for this phenomenon in terms of the additional uncertainty associated with estimating the weights. Their analysis shows that if the optimal weights are close to equal, then a simple average of competing forecasts ({\it i.e.}~assuming the weights are equal) can dominate a forecast combination with weights that are estimated from the data.
Along similar lines, \cite{Claeskens_etal_16} also provide an explanation for the forecast combination puzzle, but they explicitly treat the estimated weights as random and consider the derivation of the optimal weights and their estimation jointly.


\subsubsection{Density forecast combinations}

There is an increasing awareness of the importance of probabilistic or density forecasts, as described in Section \ref{Pred}. Thus, a recent literature has emerged on density forecast combinations or weighted linear combinations (pools) of prediction models. Density forecasts combinations were discussed in \cite{Wallis_05} and further developed by \cite{HallMitchell_07}, where the combination weights are chosen to minimize the Kullback-Leibler ``distance'' between the predicted and true but unknown density. The latter is equivalent to optimizing LPS as defined in Section \ref{Pred}. The properties of such prediction pools are examined in some detail in \cite{GewekeAmisano_11}, who show that including models that are clearly inferior to others in the pool can substantially improve prediction. Also, they illustrate that weights are not an indication of a predictive model's contribution to log score. This approach is extended by \cite{Kapetanios_etal_15}, who allow for more general specifications of the combination
weights, by letting them depend on the variable to be forecast. They specifically investigate piecewise linear weight functions and show that estimation by optimizing LPS leads to consistency and asymptotic normality\footnote{Formally, this is shown for known thresholds of the piecewise linear weights, and is conjectured to hold for unknown threshold parameters.}. They also illustrate the advantages over density forecast combinations with constant weights using simulated and real data.


\section{Applications in Economics}\label{Sec:Appl}
There is a large and rapidly growing literature where model averaging techniques are used to tackle empirical problems in economics. Before the introduction of model averaging methods, model uncertainty was typically dealt with in a less formalized manner and perhaps even simply ignored in many cases.
Without attempting to be exhaustive, this chapter briefly mentions some examples of model averaging in economic problems and highlights some instances where model averaging has provided new empirical insights. 

\subsection{Growth regressions}\label{sec:growth}

Traditionally, growth theory has been an area where many potential determinants have been suggested and empirical evidence has struggled to resolve the open-endedness of the theory (see footnote \ref{fn:openended}). Extensive  discussions of model uncertainty in the context of economic growth can be found in \cite{Brock_etal} and \cite{Temple_00}. Starting from neoclassical growth theory, \cite{Barro_91} investigates extensions by considering partial correlations and adding variables one by one to OLS regressions. By including human capital as a determinant (captured by school enrollment) he finds empirical support for the neoclassical convergence theory  ({\it i.e.}~that poorer countries tend to grow faster than richer ones). \cite{Rodrik_etal_04} distinguish three main strands of theories in the literature about drivers for growth: geography (which determines natural resources and also influences agricultural productivity and the quality of human resources), international trade (linked with market integration) and institutions (more in particular, property rights and the rule of law)\footnote{\cite{Durlauf_18} provides a very complete discussion of the empirical evidence for links between institutions and growth, concluding that institutions matter.}. Of course, there is a myriad of possible ways in which these theoretical determinants could be measured in practice, leading to a large collection of possible models in line with one or more of these theories. To make empirical implementation even more complex, only the first theoretical source of growth (geography) can be safely assumed to be exogenous, and for the other theories we need to be aware of the possibility that variables capturing these effects may well be endogenous (see Subsection \ref{sec:endogeneity}). The typical treatment of the latter issue requires us to make choices regarding instrumental variables, yet again increasing the number of potential models.
Early attempts at finding a solution include the use of
EBA (see Section \ref{Sec:NLM}) in \cite{LevineRenelt_92} who investigate the robustness of the results from linear regressions and find that very few regressors pass the extreme bounds test, while \cite{SiM_97} employs a less severe test based on the ``level of confidence'' of individual regressors averaged over models (uniformly or with weights proportional to the likelihoods). These more or less intuitive but ad-hoc approaches were precursors to a more formal treatment through BMA discussed and implemented in \cite{BrockDur01} and \cite{FLS01b}.
\cite{HendryKrolzig_04} present an application of general-to-specific modelling (see Section \ref{Sec:NLM}) in growth theory, as an alternative to BMA. However, there is a long list of applications in this area where model averaging is used, and some examples are given below, organised in terms of possible drivers of growth.

\subsubsection{Institutions}\label{sec:institutions}
As mentioned above, the theory that institutions are a crucial driver for growth is often specifically linked to property rights and the rule of law. \cite{Acemoglu_etal_01} argue that private property rights (as measured by government risk
of expropriation) are a key determinant of growth, and that the security of such property rights is crucially dependent on the colonial history. This leads them to use mortality rates of the first European settlers as an instrument for the current institutions in those countries in a 2SLS regression analysis. They find that institutions have a large effect on GDP per capita, and judge this effect to be robust with respect to adding other potential determinants to their regression model. However, these extra variables are essentially added one by one, so that the number of different models actually presented is quite limited. Also, they do not consider any trade variables. In a similar fashion, \cite{FrankelRomer_99} focus on trade as a growth driver and present empirical evidence to support that theory, but without controlling for the effects of institutions. A somewhat broader empirical framework is used in \cite{Rodrik_etal_04}, who aim to provide a comparison of the three main growth theories mentioned in the previous subsection. They use three different samples and take into account the potential endogeneity of institution and integration variables. For the former, they choose rule of law and as an integration variable, they select the ratio of trade to GDP. They use the instrument in \cite{Acemoglu_etal_01}, which they replace for the largest sample by two other instruments: the fraction of the population speaking English and the fraction speaking another European language as their first language. As an additional instrument they adopt the constructed trade share as estimated from a gravity model in \cite{FrankelRomer_99}. Geography is measured by the distance from the equator. \cite{Rodrik_etal_04} conclude that institutions are the main driver of growth between the three theories investigated. But again, this result hinges upon specific choices of variables to represent the theories, specific choices of instruments and consideration of a very limited number of models with other possible determinants. A systematic model averaging approach is implemented in \cite{Lenkoski_etal_14}, using their 2SBMA methodology as explained in Subsection \ref{sec:endogeneity}. They reexamine the data used by \cite{Rodrik_etal_04} through 2SBMA to carefully address the model and instrument uncertainty. Once they allow for additional theories in the model space, they start finding important differences with the \cite{Rodrik_etal_04} conclusions. In particular, they conclude that all three main growth theories (geography, integration and institutions) play an important part in development.
A similar 2SBMA analysis in  \cite{EicherNewiak_13} also includes intellectual property rights. Measuring intellectual property rights by the enforcement of patents, \cite{EicherNewiak_13} find evidence of strong positive effects of both intellectual and physical property rights, as well as some effects to do with geography (malaria and tropics both have a negative impact on development). The influence of trade on growth is analysed in \cite{EicherKuenzel_16}, again using the two-stage BMA approach of \cite{Lenkoski_etal_14}.
They find that sectoral export diversity serves as a crucial
growth determinant for low-income countries, and that this effect decreases with the level of
development. They also find strong evidence for institutional effects on growth.

All these model averaging results take into account many models with different instruments and different growth determinants, and thus allow for a much more complete and nuanced inference from the data than the earlier studies, more adequately reflecting the uncertainty of any conclusions drawn. In addition, the instrument constructed by \cite{Acemoglu_etal_01} has been criticised by \cite{Albouy_12}, due to the problems in obtaining reliable historical mortality data. However, a model averaging approach that takes account of the uncertainty in the instruments will simply downweigh inappropriate instruments. For example, settler mortality does not receive a lot of weight in the instrumental variables equation of \cite{Lenkoski_etal_14} for rule of law\footnote{Although settler mortality does seem to have some use an an instrument for integration.} and those of \cite{EicherNewiak_13} for patent protection and rule of law.\footnote{But when replacing patent protection by patent enforcement, settler mortality actually becomes an important instrument.} Model averaging is particularly natural and useful in the context of instrumental variables, where theory is at best indicative and data quality may be a serious issue.

\subsubsection{Energy consumption}
The question of whether energy consumption is a critical driver of economic growth is investigated in \cite{Camarero_etal_15}. This relates to an important debate in economics between competing economic theories: ecological economic theory (which considers the scarcity of resources as a limiting factor for growth) and neoclassical growth theory (where it is assumed that technological progress and substitution possibilities may serve to circumvent energy scarcity problems). There are various earlier studies that concentrate on the bivariate relationship between energy consumption and economic growth, but of course the introduction of other relevant covariates is key. In order to resolve this in a formal manner, they use the BMA framework on annual US data (both aggregate and sectoral) from 1949 to
2010, with up to 32 possible covariates. \cite{Camarero_etal_15} find that energy consumption is an important determinant of aggregate GDP growth (but their model does not investigate whether energy consumption really appears as an endogenous regressor, so that they can not assess whether there is also feedback) and also identify energy intensity, energy efficiency, the share of nuclear power and public spending as important covariates. Sectoral results support the conclusion about the importance of energy consumption, but show some variation regarding the other important determinants.

\subsubsection{Government spending and savings}
The effect of government investment versus government consumption on growth in a period of fiscal consolidation in developed economies is analysed in \cite{Jovanovic_17}. Using BMA and a dilution prior (based on the determinant of the correlation matrix), it is found that public
investment is likely to have a bigger impact on GDP than public consumption in the countries with high public debt. Also, and more controversially, the
(investment) multiplier is likely to be higher in countries with high public debt than in
countries with lower public debt. The results suggest that
fiscal consolidation should be accompanied by increased public investment.
Using the dataset from \cite{SDM} and a prior which replaces the $g$-prior in (\ref{FLSprior}) by a prior structure with a covariance for $\beta_j$ proportional to an identity matrix and independence between $\beta_j$ and $\sigma$, \cite{LeeChen_18} come to largely similar conclusions as previous studies using $g$-priors. However, they find some differences: one is that both government consumption share and government share of GDP are important drivers for growth, with a positive effect. They conjecture that the difference with the results in \cite{SDM} (finding negative effects and less importance for these variables) is a consequence of the different prior in combination with the high pairwise  correlation between these covariates.
A relatively large cross-sectional study by \cite{Blazej_etal_19} uses data for $n=168$ countries with $k=30$ possible covariates for the period 2002-2013. They find that BMA and BACE give similar results and conclude that the most important determinants are savings, initial GDP, capital formation and a location dummy for Asia and Oceania, lending support for what they call the ``Asian development model''.

\subsubsection{Natural resources}
As discussed in \cite{Frankel_12}, a surprising empirical fact is that the raw data (based on average growth of countries for 1970-2008) certainly do not suggest a positive correlation between natural resource wealth and economic growth. This is often termed the ``natural resource curse'' and \cite{Frankel_12} lists six possible mechanisms: the long-run trend of world prices for commodities; volatility in commodity prices; permanent crowding out of manufacturing, where spillover effects are thought to be concentrated; autocratic or oligarchic institutions; anarchic institutions, such as unenforceable property rights, unsustainably rapid depletion, or civil war; and cyclical expansion of the nontraded sector via the Dutch disease.

On the basis of a relatively small number of cross-country growth regressions, \cite{SachsWarner_01} conclude that the curse is real and that the most likely explanation could be through crowding out of growth-stimulating activity. They find that resource-abundant countries tend to be high-price economies and that contributes to less export-led growth in these countries.
Using BMA, \cite{ArinBraunfels_17} examine the existence of the natural resource curse focusing on the empirical links between oil rents and long-term growth. They find that oil revenues have a robust positive effect on growth. When they include interactions and treat them simply as additional covariates, they find that the positive effect can mostly be attributed to the interaction of institutional quality and oil revenues, which would suggest that institutional quality is a necessary condition for oil
revenues to have a growth-enhancing effect. However, if they use a prior that adheres to the strong heredity principle (see Section \ref{Sec:PriorModel}), they find instead that the main effect of oil rents dominates. In any case, their BMA results lead them to the conclusion ``that oil rents have a robust positive effect
on economic growth and that the resource curse is a red herring.''

\subsubsection{Other determinants}
A specific focus on the effect of measures of fiscal federalism on growth was adopted in \cite{AsatryanFeld_15}. They conclude that, after controlling
for unobserved country heterogeneity, no robust effects of federalism on growth can be found. 

\cite{Man_15} investigates whether competition in the economic and political arenas is a robust determinant
of aggregate growth, and whether there exists jointness among competition variables versus other growth determinants. This study also provides a comparison with EBA and with ``reasonable extreme bounds analysis'', which also takes the fit of the models into account. Evidence is found for the importance and positive impact on
growth of financial market competition, which appears complementary to other important growth determinants.
Competition in other areas does not emerge as a driver of economic growth.

\cite{Piribauer_16} estimates growth patterns across European regions in a spatial econometric framework, building on threshold estimation approaches \citep{Hansen_00} to account
for structural heterogeneity in the observations. The paper uses the prior structure by \cite{GeorgeMcCulloch_93,GeorgeMcCulloch_97} with SSVS (see Section \ref{sec:numerical}), and concludes that initial income, human capital endowments, infrastructure accessibility, and the age structure of the population all appear to be robust driving
forces of income growth.


\cite{Lanzafame_16} derives the natural or potential growth rates of Asian economies (using a Kalman filter on a state-space model) and investigates the determinants of potential growth rates through BMA methods (while always including some of the regressors). He finds evidence of  robust links with various aspects
of institutional quality, the technology gap with the US, trade, tertiary education, and the growth rate of the working-age population.


\subsection{Inflation and Output Forecasting}\label{sec:output}

In the context of time series modelling with ARIMA and ARFIMA models, BMA was used for posterior inference on impulse responses\footnote{The impulse response function measures the effect of a unitary shock on an economic quantity as a function of time. For a stationary process, impulse responses are the coefficients of its (infinite)  moving average representation.} for real GNP in \cite{Koop_etal_97}. This is an example where the behaviour of the quantity of interest (impulse response) is very different between the groups of models considered. ARFIMA (Autoregressive Fractionally Integrated Moving Average) models allow for long memory behaviour which is characterized by a fractional differencing parameter $\delta\in(-1,0.5)$\footnote{For this range of values of $\delta$ the process (expressed in first differences) is stationary and invertible.}. Processes with $\delta>0$ are called long memory processes, while for $\delta<0$ we have intermediate memory. The special case of $\delta=0$ corresponds to the more familiar ARIMA class, which leads to an impulse response function that tends to values on the positive real line when the time horizon goes to infinity. In contrast, ARFIMA models, lead only to limiting impulse responses of either zero (when $\delta<0$) or $\infty$ (for $\delta>0$). Even though the real interest in practice may be in finite time horizons, these limiting results of course impact on the entire impulse response function. This fact actually led \cite{Hauser_etal_99} (who use a frequentist model selection approach) to conclude that ARFIMA models are not appropriate for inference on persistence, in conflict with the earlier recommendation in \cite{DieboldRudebush_89}.
However, by using model averaging rather than model selection, we are not forced to choose either one of these models, but we can formally combine the (often necessarily disparate) inference on persistence and take proper account of the uncertainty. In \cite{Koop_etal_97}, the model space chosen is characterized by different combinations of autoregressive and moving average lag lengths (from 0 to 3) and includes both ARIMA and ARFIMA models for real US GNP. Thus, the posterior distribution of impulse responses is bimodal for (medium and long run) finite horizons and for the impulse response at infinity it contains point masses at zero and $\infty$.

\cite{CogleySargent_05} consider Bayesian averaging of three models for inflation using dynamic model weights. Another paper that uses time-varying BMA methods for inflation forecasting is \cite{vanderMaas_14}. The related strategy of dynamic model averaging, due to \cite{Raftery_etal_10} and described in Section \ref{Sec:DMA}, was used in \cite{KoopKorobilis12}.
Forecasting inflation using BMA has also been examined in \cite{EklundKarlsson}, who propose the use of so-called predictive weights in the model averaging, rather than the usual BMA based on posterior model probabilities.
\cite{Shi_16} models and forecasts quarterly US inflation and finds that BMA with
 regime switching  leads to substantial improvements in forecast performance over simple benchmark approaches
(e.g. random-walk or recursive OLS forecasts) and pure BMA or Markov switching models.
\cite{Ouysse_16} considers point and density forecasts of monthly US
inflation and output growth that are generated using principal components regression
(PCR) and BMA. A comparison between 24 BMA specifications and 2 PCR ones in an out-of-sample, 10-year rolling event
evaluation leads to the conclusion that PCR methods perform best for predicting deviations
of output and inflation from their expected paths, whereas BMA methods perform best
for predicting tail events. Thus, risk-neutral policy-makers may
prefer the PCR approach, while the BMA approach would be the best option for
a prudential, risk-averse forecaster.

\cite{Bencivelli_etal_17} investigate the use of BMA for forecasting GDP relative to simple bridge models\footnote{Bridge models relate
information published at monthly frequency to quarterly national account data, and are used for producing timely ``now-casts'' of economic activity.} and factor models. They conclude that for the Euro
area, BMA bridge models produce smaller forecast errors than a small-scale dynamic
factor model and an indirect bridge model obtained by aggregating country-specific
models.

\cite{DuctorLeiva-Leon_16} investigate the time-varying interdependence among the
economic cycles of the major world economies since the 1980's. They use a BMA panel data approach (with the model in (\ref{panelmodel}) including a time trend) to find the determinants of pairwise de-synchronization between the
business cycles of countries. They also apply WALS and find that it indicates the same main determinants as BMA.

A probit model is used for forecasting US recession periods in \cite{Aijun_etal_17}. They use a Gibbs sampler based on SSVS (but with point masses for the coefficients of the excluded regressors), and adopt a generalized double Pareto prior (which is a scale mixture of normals) for the included regression parameters along with a dilution prior over models based on the correlation between the covariates. Their empirical results on monthly U.S. data (from 1959:02 until 2009:02) with 108 potential covariates suggest the method performs well relative to the main competitors.

\subsection{VAR and DSGE modelling}\label{sec:VAR}
A popular econometric framework for jointly modelling several variables is the vector autoregressive (VAR) model.
\cite{Koop_17} provides an intuitive and accessible overview of Bayesian methods for inference with these types of models. 
BMA methodology has been applied by \cite{Garratt_etal} for probability forecasting of inflation and output growth in the context of
a small long-run structural vector error-correcting
model of the UK economy. \cite{GeorgeSunNi_08} apply BMA ideas in VARs using SSVS methods with priors which do not induce exact zero restrictions on the coefficients, as in \cite{GeorgeMcCulloch_93}.
\cite{KoopKorobilis_16} extend this to Panel VARs where the  restrictions of interest involve interdependencies between and heterogeneities
across cross-sectional units.

\cite{FeldkircherHuber_16} use a Bayesian VAR model to explore the international spillovers of expansionary US aggregate demand and supply
shocks, and of a contractionary US monetary policy shock. They use SVSS  and find evidence for significant spillovers, mostly transmitted through financial channels and with some notable cross-regional variety.

BMA methods for the more restricted dynamic stochastic general equilibrium (DSGE) models were used in \cite{StrachanvanDijk_13}, with a particular interest in the effects of investment-specific and neutral technology shocks. Evidence from US quarterly data from 1948-2009 suggests a break in the entire model structure around 1984, after which technology shocks appear to account for all stochastic trends. Investment-specific technology shocks seem more important for business cycle volatility than neutral technology shocks.

\subsection{Crises and finance}\label{sec:crises}

There is a substantial literature on early warning signals for economic crises going back many decades, illustrated by the review in \cite{FrankelSaravelos_12} and forcefully brought back to attention by the global financial
crisis of 2008-9. Unfortunately, the established early warning signals did not provide clear alerts for this recent crisis (see, for example, \cite{RoseSpiegel_11}).
\cite{Feldkircher_etal_14} focus on finding leading indicators for exchange market pressures during the crisis and their BMA results indicate that inflation plays an important aggravating role, whereas international reserves act as a mitigating factor.
Early warning signals are also investigated in \cite{Christofides_etal_EER} who consider four different crisis dimensions: banking, balance of payments, exchange rate pressure, and recession and use balanced samples in a BMA analysis. They do not identify any single early warning signal for all dimensions of the 2008 crisis, but
find that the importance of such signals is specific to the particular dimension of the crisis being examined. They argue that the  consensus in the previous literature about early warning signals (such as ``foreign currency reserves'' and ``exchange rate overvaluations'' mentioned in, for example, \cite{FrankelSaravelos_12}) hinges critically on the fact that many earlier analyses only considered models that were in line with a particular theory. Taking theory uncertainty into account in a formal statistical framework, the conclusions change.

Following the work of \cite{RoseSpiegel_11} and the earlier BMA approach of \cite{Giannone_etal_11}, \cite{Feldkircher14} uses BMA to identify the main macroeconomic and financial
market conditions that help explain the real economic effects of the
crisis of 2008-9. He finds that countries with strong pre-crisis growth in credit and/or in real activity tended to be less resilient. \cite{Ho_15} investigates the causes of the crisis, using BMA, BACE and the approach of \cite{Hoeting_etal_96} (see Section \ref{Sec:Outlier}) to deal with outliers, and finds that the three methods lead to broadly similar results. The same question about the determinants of the 2008 crisis was addressed in \cite{Chen_etal_17}, who use a hierarchical prior structure with groups of variables (grouped according to a common theory about the origins of the crisis) and individual variables within each group. They use BMA to deal with uncertainty at both levels and find that ``financial policies and trade linkages are the most relevant groups with regard to the relative macroeconomic performance of different countries during the crisis. Within the selected groups, a
number of pre-existing financial proxies, along with measures of trade linkages, were significantly correlated
with real downturns during the crisis. Controlling for both variable uncertainty and group uncertainty, our
group variable selection approach is able to identify more variables that are significantly correlated with crisis
intensity than those found in past studies that select variables individually.''

The drivers of financial contagion after currency crises were investigated through BMA methods in \cite{Dasgupta_etal_11}. They use a probit model for the occurrence of a currency crisis in 54 to 71 countries for four years in the 1990s and find that institutional similarity is an important predictor of financial contagion during emerging market crises.
\cite{Puy_16} investigates the global and regional dynamics in equity and bond flows, using data on portfolio investments from international mutual funds. In addition, he finds strong evidence of global contagion.
To assess the determinants of contagion, he regresses the fraction of variance of equity and bond funding attributable to the world factor
on a set of 14 structural variables, using both WALS and BMA. Both point towards distance and political risk as robust drives of contagion, \cite{Puy_16} concludes that ``sudden surges/stops tend to strike fragile countries, {\it i.e.}~emerging markets with unstable political systems and poor connection to the main financial centers.''

\cite{Moral-BenitoRoehn_16} explore the relationship
between financial market regulation and current account balances. They use a dynamic panel model and combine the BMA methodology with a
likelihood-based estimator that accommodates both persistence and unobserved heterogeneity. In their investigation of the determinants of current account balances, \cite{Desbordes_etal_18} highlight ``three features which are
likely to be shared by many panel data applications: high model uncertainty, presence of slope heterogeneity, and potential divergence in short-run and long-run effects''.

The use of BMA in forecasting exchange rates by \cite{Wright_08} leads to the conclusion that BMA provides slightly better out-of-sample forecasts (measured by mean squared prediction errors) than the traditional random walk benchmark. This is confirmed by \cite{Ribeiro_17}, who also argues that a bootstrap-based method, called bumping, performs even better.
\cite{Iyke_15} analyses the real exchange rate in Mauritius using BMA. Different priors are adopted, including empirical Bayes. There are attempts to control for multicollinearity in the macro determinants using three competing model priors incorporating dilution, among which the tessellation prior and the weak heredity prior (see Section \ref{Sec:PriorModel}).
\cite{AdlerGrisse_17} examine behavioral equilibrium exchange rates
models, which relate a long-run cointegration relationship between real exchange rates to fundamental macroeconomic variables, in a panel regression across currencies. They use BACE to deal with model uncertainty and find that some variables (central bank reserves, government consumption, private credit, real interest rates and the terms of trade) are robustly linked with real exchange rates. The introduction of fixed country effects in the models greatly improves the fit to real exchange rates over time.

BMA applied to a meta-analysis is used by \cite{ZigraiovaHavranek_16} to investigate the relationship between bank
competition and financial stability. They find some evidence of publication bias\footnote{Generally, this is the situation that the probability of a result being reported in the literature ({\it i.e.}~of the paper being published) depends on the  sign or
statistical significance of the estimated effect. In this case, the authors found some evidence that some authors
of primary studies tend to discard estimates inconsistent with the competition-fragility hypothesis, one of the two main hypotheses in this area. For an in-depth discussion of publication bias and meta-analyses in economics, see \cite{ChristensenMiguel_18}.} but encounter no clear link between  bank competition and stability, even
when correcting for publication bias and potential misspecifications.

\cite{DevereuxDwyer_16} examine the output costs associated with 150 banking crises
using cross country data for the years after 1970. They use BMA to identify important determinants of output changes after crises and conclude that for high-income countries
the behavior of real GDP after a banking crisis is most closely associated with prior economic conditions, where above-average changes in credit tend to be associated with  larger expected decreases in real GDP. For low-income economies, the existence of a stock market and deposit insurance are linked
with quicker recovery of real GDP.

\cite{PelsterVilsmeier_16} use Bayesian Model Averaging to
assess the pricing-determinants of credit default swaps. They use an autoregressive distributed lag model with time-invariant fixed effects and approximate posterior model probabilities on the basis of smoothed AIC. They conclude that credit default swaps price dynamics can be mainly explained by factors describing firms' sensitivity to extreme market movements, in particular variables measuring tail dependence (based on so-called dynamic
copula models).

\cite{Horvath_etal_17} explore the determinants of financial development as measured by  financial depth (both
for banks and stock markets),  the efficiency of financial intermediaries (both for banks and
stock markets),  financial stability and access to finance. They use BMA to analyse financial development in 80 countries using nearly
40 different explanatory variables and find that the rule of law is a major factor in
influencing financial development regardless of the measure used. In addition, they conclude that the level of economic development matters and that greater wealth inequality is associated with greater stock market depth, although it does not matter for
the development of the banking sector or for the efficiency of stock markets and banks.

The determinants of US monetary policy are investigated in \cite{WolfelWeber_17}, who conclude from a BMA analysis that over the long-run (1960-2014) the important variables in explaining the Federal Funds Rate are inflation, unemployment rates and long-term interest rates. Using samples starting in 1973 (post Bretton-Woods) and 1982 (real-time data),
the fiscal deficit and monetary aggregates were also found to be relevant. \cite{WolfelWeber_17} also account for parameter instability through the introduction of an unknown number of structural breaks and find strong support
for models with such breaks, although they conclude that there is less evidence for structural break since the 1990s.

\cite{WatsonDeller_17} consider the relationship between economic diversity and unemployment in the light of the economic shocks provided by the recent ``Great Recession''. They use a spatial BMA model allowing for spatial spillover effects on data from US counties with a Herfindahl diversity index computed across 87 different sectors. They conclude that increased economic diversity within the county itself is associated with significantly reduced unemployment rates across all years of the sample (2007-2014). The economic diversity of neighbours is only strongly associated with reduced unemployment rates at the height of the Great Recession.

\cite{Ng_etal_16} investigate the relevance of social capital in stock market development using BMA methods
and conclude that trust is a robust and positive determinant of stock market depth and liquidity.

BMA was used to identify the leading indicators of financial stress in 25 OECD countries by \cite{Vasicek_etal_17}. They find that financial stress is difficult to predict out of sample, either modelling all countries at the same time (as a panel) or individually.

\subsection{Production modelling}
For production or cost modelling through stochastic frontier models, there exists a large amount of uncertainty regarding the distributional assumptions, particularly for the one-sided errors that capture inefficiencies. There is no theoretical guidance and a number of different distributions are used in empirical work. Given the typical lack of observations per firm (just one in cross-sectional analyses), the distribution used has a potentially large effect on the inference on firm efficiencies\footnote{For example, the exponential inefficiency distribution is a popular choice, as it is relatively easy to work with, but because of its shape (with a lot of mass near zero) it will tend to lead to a cluster of highly efficient firms; this contrasts with other, more flexible, distributions, such as the gamma or generalized gamma distributions or mixtures of distributions; see \cite{GriffinSteel_08}.}, a key aspect of interest in this type of modelling.

Bayesian methods were introduced in this context by \cite{Broeck_etal_94}. They deal with the uncertainty regarding the specification of the inefficiency distribution through BMA, so that inference on all aspects of interest (such as elasticities, returns to scale, firm-specific posterior and predictive efficiencies, etc.) is appropriately averaged over the models considered. Other, more ad-hoc, approaches to providing model averaged estimates for some features
of productivity or inefficiency appear, for example, in \cite{Sickles_05}, who takes simple averages of
technical efficiency estimates for US banks across a range of alternative stochastic frontier
panel data models, with the aim of providing ``a clear and informative summary''. This is formalized in \cite{Parmeter_etal_16}, who develop FMA methods for (cross-sectional) stochastic frontier models. In particular, they consider models that allow for the variance of the symmetric model error and the parameters of the inefficiency distribution to depend on possible covariates, and the uncertainty resides in which are the appropriate sets of covariates in   these distributions and which covariates should be included in the frontier itself (typically functions of inputs for a production frontier, but there could be other covariates, such as time in a panel setup, etc.). One method averages inefficiencies and works for nested structures, and another is a $J$-fold cross-validation method similar to JMA, which can be used when inefficiency distributions are not nested.

\cite{McKenzie_16} considers three different stochastic frontier models with varying degrees of flexibility in the dynamics of productivity change and technological growth, and uses BMA 
to conduct inference on  productivity growth of railroads.

Cost efficiency in the Polish electricity distribution sector is analysed through Bayesian methods in \cite{MakielaOsiewalski_18}. Given the complexity of the models involved, they do not run a chain over model space but treat models separately, using the so-called corrected arithmetic mean estimator for the marginal likelihoods (this involves importance sampling after trimming the parameter space). They use BMA in carefully chosen model spaces that address uncertainty regarding the included covariates and also regarding the existence of firm inefficiencies.

\subsection{Other applications}

\cite{Havranek_et_al_15} use BMA in a meta-analysis of intertemporal substitution in consumption. \cite{HavranekSokolova_16} investigate the mean excess sensitivity reported in studies estimating consumption Euler equations. Using BMA methods, they control for 48 variables related to
the context in which estimates are obtained in a sample of 2,788 estimates reported in 133 published studies. Reported mean excess sensitivity seems materially affected by demographic variables, publication bias and liquidity constraints and they conclude that the permanent income hypothesis seems a pretty good approximation of the actual
behavior of the average consumer. \cite{Havranek_etal_17} consider estimates of habit formation in consumption in 81 published studies and try and relate differences in the estimates to various characteristics of the studies. They use BMA (with MC$^3$) and FMA\footnote{Here they follow the approach suggested by \cite{AminiParmeter}, who build on \cite{Magnus_etal_10} and use orthogonalization of the covariate space, thus reducing the number of models that need to be estimated from $2^k$ to $k$. In individual regressions they use inverse-variance weights to account for the estimated dependent variable issue.} and find broadly similar results using both methods. Another example of the use of BMA in meta-analysis is \cite{Philips_16} who investigates political budget cycles and finds support for some of the context-conditional theories in that literature.

The determinants of export diversification are examined in \cite{JetterHassan_15} who conclude that Primary school enrollment has a robust positive effect on export diversification,
whereas the share of natural resources in gross domestic product lowers
diversification levels. Using the IVBMA approach of \cite{KarlLenkoski_12} (see Subsection \ref{sec:endogeneity}) they find that these findings are robust to accounting for endogeneity.

\cite{Kourtellos_etal_16} use BMA methods to investigate the variation in intergenerational spatial mobility across commuter zones in the US using model priors based on the dilution idea. Their  results show substantial evidence of heterogeneity, which suggests exploring nonlinearities in the
spatial mobility process.

Returns to education have been examined through BMA in \cite{TobiasLi}. 
\cite{Koop_et_al_12} use their instrumental variables BMA method in this context.
\cite{Cordero_etal_16} use BMA methods to assess the determinants of cognitive and non-cognitive educational outcomes in Spain. 
The link between tuition fees and demand for higher education has been investigated in \cite{Havranek_etal_18} in a meta-analysis framework. After accounting for publication bias, they conclude that the mean effect of tuition fees on enrolment is close to zero. BMA and FMA approaches (the latter using the orthogonal transformation of \cite{AminiParmeter}) lead to very similar results and also indicate that enrolment of male students and students at private universities does tend to decrease with tuition fees.

\cite{Daude_etal_16} investigate the drivers of productive capabilities (which are important for growth) using BACE based on bias-corrected least squares
dummy variable estimates \citep{Kiviet_95} in a dynamic panel context with country-specific effects.

Through spatial BMA, \cite{Oberdabernig_etal_16} examine democracy determinants  and find that spatial spillovers are important even after controlling for a large number of geographical covariates, using a student-$t$ version of the SAR model (see subsection \ref{sec:spatial}) with fixed degrees of freedom. Also employing a model with spatial effects,
\cite{HortasRios_16} examine the main drivers of urban income inequality using Spanish municipal data.

\cite{Cohen_etal_16} investigate the social acceptance of power transmission lines using an EU survey. An ordered probit model was used to model the level of acceptance and the fixed country effects of that regression were then used as dependent variables in a BMA analysis, to further explain the heterogeneity between the 27 countries covered in the survey.


In order to identify the main determinants of corruption, \cite{JetterParmeter_17} apply IVBMA  to corruption data with a large number of endogenous regressors, using lagged values as instruments. They conclude that institutional characteristics ({\it e.g.}~rule of law, government effectiveness and urbanization rate) and the extent of primary schooling emerge as important predictors, while they find little support for historical, time-invariant cultural, and geographic determinants. 

\cite{Pham_17} investigates the impact of different globalization dimensions (both
economic and non-economic) on the informal sector and shadow economy  in developing countries. The methodology of \cite{Leon-GonzalezMontolio_15} is used to deal with endogenous regressors as well as country-specific fixed effects.

The effect of the abundance of resources on the efficiency of resource usage is explored in \cite{Hartwell_16}. This paper considers 130 countries over various time frames from 1970 to 2011, both resource-abundant and resource-scarce, to ascertain a link between abundance of resources and less efficient usage of those resources. Efficiency is measured by {\it e.g.}~gas or oil consumption per unit of GDP, and three-stage least squares estimates are obtained for a system of equations. Model averaging is then conducted according to WALS. The paper concludes that for resource-abundant countries, the improvement of property rights will lead to a more environmentally sustainable resource usage.

\cite{WeiCao_17} use dynamic model averaging (DMA) to
forecast the growth rate of house prices in 30 major Chinese cities. They use the MCS test (see Section \ref{Sec:NLM}) to conclude
 that DMA achieves significantly higher forecasting accuracy than other models in
both the recursive and rolling forecasting modes. They find that the importance of
predictors for Chinese house prices varies substantially over time and that the
Google search index for house prices has recently surpassed the forecasting
ability of traditional macroeconomic variables. Housing prices in Hong Kong were analysed in \cite{Magnus_etal_11} using a GLS version of WALS.

Robust determinants of bilateral trade are investigated in \cite{Chen_etal_16}, using their LIBMA methodology (see Section \ref{Sec:Panel}). They find evidence of trade persistence and of important roles for  the exchange rate regime, several of the traditional ``core'' variables of the trade gravity model as well as trade creation and diversion through trade agreements. They stress that neglecting to address model uncertainty, dynamics, and endogeneity simultaneously would lead to quite different conclusions.


\section{Software and resources}\label{Sec:software}
The free availability of software is generally very important for the adoption of methodology by applied users. There are a number of publicly available computational resources for conducting BMA. Early contributions are the code by \cite{Raftery_etal_97} (now published as an {\tt R} package in \cite{Raftery_etal_soft}) and the {\tt Fortran} code used by \cite{FLS01a}.

Recently, a number of {\tt R}-packages have been created, in particular the frequently used BMS package \citep{webBMS16}. 
Details about BMS are given in \cite{FZ_15}. Two other well-known
{\tt R}-packages are  BAS \citep{Cly17}, explained in \cite{Clydeetal11},
and BayesVarSel \citep{GarFor15}, described in \cite{GarFor17}. When endogenous regressors are suspected, the {\tt R}-package ivbma \citep*{ivbma}
implements the method of \cite{KarlLenkoski_12}. For situations where we wish to allow for flexible nonlinear effects of the regressors, inference for  (generalized) additive models as in \cite{SabanesHeld_11} and \cite{Sabanes_etal_15} can be conducted by the packages glmBfp \citep{glmBfp} on CRAN and hypergsplines \citep{hypergsplines} on R-Forge, respectively. For dynamic models, an efficient implementation of the DMA methodology of \cite{Raftery_etal_10} is provided in the {\tt R} package eDMA \citep{eDMA}, as described in \cite{CataniaNonejad_17}. This software uses parallel computing if shared memory multiple processors hardware is available. The model confidence set approach (as described in Section \ref{Sec:NLM}) can be implemented through the {\tt R} package MCS \citep{MCS} as described in \cite{BernardiCatania_17}.
Finally, the {\tt R} packages MuMIn \citep{MuMIn} and AICcmodavg \citep{AICcmodavg}
contain a wide range of different information-theoretic model selection and FMA methods.

In addition, code exists in other computing environments; for example \cite{LeSage_15} describes Matlab code for BMA with spatial models.
\cite{BlazejowskiKwiatkowski_15,BlazeKwiat_18} present packages that implements BMA (including jointness measures) and BACE in gretl.\footnote{Gretl is a free, open-source software (written in C) for econometric analysis with a graphical user interface.}
\cite{PerrakisNtzoufras_18} describe an implementation in WinBUGS (using Gibbs sampling over all parameters and the models) for BMA under hyperpriors on $g$.

Using the BMS package \cite{AminiParmeter} successfully replicate the BMA results of \cite{FLS01b}, \cite{MasanPapa_08} and \cite{DW_09}.
\cite{Forte_etal_17} provide a systematic review of {\tt R}-packages publicly available in CRAN for Bayesian model selection and model averaging in normal linear regression models. In particular, they examine in detail the packages BAS, BayesFactor \citep{MorRou15}, BayesVarSel, BMS and mombf \citep{Rossell14} and highlight differences in priors that can be accommodated (within the class described in (\ref{FLSprior})), numerical implementation and posterior summaries provided. All packages lead to very similar results on a number of real data sets, and generally provide reliable inference within 10 minutes of running time on a simple PC for problems up to $k=100$ or so covariates. They find that BAS is overall faster than the other packages considered  but with a very high cost in terms of memory requirements and, overall, they recommend BAS with estimation based on model visit frequencies\footnote{The BAS package also has the option to use the sampling method
(without replacement) called Bayesian Adaptive Sampling (BAS) described in \cite{Clydeetal11}, which is based on renormalization and leads to less accurate estimates in line with the comments in Section \ref{sec:numerical}.}. If memory restrictions are an issue (for moderately large $k$ or long runs) then BayesVarSel is a good choice for small or moderate values of $n$, while BMS is preferable when $n$ is large.

A number of researchers have made useful BMA and FMA resources freely available:
\begin{itemize}
\item Clyde: \url{http://stat.duke.edu/~clyde/software} for BAS and her papers can be found at \url{http://www2.stat.duke.edu/~clyde/research/}.
\item Feldkircher and Zeugner: \url{http://bms.zeugner.eu/resources/} a dedicated BMA resource page with lots of free software and introductory material.
\item Hansen: \url{https://www.ssc.wisc.edu/~bhansen/progs/progs_ma.html} contains code  (in R, Matlab, Gauss and STATA) implementing MMA and JMA.
\item Magnus: \url{http://www.janmagnus.nl/items/WALS.pdf} for MATLAB and Stata implementations of WALS, described in \cite{DeLucaMagnus_11}.
\item Raftery: \url{https://www.stat.washington.edu/raftery/Research/bma.html} for his BMA papers and \url{http://www.stat.washington.edu/raftery/software.html} for software and data.
\item Steel:  \url{http://www.warwick.ac.uk/go/msteel/steel_homepage/bma} has BMA papers that I contributed to as well as code (Fortran) and data.

\end{itemize}

\section{Conclusions}

Model uncertainty is a pervasive (and often not fully recognized) problem in economic applications. It is important to realize that simply ignoring the problem (and sticking to a single model, as is the traditional approach) is clearly not a solution, and has the dangerous consequence of presenting conclusions with an excess of confidence, since the results do not take into account the host of other possible models that could have been tried. Even if we do acknowledge the existence of other models, the use of model selection techniques and presenting our inference conditional upon the single chosen model typically leads to an underestimation of our uncertainty and can induce important biases\footnote{Model selection can only be a valid strategy in certain circumstances, for example if we are interested in the identity of the ``true'' model and if the latter is part of the model space or if all other models in the appropriate model space are massively dominated by a single model. Both situations are unlikely to occur in economic applications.}. Of course, for applying model averaging it is key that we define an appropriate model space, which adequately reflects all the uncertainty in the problem, for example in terms of theory, implementation or modelling assumptions.

In my view, the proliferation of ``empirically supported'' theories in economics, for example, on what are key drivers for growth, or which indicators can serve as early warning signals for financial crises, is to an important extent due to the tendency of investigators to focus their empirical analyses on models that are largely in line with their own preferred theories. This means that debates often remain unresolved, and I firmly believe that model averaging with appropriately chosen model spaces, encompassing (at least) all available theories, can provide very compelling evidence in such situations. I interpret the rapidly growing use of model averaging methods in economics as a recognition of how much can be gained by adopting principled approaches to the resolution of model uncertainty.

Model averaging methods broadly come in two flavours: Bayesian (BMA) and frequentist (FMA).
The choice between BMA versus FMA is to some extent a matter of taste and may depend on the particular focus and aims of the investigation. For this author, the unambiguous, finite sample nature of BMA makes it particularly attractive for use in situations of model uncertainty. Also, the availability of posterior inclusion probabilities for the regressors and  model probabilities (which also allows for model selection if required) seem to be clear advantages of BMA. In addition, BMA is easily implemented in very large model spaces and possesses some attractive properties (as detailed in Subsection \ref{sec:propertiesBMA}).

Clearly, priors matter for BMA and it is crucial to be aware of this. Looking for solutions that do not depend on prior assumptions at all can realistically only be achieved on the surface by hiding the implicit prior assumptions underneath. I believe it is much preferable to be explicit about the prior assumptions and the recent research in prior sensitivity can serve to highlight which aspects of the prior are particularly critical for the results and how we can ``robustify'' our prior choices. A recommended way to do this is through the use of hyperpriors on hyperparameters such as $w$ and $g$, given a prior structure such as the one in (\ref{FLSprior}). We can then, typically, make reasonable choices for our robustified priors by eliciting simple quantities, such as prior mean model size. The resulting prior avoids being unintentionally informative and has the extra advantage of making the analysis more adaptive to the data. For example, in cases of weak or unreliable data it will tend to favour smaller values of $g$, avoiding unwarranted precise distinctions between models. This may well lead to larger model sizes, but that can easily be counteracted by choosing a prior on the model space that is centered over smaller models.


Sensitivity analysis (over a range of different priors and different sampling models) is indispensable if we want to convince our colleagues, clients and policy makers. Providing an explicit mapping from these many assumptions to the main results is a key aspect of careful applied research, and should not be neglected. There are many things that theory and prior desiderata can tell us, but there will always remain a lot that is up to the user, and then it is important to try and capture a wide array of possible reasonable assumptions underlying the analysis. In essence, this is the key message of model averaging and we should take it to heart whenever we do empirical research, certainly in non-experimental sciences such as economics.

I believe that model averaging is well on its way to take its rightful place in the toolbox of the economist, as a powerful methodology for the resolution of uncertainty. The free availability of well-documented and efficient software should stimulate its wider adoption in the profession. Besides formally accounting for uncertainty (as  expressed by our choice of model space) in our inference and policy conclusions, it can, in my view, also contribute to constructive and focused communication within economics. In particular, model averaging creates an excellent unifying platform to highlight and explain the reasons for differences in empirical findings, through {\it e.g.}~differences in model space, priors in BMA or weights in FMA.


\bibliographystyle{chicago}
\bibliography{References}

\begin{thebibliography}{}

\bibitem[\protect\citeauthoryear{Acemoglu, Johnson, and Robinson}{Acemoglu
  et~al.}{2001}]{Acemoglu_etal_01}
Acemoglu, D., S.~Johnson, and J.~Robinson (2001).
\newblock The colonial origins of comparative development: An empirical
  investigation.
\newblock {\em American Economic Review\/}~{\em 91}, 1369--1401.

\bibitem[\protect\citeauthoryear{Adler and Grisse}{Adler and
  Grisse}{2017}]{AdlerGrisse_17}
Adler, K. and C.~Grisse (2017).
\newblock Thousands of {BEER}s: Take your pick.
\newblock {\em Review of International Economics\/}~{\em 25}, 1078--104.

\bibitem[\protect\citeauthoryear{Aijun, Ju, Hongqiang, and Jinguan}{Aijun
  et~al.}{2018}]{Aijun_etal_17}
Aijun, Y., X.~Ju, Y.~Hongqiang, and L.~Jinguan (2018).
\newblock Sparse {B}ayesian variable selection in probit model for forecasting
  {U.S.}~recessions using a large set of predictors.
\newblock {\em Computational Economics\/}~{\em 51}, 1123--38.

\bibitem[\protect\citeauthoryear{Albert and Chib}{Albert and
  Chib}{1993}]{AlbertChib_93}
Albert, J.~H. and S.~Chib (1993).
\newblock {B}ayesian analysis of binary and polychotomous response data.
\newblock {\em Journal of the American Statistical Association\/}~{\em 88},
  669--79.

\bibitem[\protect\citeauthoryear{Albouy}{Albouy}{2012}]{Albouy_12}
Albouy, D. (2012).
\newblock The colonial origins of comparative development: An empirical
  investigation: Comment.
\newblock {\em American Economic Review\/}~{\em 102}, 3059--3076.

\bibitem[\protect\citeauthoryear{Alvarez and Arellano}{Alvarez and
  Arellano}{2003}]{AlvarezArellano03}
Alvarez, J. and M.~Arellano (2003).
\newblock The time series and cross-section asymptotics of dynamic panel data
  estimators.
\newblock {\em Econometrica\/}~{\em 71}, 1121--59.

\bibitem[\protect\citeauthoryear{Amini and Parmeter}{Amini and
  Parmeter}{2011}]{Amini_Parmeter_11}
Amini, S. and C.~Parmeter (2011).
\newblock Bayesian model averaging in {R}.
\newblock {\em Journal of Economic and Social Measurement\/}~{\em 36}, 253--87.

\bibitem[\protect\citeauthoryear{Amini and Parmeter}{Amini and
  Parmeter}{2012}]{AminiParmeter}
Amini, S.~M. and C.~F. Parmeter (2012).
\newblock Comparison of model averaging techniques: Assessing growth
  determinants.
\newblock {\em Journal of Applied Econometrics\/}~{\em 27}, 870--76.

\bibitem[\protect\citeauthoryear{Arin and Braunfels}{Arin and
  Braunfels}{2018}]{ArinBraunfels_17}
Arin, K. and E.~Braunfels (2018).
\newblock The resource curse revisited: A {B}ayesian model averaging approach.
\newblock {\em Energy Economics\/}~{\em 70}, 170--8.

\bibitem[\protect\citeauthoryear{Asatryan and Feld}{Asatryan and
  Feld}{2015}]{AsatryanFeld_15}
Asatryan, Z. and L.~Feld (2015).
\newblock Revisiting the link between growth and federalism: A {B}ayesian model
  averaging approach.
\newblock {\em Journal of Comparative Economics\/}~{\em 43}, 772--81.

\bibitem[\protect\citeauthoryear{Atchad\'e and Rosenthal}{Atchad\'e and
  Rosenthal}{2005}]{AtchadeRosenthal_05}
Atchad\'e, Y. and J.~Rosenthal (2005).
\newblock On adaptive {M}arkov chain {M}onte {C}arlo algorithms.
\newblock {\em Bernoulli\/}~{\em 11}, 815--28.

\bibitem[\protect\citeauthoryear{Barbieri and Berger}{Barbieri and
  Berger}{2004}]{BarbieriBerger}
Barbieri, M. and J.~Berger (2004).
\newblock Optimal predictive model selection.
\newblock {\em Annals of Statistics\/}~{\em 32}, 870--97.

\bibitem[\protect\citeauthoryear{Barro}{Barro}{1991}]{Barro_91}
Barro, R. (1991).
\newblock Economic growth in a cross section of countries.
\newblock {\em Quarterly Journal of Economics\/}~{\em 106}, 407--44.

\bibitem[\protect\citeauthoryear{Barto\'n}{Barto\'n}{2016}]{MuMIn}
Barto\'n, K. (2016).
\newblock {MuMIn} - {R} package for model selection and multi-model inference.
\newblock \url{http://mumin.r-forge.r-project.org/}.

\bibitem[\protect\citeauthoryear{Bates and Granger}{Bates and
  Granger}{1969}]{BatesGranger}
Bates, J. and C.~Granger (1969).
\newblock The combination of forecasts.
\newblock {\em Operations Research Quarterly\/}~{\em 20}, 451--68.

\bibitem[\protect\citeauthoryear{Bayarri, Berger, Forte, and
  Garc\'{\i}a-Donato}{Bayarri et~al.}{2012}]{Bayarri_etal_12}
Bayarri, M.-J., J.~Berger, A.~Forte, and G.~Garc\'{\i}a-Donato (2012).
\newblock Criteria for {B}ayesian model choice with application to variable
  selection.
\newblock {\em Annals of Statistics\/}~{\em 40}, 1550--77.

\bibitem[\protect\citeauthoryear{Bencivelli, Marcellino, and
  Moretti}{Bencivelli et~al.}{2017}]{Bencivelli_etal_17}
Bencivelli, L., M.~Marcellino, and G.~Moretti (2017).
\newblock Forecasting economic activity by {B}ayesian bridge model averaging.
\newblock {\em Empirical Economics\/}~{\em 53}, 21--40.

\bibitem[\protect\citeauthoryear{Benjamini and Hochberg}{Benjamini and
  Hochberg}{1995}]{Benjamini_H_95}
Benjamini, Y. and Y.~Hochberg (1995).
\newblock Controlling the false discovery rate: a practical and powerful
  approach to multiple testing.
\newblock {\em Journal of the Royal Statistical Society, Series B\/}~{\em 57},
  289--300.

\bibitem[\protect\citeauthoryear{Berger, Garc\'{\i}a-Donato,
  Mart\'{\i}nez-Beneito, and Pe\~na}{Berger et~al.}{2016}]{Berger_etal_16}
Berger, J., G.~Garc\'{\i}a-Donato, M.~Mart\'{\i}nez-Beneito, and V.~Pe\~na
  (2016).
\newblock Bayesian variable selection in high dimensional problems without
  assumptions on prior model probabilities.
\newblock technical report arXiv:1607.02993v1.

\bibitem[\protect\citeauthoryear{Berger and Pericchi}{Berger and
  Pericchi}{1996}]{BergerPericchi1996}
Berger, J. and L.~Pericchi (1996).
\newblock The intrinsic {B}ayes factor for model selection and prediction.
\newblock {\em Journal of the American Statistical Association\/}~{\em 91},
  109--22.

\bibitem[\protect\citeauthoryear{Berger and Pericchi}{Berger and
  Pericchi}{2001}]{BergerPericchi_01}
Berger, J. and L.~Pericchi (2001).
\newblock Objective {B}ayesian methods for model selection: Introduction and
  comparison.
\newblock In P.~Lahiri (Ed.), {\em Model Selection}, Institute of Mathematical
  Statistics Lecture Notes - Monograph Series 38, Beachwood, OH: IMS, pp.\
  135--207.

\bibitem[\protect\citeauthoryear{Bernardi and Catania}{Bernardi and
  Catania}{2018}]{BernardiCatania_17}
Bernardi, M. and L.~Catania (2018).
\newblock The model confidence set package for {R}.
\newblock {\em International Journal of Computational Economics and
  Econometrics\/}~{\em 8}, 144--158.

\bibitem[\protect\citeauthoryear{Bernardo and Smith}{Bernardo and
  Smith}{1994}]{BernardoSmith_94}
Bernardo, J. and A.~Smith (1994).
\newblock {\em Bayesian Theory}.
\newblock Chichester: Wiley.

\bibitem[\protect\citeauthoryear{B{\l}a\.{z}ejowski, Kufel, and
  Kwiatkowski}{B{\l}a\.{z}ejowski et~al.}{2018}]{Blazejowski_etal_18}
B{\l}a\.{z}ejowski, M., P.~Kufel, and J.~Kwiatkowski (2018).
\newblock Model simplification and variable selection: A replication of the
  {UK} inflation model by {H}endry (2001).
\newblock MPRA Paper 88745.

\bibitem[\protect\citeauthoryear{B{\l}a\.{z}ejowski and
  Kwiatkowski}{B{\l}a\.{z}ejowski and
  Kwiatkowski}{2015}]{BlazejowskiKwiatkowski_15}
B{\l}a\.{z}ejowski, M. and J.~Kwiatkowski (2015).
\newblock Bayesian model averaging and jointness measures for gretl.
\newblock {\em Journal of Statistical Software\/}~{\em 68\/}(5).

\bibitem[\protect\citeauthoryear{B{\l}a\.{z}ejowski and
  Kwiatkowski}{B{\l}a\.{z}ejowski and Kwiatkowski}{2018}]{BlazeKwiat_18}
B{\l}a\.{z}ejowski, M. and J.~Kwiatkowski (2018).
\newblock Bayesian averaging of classical estimates ({BACE}) for gretl.
\newblock Gretl Working Paper~6, Universit{\`a} Politecnica delle Marche.

\bibitem[\protect\citeauthoryear{B{\l}a\.{z}ejowski, Kwiatkowski, and
  Gazda}{B{\l}a\.{z}ejowski et~al.}{2019}]{Blazej_etal_19}
B{\l}a\.{z}ejowski, M., J.~Kwiatkowski, and J.~Gazda (2019).
\newblock Sources of economic growth: A global perspective.
\newblock {\em Sustainability\/}~{\em 11\/}(275).

\bibitem[\protect\citeauthoryear{Brock and Durlauf}{Brock and
  Durlauf}{2001}]{BrockDur01}
Brock, W. and S.~Durlauf (2001).
\newblock Growth empirics and reality.
\newblock {\em World Bank Economic Review\/}~{\em 15}, 229--72.

\bibitem[\protect\citeauthoryear{Brock and Durlauf}{Brock and
  Durlauf}{2015}]{BrockDurlauf_15}
Brock, W. and S.~Durlauf (2015).
\newblock On sturdy policy evaluation.
\newblock {\em Journal of Legal Studies\/}~{\em 44}, S447--73.

\bibitem[\protect\citeauthoryear{Brock, Durlauf, and West}{Brock
  et~al.}{2003}]{Brock_etal}
Brock, W., S.~Durlauf, and K.~West (2003).
\newblock Policy evaluation in uncertain economic environments.
\newblock {\em Brookings Papers of Economic Activity\/}~{\em 1}, 235--322 (with
  discussion).

\bibitem[\protect\citeauthoryear{Brown, Vannucci, and Fearn}{Brown
  et~al.}{1998}]{Brown_etal_98}
Brown, P., M.~Vannucci, and T.~Fearn (1998).
\newblock Bayesian wavelength selection in multicomponent analysis.
\newblock {\em Journal of Chemometrics\/}~{\em 12}, 173--82.

\bibitem[\protect\citeauthoryear{Brown, Fearn, and Vannucci}{Brown
  et~al.}{1999}]{Brown_etal_99}
Brown, P.~J., T.~Fearn, and M.~Vannucci (1999).
\newblock The choice of variables in multivariate regression: A non-conjugate
  {B}ayesian decision theory approach.
\newblock {\em Biometrika\/}~{\em 86}, 635--48.

\bibitem[\protect\citeauthoryear{Buckland, Burnham, and Augustin}{Buckland
  et~al.}{1997}]{Buckland_etal_97}
Buckland, S., K.~Burnham, and N.~Augustin (1997).
\newblock Model selection: an integral part of inference.
\newblock {\em Biometrics\/}~{\em 53}, 603--18.

\bibitem[\protect\citeauthoryear{Burnham and Anderson}{Burnham and
  Anderson}{2002}]{BurnhamAnderson_02}
Burnham, K. and D.~Anderson (2002).
\newblock {\em Model selection and multimodel inference: a practical
  information-theoretic approach (2nd ed.)}.
\newblock New York: Springet.

\bibitem[\protect\citeauthoryear{Camarero, Forte, Garc\'{\i}a-Donato, Mendoza,
  and Ordo\~nez}{Camarero et~al.}{2015}]{Camarero_etal_15}
Camarero, M., A.~Forte, G.~Garc\'{\i}a-Donato, Y.~Mendoza, and J.~Ordo\~nez
  (2015).
\newblock Variable selection in the analysis of energy consumption-growth
  nexus.
\newblock {\em Energy Economics\/}~{\em 52}, 2017--16.

\bibitem[\protect\citeauthoryear{Castillo, Schmidt-Hieber, and van~der
  Vaart}{Castillo et~al.}{2015}]{Castillo_etal_15}
Castillo, I., J.~Schmidt-Hieber, and A.~van~der Vaart (2015).
\newblock Bayesian linear regression with sparse priors.
\newblock {\em Annals of Statistics\/}~{\em 43}, 1986--2018.

\bibitem[\protect\citeauthoryear{Catania and Bernardi}{Catania and
  Bernardi}{2017}]{MCS}
Catania, L. and M.~Bernardi (2017).
\newblock {MCS}: Model confidence set procedure, {R} package.
\newblock \url{https://cran.r-project.org/web/packages/MCS}.

\bibitem[\protect\citeauthoryear{Catania and Nonejad}{Catania and
  Nonejad}{2017}]{eDMA}
Catania, L. and N.~Nonejad (2017).
\newblock e{DMA}: Dynamic model averaging with grid search, {R} package.
\newblock \url{https://cran.r-project.org/web/packages/eDMA}.

\bibitem[\protect\citeauthoryear{Catania and Nonejad}{Catania and
  Nonejad}{2018}]{CataniaNonejad_17}
Catania, L. and N.~Nonejad (2018).
\newblock Dynamic model averaging for practitioners in economics and finance:
  The e{DMA} package.
\newblock {\em Journal of Statistical Software\/}~{\em 84}.

\bibitem[\protect\citeauthoryear{Charitidou, Fouskakis, and
  Ntzoufras}{Charitidou et~al.}{2018}]{Charitidou_etal_18}
Charitidou, E., D.~Fouskakis, and I.~Ntzoufras (2018).
\newblock Objective {B}ayesian transformation and variable selection using
  default {B}ayes factors.
\newblock {\em Statistics and Computing\/}~{\em 28}, 579--94.

\bibitem[\protect\citeauthoryear{Chatfield}{Chatfield}{1995}]{Chatfield}
Chatfield, C. (1995).
\newblock Model uncertainty, data mining and statistical inference.
\newblock {\em Journal of the Royal Statistical Society, Series A\/}~{\em 158},
  419--66 (with discussion).

\bibitem[\protect\citeauthoryear{Chen, Mirestean, and Tsangarides}{Chen
  et~al.}{2018}]{Chen_etal_16}
Chen, H., A.~Mirestean, and C.~G. Tsangarides (2018).
\newblock Bayesian model averaging for dynamic panels with an application to a
  trade gravity model.
\newblock {\em Econometric Reviews\/}~{\em 37}, 777--805.

\bibitem[\protect\citeauthoryear{Chen and Ibrahim}{Chen and
  Ibrahim}{2003}]{ChenIbrahim_03}
Chen, M. and J.~Ibrahim (2003).
\newblock Conjugate priors for generalized linear models.
\newblock {\em Statistica Sinica\/}~{\em 13}, 461--76.

\bibitem[\protect\citeauthoryear{Chen, Chen, Chu, and Lee}{Chen
  et~al.}{2017}]{Chen_etal_17}
Chen, R.-B., Y.-C. Chen, C.-H. Chu, and K.-J. Lee (2017).
\newblock On the determinants of the 2008 financial crisis: {A} {B}ayesian
  approach to the selection of groups and variables.
\newblock {\em Studies in Nonlinear Dynamics \& Econometrics\/}~{\em 21},
  article 20160107.

\bibitem[\protect\citeauthoryear{Cheng and Hansen}{Cheng and
  Hansen}{2015}]{Cheng_Hansen_15}
Cheng, X. and B.~Hansen (2015).
\newblock Forecasting with factor-augmented regression: A frequentist model
  averaging approach.
\newblock {\em Journal of Econometrics\/}~{\em 186}, 280--93.

\bibitem[\protect\citeauthoryear{Chib}{Chib}{1995}]{chib1995}
Chib, S. (1995).
\newblock Marginal likelihood from the {G}ibbs output.
\newblock {\em Journal of the American Statistical Association\/}~{\em 90},
  1313--1321.

\bibitem[\protect\citeauthoryear{Chib}{Chib}{2011}]{Chib_11}
Chib, S. (2011).
\newblock Introduction to simulation and {MCMC} methods.
\newblock In J.~Geweke, G.~Koop, and H.~van Dijk (Eds.), {\em The Oxford
  Handbook of Bayesian Econometrics}, Oxford: Oxford University Press, pp.\
  183--217.

\bibitem[\protect\citeauthoryear{Chipman, Hamada, and Wu}{Chipman
  et~al.}{1997}]{Chipman_etal_97}
Chipman, H., M.~Hamada, and C.~Wu (1997).
\newblock A {B}ayesian variable selection approach for analyzing designed
  experiments with complex aliasing.
\newblock {\em Technometrics\/}~{\em 39}, 372--81.

\bibitem[\protect\citeauthoryear{Christensen and Miguel}{Christensen and
  Miguel}{2018}]{ChristensenMiguel_18}
Christensen, G. and E.~Miguel (2018).
\newblock Transparency, reproducibility, and the credibility of economics
  research.
\newblock {\em Journal of Economic Literature\/}~{\em 56}, 920--80.

\bibitem[\protect\citeauthoryear{Christofides, Eicher, and
  Papageorgiou}{Christofides et~al.}{2016}]{Christofides_etal_EER}
Christofides, C., T.~Eicher, and C.~Papageorgiou (2016).
\newblock Did established early warning signals predict the 2008 crises?
\newblock {\em European Economic Review\/}~{\em 81}, 103--114.
\newblock Special issue on ``Model Uncertainty in Economics''.

\bibitem[\protect\citeauthoryear{Ciccone and Jaroci\'nski}{Ciccone and
  Jaroci\'nski}{2010}]{CicconeJaros_10}
Ciccone, A. and M.~Jaroci\'nski (2010).
\newblock Determinants of economic growth: Will data tell?
\newblock {\em American Economic Journal: Macroeconomics\/}~{\em 2}, 222--46.

\bibitem[\protect\citeauthoryear{Claeskens, Croux, and van
  Kerckhoven}{Claeskens et~al.}{2006}]{Claeskens_etal_06}
Claeskens, G., C.~Croux, and J.~van Kerckhoven (2006).
\newblock Variable selection for logistic regression using a prediction-focused
  information criterion.
\newblock {\em Biometrics\/}~{\em 62}, 972--9.

\bibitem[\protect\citeauthoryear{Claeskens and Hjort}{Claeskens and
  Hjort}{2003}]{ClaeskensHjort_03}
Claeskens, G. and N.~Hjort (2003).
\newblock The focused information criterion.
\newblock {\em Journal of the American Statistical Association\/}~{\em 98},
  900--45 (with discussion).

\bibitem[\protect\citeauthoryear{Claeskens, Magnus, Vasnev, and Wang}{Claeskens
  et~al.}{2015}]{Claeskens_etal_16}
Claeskens, G., J.~Magnus, A.~Vasnev, and W.~Wang (2015).
\newblock The forecast combination puzzle: A simple theoretical explanation.
\newblock {\em International Journal of Forecasting\/}~{\em 32}, 754--62.

\bibitem[\protect\citeauthoryear{Clyde}{Clyde}{2017}]{Cly17}
Clyde, M. (2017).
\newblock {BAS}: {B}ayesian adaptive sampling for {B}ayesian model averaging,
  {R} package version 1.4.3.
\newblock \url{https://cran.r-project.org/web/packages/BAS}.

\bibitem[\protect\citeauthoryear{Clyde and George}{Clyde and
  George}{2004}]{ClydeGeorge04}
Clyde, M. and E.~George (2004).
\newblock Model uncertainty.
\newblock {\em Statistical Science\/}~{\em 19}, 81--94.

\bibitem[\protect\citeauthoryear{Clyde, Ghosh, and Littman}{Clyde
  et~al.}{2011}]{Clydeetal11}
Clyde, M., J.~Ghosh, and M.~Littman (2011).
\newblock Bayesian adaptive sampling for variable selection and model
  averaging.
\newblock {\em Journal of Computational and Graphical Statistics\/}~{\em 20},
  80--101.

\bibitem[\protect\citeauthoryear{Clyde and Iversen}{Clyde and
  Iversen}{2013}]{ClydeIversen_13}
Clyde, M. and E.~Iversen (2013).
\newblock Bayesian model averaging in the {M}-open framework.
\newblock In P.~Damien, P.~Dellaportas, N.~Polson, and D.~Stephens (Eds.), {\em
  Bayesian Theory and Applications}, Oxford: Oxford University Press, pp.\
  483--98.

\bibitem[\protect\citeauthoryear{Cogley and Sargent}{Cogley and
  Sargent}{2005}]{CogleySargent_05}
Cogley, T. and T.~Sargent (2005).
\newblock The conquest of {US} inflation: Learning and robustness to model
  uncertainty.
\newblock {\em Review of Economic Dynamics\/}~{\em 8}, 528--63.

\bibitem[\protect\citeauthoryear{Cohen, Moeltner, Reichl, and
  Schmidthaler}{Cohen et~al.}{2016}]{Cohen_etal_16}
Cohen, J., K.~Moeltner, J.~Reichl, and M.~Schmidthaler (2016).
\newblock An empirical analysis of local opposition to new transmission lines
  across the {EU}-27.
\newblock {\em The Energy Journal\/}~{\em 37}, 59--82.

\bibitem[\protect\citeauthoryear{Cordero, Mu\~niz, and Polo}{Cordero
  et~al.}{2016}]{Cordero_etal_16}
Cordero, J., M.~Mu\~niz, and C.~Polo (2016).
\newblock The determinants of cognitive and non-cognitive educational outcomes:
  empirical evidence in {S}pain using a {B}ayesian approach.
\newblock {\em Applied Economics\/}~{\em 48}, 3355--72.

\bibitem[\protect\citeauthoryear{Crespo~Cuaresma}{Crespo~Cuaresma}{2011}]{Cuaresma_11}
Crespo~Cuaresma, J. (2011).
\newblock How different is {A}frica? a comment on {M}asanjala and
  {P}apageorgiou.
\newblock {\em Journal of Applied Econometrics\/}~{\em 26}, 1041--47.

\bibitem[\protect\citeauthoryear{Crespo~Cuaresma, Doppelhofer, Huber, and
  Piribauer}{Crespo~Cuaresma et~al.}{2018}]{Cuaresma_etal_17}
Crespo~Cuaresma, J., G.~Doppelhofer, F.~Huber, and P.~Piribauer (2018).
\newblock Human capital accumulation and long-term income growth projections
  for {E}uropean regions.
\newblock {\em Journal of Regional Science\/}~{\em 58}, 81--99.

\bibitem[\protect\citeauthoryear{Crespo~Cuaresma and
  Feldkircher}{Crespo~Cuaresma and Feldkircher}{2013}]{CuaresmaFeldkircher_13}
Crespo~Cuaresma, J. and M.~Feldkircher (2013).
\newblock Spatial filtering, model uncertainty and the speed of income
  convergence in {E}urope.
\newblock {\em Journal of Applied Econometrics\/}~{\em 28}, 720--741.

\bibitem[\protect\citeauthoryear{Crespo~Cuaresma, Gr{\"u}n, Hofmarcher, Humer,
  and Moser}{Crespo~Cuaresma et~al.}{2016}]{Crespo_etal_16}
Crespo~Cuaresma, J., B.~Gr{\"u}n, P.~Hofmarcher, S.~Humer, and M.~Moser (2016).
\newblock Unveiling covariate inclusion structures in economic growth
  regressions using latent class analysis.
\newblock {\em European Economic Review\/}~{\em 81}, 189--202.
\newblock Special issue on ``Model Uncertainty in Economics''.

\bibitem[\protect\citeauthoryear{Cui and George}{Cui and
  George}{2008}]{CuiGeorge_08}
Cui, W. and E.~George (2008).
\newblock Empirical {B}ayes vs.~fully {B}ayes variable selection.
\newblock {\em Journal of Statistical Planning and Inference\/}~{\em 138},
  888--900.

\bibitem[\protect\citeauthoryear{Dasgupta, Leon-Gonzalez, and
  Shortland}{Dasgupta et~al.}{2011}]{Dasgupta_etal_11}
Dasgupta, A., R.~Leon-Gonzalez, and A.~Shortland (2011).
\newblock Regionality revisited: An examination of the direction of spread of
  currency crises.
\newblock {\em Journal of International Money and Finance\/}~{\em 30}, 831--48.

\bibitem[\protect\citeauthoryear{Daude, Nagengast, and Perea}{Daude
  et~al.}{2016}]{Daude_etal_16}
Daude, C., A.~Nagengast, and J.~Perea (2016).
\newblock Productive capabilities: An empirical analysis of their drivers.
\newblock {\em The Journal of International Trade \& Economic
  Development\/}~{\em 25}, 504--35.

\bibitem[\protect\citeauthoryear{De~Luca, Magnus, and Peracchi}{De~Luca
  et~al.}{2018}]{DeLuca_etal_18}
De~Luca, G., J.~R. Magnus, and F.~Peracchi (2018).
\newblock Weighted-average least squares estimation of generalized linear
  models.
\newblock {\em Journal of Econometrics\/}~{\em 204}, 1--17.

\bibitem[\protect\citeauthoryear{Dearmon and Smith}{Dearmon and
  Smith}{2016}]{DearmonSmith_16}
Dearmon, J. and T.~Smith (2016).
\newblock Gaussian process regression and {B}ayesian model averaging: An
  alternative approach to modeling spatial phenomena.
\newblock {\em Geographical Analysis\/}~{\em 48}, 82--111.

\bibitem[\protect\citeauthoryear{Deckers and Hanck}{Deckers and
  Hanck}{2014}]{DeckersHanck_14}
Deckers, T. and C.~Hanck (2014).
\newblock Variable selection in cross-section regressions: Comparisons and
  extensions.
\newblock {\em Oxford Bulletin of Economics and Statistics\/}~{\em 76},
  841--73.

\bibitem[\protect\citeauthoryear{Desbordes, Koop, and Vicard}{Desbordes
  et~al.}{2018}]{Desbordes_etal_18}
Desbordes, R., G.~Koop, and V.~Vicard (2018).
\newblock One size does not fit all... panel data: {B}ayesian model averaging
  and data poolability.
\newblock {\em Economic Modelling\/}~{\em 75}, 364--76.

\bibitem[\protect\citeauthoryear{Devereux and Dwyer}{Devereux and
  Dwyer}{2016}]{DevereuxDwyer_16}
Devereux, J. and G.~Dwyer (2016).
\newblock What determines output losses after banking crises?
\newblock {\em Journal of International Money and Finance\/}~{\em 69}, 69--94.

\bibitem[\protect\citeauthoryear{DiCiccio, Kass, Raftery, and
  Wasserman}{DiCiccio et~al.}{1997}]{DiCiccio_etal_97}
DiCiccio, D., R.~Kass, A.~Raftery, and L.~Wasserman (1997).
\newblock Computing {B}ayes factors by combining simulation and asymptotic
  approximations.
\newblock {\em Journal of the American Statistical Association\/}~{\em 92},
  903--15.

\bibitem[\protect\citeauthoryear{Diebold}{Diebold}{1991}]{Diebold_91}
Diebold, F. (1991).
\newblock A note on {B}ayesian forecast combinatton procedures.
\newblock In A.~Westlund and P.~Hackl (Eds.), {\em Economic structural change:
  Analysis and forecasting}, New York: Springer Verlag, pp.\  225--32.

\bibitem[\protect\citeauthoryear{Diebold and Pauly}{Diebold and
  Pauly}{1990}]{DieboldPauly_90}
Diebold, F. and P.~Pauly (1990).
\newblock The use of prior information in forecast combination.
\newblock {\em International Journal of Forecasting\/}~{\em 6}, 503--8.

\bibitem[\protect\citeauthoryear{Diebold and Rudebusch}{Diebold and
  Rudebusch}{1989}]{DieboldRudebush_89}
Diebold, F. and G.~Rudebusch (1989).
\newblock Long memory and persistence in aggregate output.
\newblock {\em Journal of Monetary Economics\/}~{\em 24}, 189--209.

\bibitem[\protect\citeauthoryear{Domingos}{Domingos}{2000}]{Domingos_00}
Domingos, P. (2000).
\newblock Bayesian averaging of classifiers and the overfitting problem.
\newblock In {\em Proceedings of the Seventeenth International Conference on
  Machine Learning}, Stanford, CA, pp.\  223--30.

\bibitem[\protect\citeauthoryear{Doppelhofer, Moe~Hansen, and
  Weeks}{Doppelhofer et~al.}{2016}]{Doppelhofer_etal_16}
Doppelhofer, G., O.-P. Moe~Hansen, and M.~Weeks (2016).
\newblock Determinants of long-term economic growth redux: A measurement error
  model averaging ({MEMA}) approach.
\newblock Working paper 19/16, Norwegian School of Economics.

\bibitem[\protect\citeauthoryear{Doppelhofer and Weeks}{Doppelhofer and
  Weeks}{2009}]{DW_09}
Doppelhofer, G. and M.~Weeks (2009).
\newblock Jointness of growth determinants.
\newblock {\em Journal of Applied Econometrics\/}~{\em 24}, 209--44.

\bibitem[\protect\citeauthoryear{Doppelhofer and Weeks}{Doppelhofer and
  Weeks}{2011}]{DoppelhoferWeeks_11}
Doppelhofer, G. and M.~Weeks (2011).
\newblock Robust growth determinants.
\newblock Working Paper in Economics 1117, University of Cambridge.

\bibitem[\protect\citeauthoryear{Dormann, Calabrese, Guillera-Arroita,
  Matechou, Bahn, Barto{\'n}, Beale, Ciuti, Elith, Gerstner, Guelat, Keil,
  Lahoz-Monfort, Pollock, Reineking, Roberts, Schr{\"o}der, Thuiller, Warton,
  Wintle, Wood, W{\"u}est, and Hartig}{Dormann et~al.}{2018}]{Dorman_etal_18}
Dormann, C.~F., J.~M. Calabrese, G.~Guillera-Arroita, E.~Matechou, V.~Bahn,
  K.~Barto{\'n}, C.~M. Beale, S.~Ciuti, J.~Elith, K.~Gerstner, J.~Guelat,
  P.~Keil, J.~J. Lahoz-Monfort, L.~J. Pollock, B.~Reineking, D.~R. Roberts,
  B.~Schr{\"o}der, W.~Thuiller, D.~I. Warton, B.~A. Wintle, S.~N. Wood, R.~O.
  W{\"u}est, and F.~Hartig (2018).
\newblock Model averaging in ecology: a review of {B}ayesian,
  information-theoretic, and tactical approaches for predictive inference.
\newblock {\em Ecological Monographs\/}, forthcoming.

\bibitem[\protect\citeauthoryear{Drachal}{Drachal}{2016}]{Drachal_16}
Drachal, K. (2016).
\newblock Forecasting spot oil price in a dynamic model averaging framework -
  have the determinants changed over time?
\newblock {\em Energy Economics\/}~{\em 60}, 35--46.

\bibitem[\protect\citeauthoryear{Draper}{Draper}{1995}]{Draper}
Draper, D. (1995).
\newblock Assessment and propagation of model uncertainty.
\newblock {\em Journal of the Royal Statistical Society, Series B\/}~{\em 57},
  45--97 (with discussion).

\bibitem[\protect\citeauthoryear{Draper and Fouskakis}{Draper and
  Fouskakis}{2000}]{DraperFouskakis_00}
Draper, D. and D.~Fouskakis (2000).
\newblock A case study of stochastic optimization in health policy: Problem
  formulation and preliminary results.
\newblock {\em Journal of Global Optimization\/}~{\em 18}, 399--416.

\bibitem[\protect\citeauthoryear{Ductor and Leiva-Leon}{Ductor and
  Leiva-Leon}{2016}]{DuctorLeiva-Leon_16}
Ductor, L. and D.~Leiva-Leon (2016).
\newblock Dynamics of global business cycle interdependence.
\newblock {\em Journal of International Economics\/}~{\em 102}, 110--27.

\bibitem[\protect\citeauthoryear{Dupuis and Robert}{Dupuis and
  Robert}{2003}]{DupuisRobert_03}
Dupuis, J. and C.~Robert (2003).
\newblock Variable selection in qualitative models via an entropic explanatory
  power.
\newblock {\em Journal of Statistical Planning and Inference\/}~{\em 111},
  77--94.

\bibitem[\protect\citeauthoryear{Durlauf}{Durlauf}{2018}]{Durlauf_18}
Durlauf, S. (2018).
\newblock Institutions, development and growth: Where does evidence stand?
\newblock Working Paper WP18/04.1, Economic Development \& Institutions.

\bibitem[\protect\citeauthoryear{Durlauf, Fu, and Navarro}{Durlauf
  et~al.}{2012}]{Durlauf_etal_12b}
Durlauf, S., C.~Fu, and S.~Navarro (2012).
\newblock Assumptions matter: Model uncertainty and the deterrent effect of
  capital punishment.
\newblock {\em American Economic Review: Papers and Proceedings\/}~{\em 102},
  487--92.

\bibitem[\protect\citeauthoryear{Durlauf, Kourtellos, and Tan}{Durlauf
  et~al.}{2008}]{Durlauf_etal_08}
Durlauf, S., A.~Kourtellos, and C.~Tan (2008).
\newblock Are any growth theories robust?
\newblock {\em Economic Journal\/}~{\em 118}, 329--46.

\bibitem[\protect\citeauthoryear{Durlauf, Kourtellos, and Tan}{Durlauf
  et~al.}{2012}]{Durlauf_etal_11}
Durlauf, S., A.~Kourtellos, and C.~Tan (2012).
\newblock Is {G}od in the details? a reexamination of the role of religion in
  economic growth.
\newblock {\em Journal of Applied Econometrics\/}~{\em 27}, 1059--75.

\bibitem[\protect\citeauthoryear{Eicher and Newiak}{Eicher and
  Newiak}{2013}]{EicherNewiak_13}
Eicher, T. and M.~Newiak (2013).
\newblock Intellectual property rights as development determinants.
\newblock {\em Canadian Journal of Economics\/}~{\em 46}, 4--22.

\bibitem[\protect\citeauthoryear{Eicher, Papageorgiou, and Raftery}{Eicher
  et~al.}{2011}]{Eicheretal11}
Eicher, T., C.~Papageorgiou, and A.~Raftery (2011).
\newblock Default priors and predictive performance in {B}ayesian model
  averaging, with application to growth determinants.
\newblock {\em Journal of Applied Econometrics\/}~{\em 26}, 30--55.

\bibitem[\protect\citeauthoryear{Eicher and Kuenzel}{Eicher and
  Kuenzel}{2016}]{EicherKuenzel_16}
Eicher, T.~S. and D.~J. Kuenzel (2016).
\newblock The elusive effects of trade on growth: Export diversity and economic
  take-off.
\newblock {\em Canadian Journal of Economics\/}~{\em 49}, 264--295.

\bibitem[\protect\citeauthoryear{Eklund and Karlsson}{Eklund and
  Karlsson}{2007}]{EklundKarlsson}
Eklund, J. and S.~Karlsson (2007).
\newblock Forecast combination and model averaging using predictive measures.
\newblock {\em Econometric Reviews\/}~{\em 26}, 329--63.

\bibitem[\protect\citeauthoryear{Fan and Li}{Fan and Li}{2001}]{FanLi01}
Fan, J. and R.~Li (2001).
\newblock Variable selection via nonconcave penalized likelihood and its oracle
  properties.
\newblock {\em Journal of the American Statistical Association\/}~{\em 96},
  1348--60.

\bibitem[\protect\citeauthoryear{Feldkircher}{Feldkircher}{2014}]{Feldkircher14}
Feldkircher, M. (2014).
\newblock The determinants of vulnerability to the global financial crisis 2008
  to 2009: Credit growth and other sources of risk.
\newblock {\em Journal of International Money and Finance\/}~{\em 43}, 19--49.

\bibitem[\protect\citeauthoryear{Feldkircher, Horvath, and Rusnak}{Feldkircher
  et~al.}{2014}]{Feldkircher_etal_14}
Feldkircher, M., R.~Horvath, and M.~Rusnak (2014).
\newblock Exchange market pressures during the financial crisis: A {B}ayesian
  model averaging evidence.
\newblock {\em Journal of International Money and Finance\/}~{\em 40}, 21--41.

\bibitem[\protect\citeauthoryear{Feldkircher and Huber}{Feldkircher and
  Huber}{2016}]{FeldkircherHuber_16}
Feldkircher, M. and F.~Huber (2016).
\newblock The international transmission of {US} shocks - evidence from
  {B}ayesian global vector autoregressions.
\newblock {\em European Economic Review\/}~{\em 81}, 167--88.
\newblock Special issue on ``Model Uncertainty in Economics''.

\bibitem[\protect\citeauthoryear{Feldkircher and Zeugner}{Feldkircher and
  Zeugner}{2009}]{FZ09}
Feldkircher, M. and S.~Zeugner (2009).
\newblock Benchmark priors revisited: On adaptive shrinkage and the supermodel
  effect in {B}ayesian model averaging.
\newblock Working Paper 09/202, IMF.

\bibitem[\protect\citeauthoryear{Feldkircher and Zeugner}{Feldkircher and
  Zeugner}{2012}]{FeldkircherZeugner_12}
Feldkircher, M. and S.~Zeugner (2012).
\newblock The impact of data revisions on the robustness of growth
  determinants: A note on `determinants of economic growth. will data tell'?
\newblock {\em Journal of Applied Econometrics\/}~{\em 27}, 686--94.

\bibitem[\protect\citeauthoryear{Feldkircher and Zeugner}{Feldkircher and
  Zeugner}{2014}]{webBMS16}
Feldkircher, M. and S.~Zeugner (2014).
\newblock R-package {BMS}: {B}ayesian {M}odel {A}veraging in {R}.
\newblock \url{http://bms.zeugner.eu}.

\bibitem[\protect\citeauthoryear{Fern\'andez, Ley, and Steel}{Fern\'andez
  et~al.}{2001a}]{FLS01a}
Fern\'andez, C., E.~Ley, and M.~Steel (2001a).
\newblock Benchmark priors for {B}ayesian model averaging.
\newblock {\em Journal of Econometrics\/}~{\em 100}, 381--427.

\bibitem[\protect\citeauthoryear{Fern\'andez, Ley, and Steel}{Fern\'andez
  et~al.}{2001b}]{FLS01b}
Fern\'andez, C., E.~Ley, and M.~Steel (2001b).
\newblock Model uncertainty in cross-country growth regressions.
\newblock {\em Journal of Applied Econometrics\/}~{\em 16}, 563--76.

\bibitem[\protect\citeauthoryear{Fletcher}{Fletcher}{2018}]{Fletcher_18}
Fletcher, D. (2018).
\newblock {\em Model Averaging}.
\newblock Heidelberg: Springer Nature.

\bibitem[\protect\citeauthoryear{Forte, Garc\'{\i}a-Donato, and Steel}{Forte
  et~al.}{2018}]{Forte_etal_17}
Forte, A., G.~Garc\'{\i}a-Donato, and M.~Steel (2018).
\newblock Methods and tools for {B}ayesian variable selection and model
  averaging in normal linear regression.
\newblock {\em International Statistical Review\/}~{\em 86}, 237--58.

\bibitem[\protect\citeauthoryear{Foster and George}{Foster and
  George}{1994}]{FosterGeorge_94}
Foster, D. and E.~George (1994).
\newblock The risk inflation criterion for multiple regression.
\newblock {\em Annals of Statistics\/}~{\em 22}, 1947--75.

\bibitem[\protect\citeauthoryear{Fouskakis and Ntzoufras}{Fouskakis and
  Ntzoufras}{2016a}]{FouskakisNtzoufras_16}
Fouskakis, D. and I.~Ntzoufras (2016a).
\newblock Limiting behavior of the {J}effreys power-expected-posterior {B}ayes
  factor in {G}aussian linear models.
\newblock {\em Brazilian Journal of Probability and Statistics\/}~{\em 30},
  299--320.

\bibitem[\protect\citeauthoryear{Fouskakis and Ntzoufras}{Fouskakis and
  Ntzoufras}{2016b}]{FouskakisNtzoufras_16a}
Fouskakis, D. and I.~Ntzoufras (2016b).
\newblock Power-conditional-expected priors: {U}sing $g$-priors with random
  imaginary data for variable selection.
\newblock {\em Journal of Computational and Graphical Statistics\/}~{\em 25},
  647--64.

\bibitem[\protect\citeauthoryear{Fouskakis, Ntzoufras, and Perrakis}{Fouskakis
  et~al.}{2018}]{Perrakis_etal_15}
Fouskakis, D., I.~Ntzoufras, and K.~Perrakis (2018).
\newblock Power-expected-posterior priors for generalized linear models.
\newblock {\em Bayesian Analysis\/}~{\em 13}, 721--48.

\bibitem[\protect\citeauthoryear{Fragoso, Bertoli, and Louzada}{Fragoso
  et~al.}{2018}]{FragosoNeto_15}
Fragoso, T., W.~Bertoli, and F.~Louzada (2018).
\newblock Bayesian model averaging: A systematic review and conceptual
  classification.
\newblock {\em International Statistical Review\/}~{\em 86}, 1--28.

\bibitem[\protect\citeauthoryear{Frankel}{Frankel}{2012}]{Frankel_12}
Frankel, J. (2012).
\newblock The natural resource curse: A survey of diagnoses and some
  prescriptions.
\newblock In R.~Arezki, C.~Pattillo, M.~Quintyn, and M.~Zhu (Eds.), {\em
  Commodity Price Volatility and Inclusive Growth in Low-Income Countries},
  Washington, D.C.: International Monetary Fund, pp.\  7--34.

\bibitem[\protect\citeauthoryear{Frankel and Romer}{Frankel and
  Romer}{1999}]{FrankelRomer_99}
Frankel, J. and D.~Romer (1999).
\newblock Does trade cause growth?
\newblock {\em American Economic Review\/}~{\em 89}, 379--399.

\bibitem[\protect\citeauthoryear{Frankel and Saravelos}{Frankel and
  Saravelos}{2012}]{FrankelSaravelos_12}
Frankel, J. and G.~Saravelos (2012).
\newblock Can leading indicators assess country vulnerability? {E}vidence from
  the 2008-09 global financial crisis.
\newblock {\em Journal of International Economics\/}~{\em 87}, 216--31.

\bibitem[\protect\citeauthoryear{Furnival and Wilson}{Furnival and
  Wilson}{1974}]{FurnivalWilson}
Furnival, G. and R.~Wilson (1974).
\newblock Regressions by leaps and bounds.
\newblock {\em Technometrics\/}~{\em 16}, 499--511.

\bibitem[\protect\citeauthoryear{Garc\'{i}a-Donato and Forte}{Garc\'{i}a-Donato
  and Forte}{2015}]{GarFor15}
Garc\'{i}a-Donato, G. and A.~Forte (2015).
\newblock {BayesVarSel}: {B}ayes factors, model choice and variable selection
  in linear models, {R} package version 1.6.1.
\newblock \url{http://CRAN.R-project.org/package=BayesVarSel}.

\bibitem[\protect\citeauthoryear{Garc\'{\i}a-Donato and
  Forte}{Garc\'{\i}a-Donato and Forte}{2018}]{GarFor17}
Garc\'{\i}a-Donato, G. and A.~Forte (2018).
\newblock Bayesian testing, variable selection and model averaging in linear
  models using {{R}} with {B}ayes{V}ar{S}el.
\newblock {\em R Journal\/}~{\em 10}, 155--74.

\bibitem[\protect\citeauthoryear{Garc{\'{\i}}a-Donato and
  Mart{\'{\i}}nez-Beneito}{Garc{\'{\i}}a-Donato and
  Mart{\'{\i}}nez-Beneito}{2013}]{GarciaDonato}
Garc{\'{\i}}a-Donato, G. and M.~Mart{\'{\i}}nez-Beneito (2013).
\newblock On sampling strategies in {B}ayesian variable selection problems with
  large model spaces.
\newblock {\em Journal of the American Statistical Association\/}~{\em 108},
  340--52.

\bibitem[\protect\citeauthoryear{Garratt, Lee, Pesaran, and Shin}{Garratt
  et~al.}{2003}]{Garratt_etal}
Garratt, A., K.~Lee, M.~Pesaran, and Y.~Shin (2003).
\newblock Forecasting uncertainties in macroeconometric modelling: An
  application to the {UK} economy.
\newblock {\em Journal of the American Statistical Association\/}~{\em 98},
  829--38.

\bibitem[\protect\citeauthoryear{Gelfand and Ghosh}{Gelfand and
  Ghosh}{1998}]{GelfandGhosh_98}
Gelfand, A. and S.~Ghosh (1998).
\newblock Model choice: A minimum posterior predictive loss approach.
\newblock {\em Biometrika\/}~{\em 85}, 1--11.

\bibitem[\protect\citeauthoryear{George}{George}{1999a}]{George_99b}
George, E. (1999a).
\newblock Comment on ``{B}ayesian model averaging: {A} tutorial'' by
  {J.~H}oeting, {D.~M}adigan, {A.~R}aftery and {C.~V}olinksy.
\newblock {\em Statistical Science\/}~{\em 14}, 409--12.

\bibitem[\protect\citeauthoryear{George}{George}{1999b}]{George_99}
George, E. (1999b).
\newblock Discussion of ``{B}ayesian model averaging and model search
  strategies'' by {M}. {C}lyde.
\newblock In J.~Bernardo, J.~Berger, A.~Dawid, and A.~Smith (Eds.), {\em
  Bayesian Statistics 6}, Oxford: Oxford University Press, pp.\  175--7.

\bibitem[\protect\citeauthoryear{George}{George}{2010}]{George_10}
George, E. (2010).
\newblock Dilution priors: Compensating for model space redundancy.
\newblock In J.~Berger, T.~Cai, and I.~Johnstone (Eds.), {\em Borrowing
  Strength: Theory Powering Applications}, Institute of Mathematical Statistics
  - Collections, Vol.~6, Beachwood, OH: IMS, pp.\  158--65.

\bibitem[\protect\citeauthoryear{George and Foster}{George and
  Foster}{2000}]{GeorgeFoster_00}
George, E. and D.~Foster (2000).
\newblock Calibration and empirical {B}ayes variable selection.
\newblock {\em Biometrika\/}~{\em 87}, 731--47.

\bibitem[\protect\citeauthoryear{George and McCulloch}{George and
  McCulloch}{1993}]{GeorgeMcCulloch_93}
George, E. and R.~McCulloch (1993).
\newblock Variable selection via {G}ibbs sampling.
\newblock {\em Journal of the American Statistical Association\/}~{\em 88},
  881--89.

\bibitem[\protect\citeauthoryear{George and McCulloch}{George and
  McCulloch}{1997}]{GeorgeMcCulloch_97}
George, E. and R.~McCulloch (1997).
\newblock Approaches for {B}ayesian variable selection.
\newblock {\em Statistica Sinica\/}~{\em 7}, 339--73.

\bibitem[\protect\citeauthoryear{George, Sun, and Ni}{George
  et~al.}{2008}]{GeorgeSunNi_08}
George, E., D.~Sun, and S.~Ni (2008).
\newblock Bayesian stochastic search for {VAR} model restrictions.
\newblock {\em Journal of Econometrics\/}~{\em 142}, 553--80.

\bibitem[\protect\citeauthoryear{Geweke and Amisano}{Geweke and
  Amisano}{2011}]{GewekeAmisano_11}
Geweke, J. and G.~Amisano (2011).
\newblock Optimal prediction pools.
\newblock {\em Journal of Econometrics\/}~{\em 164}, 130--41.

\bibitem[\protect\citeauthoryear{Ghosh and Ghattas}{Ghosh and
  Ghattas}{2015}]{GhoshGattas_16}
Ghosh, J. and A.~Ghattas (2015).
\newblock Bayesian variable selection under collinearity.
\newblock {\em The American Statistician\/}~{\em 69}, 165--73.

\bibitem[\protect\citeauthoryear{Giannone, Lenza, and Primiceri}{Giannone
  et~al.}{2018}]{Giannone_etal_18}
Giannone, D., M.~Lenza, and G.~Primiceri (2018).
\newblock Economic predictions with big data: The illusion of sparsity.
\newblock Staff Reports 847, Federal Reserve Bank of New York.

\bibitem[\protect\citeauthoryear{Giannone, Lenza, and Reichlin}{Giannone
  et~al.}{2011}]{Giannone_etal_11}
Giannone, D., M.~Lenza, and L.~Reichlin (2011).
\newblock Market freedom and the global recession.
\newblock {\em IMF Economic Review\/}~{\em 59}, 111--35.

\bibitem[\protect\citeauthoryear{Gneiting and Raftery}{Gneiting and
  Raftery}{2007}]{GneitingRaft_07}
Gneiting, T. and A.~Raftery (2007).
\newblock Strictly proper scoring rules, prediction and estimation.
\newblock {\em Journal of the American Statistical Association\/}~{\em 102},
  359--78.

\bibitem[\protect\citeauthoryear{Good}{Good}{1952}]{Good_52}
Good, I. (1952).
\newblock Rational decisions.
\newblock {\em Journal of the Royal Statistical Society. B\/}~{\em 14},
  107--114.

\bibitem[\protect\citeauthoryear{Granger}{Granger}{1989}]{Granger_89}
Granger, C. (1989).
\newblock Combining forecasts - twenty years later.
\newblock {\em Journal of Forecasting\/}~{\em 8}, 167--73.

\bibitem[\protect\citeauthoryear{Gravestock and Saban\'es~Bov\'e}{Gravestock
  and Saban\'es~Bov\'e}{2017}]{glmBfp}
Gravestock, I. and D.~Saban\'es~Bov\'e (2017).
\newblock {glmBfp}: {B}ayesian fractional polynomials for {GLM}s, {R} package
  version 0.0-51.
\newblock \url{https://cran.r-project.org/package=glmBfp}.

\bibitem[\protect\citeauthoryear{Griffin, {\L}atuszy\'nski, and Steel}{Griffin
  et~al.}{2017}]{Griffin_etal_17}
Griffin, J., K.~{\L}atuszy\'nski, and M.~Steel (2017).
\newblock In search of lost (mixing) time: Adaptive {MCMC} schemes for
  {B}ayesian variable selection with very large $p$.
\newblock technical report arXiv:1708.05678, University of Warwick.

\bibitem[\protect\citeauthoryear{Griffin and Steel}{Griffin and
  Steel}{2008}]{GriffinSteel_08}
Griffin, J. and M.~Steel (2008).
\newblock Flexible mixture modelling of stochastic frontiers.
\newblock {\em Journal of Productivity Analysis\/}~{\em 29}, 33--50.

\bibitem[\protect\citeauthoryear{Hall and Mitchell}{Hall and
  Mitchell}{2007}]{HallMitchell_07}
Hall, S. and J.~Mitchell (2007).
\newblock Combining density forecasts.
\newblock {\em International Journal of Forecasting\/}~{\em 23}, 1--13.

\bibitem[\protect\citeauthoryear{Hanck}{Hanck}{2016}]{Hanck_16}
Hanck, C. (2016).
\newblock I just ran two trillion regressions.
\newblock {\em Economics Bulletin\/}~{\em 36}, 2017--42.

\bibitem[\protect\citeauthoryear{Hansen}{Hansen}{2000}]{Hansen_00}
Hansen, B. (2000).
\newblock Sample splitting and threshold estimation.
\newblock {\em Econometrica\/}~{\em 68}, 575--603.

\bibitem[\protect\citeauthoryear{Hansen}{Hansen}{2007}]{Hansen_07}
Hansen, B. (2007).
\newblock Least squares model averaging.
\newblock {\em Econometrica\/}~{\em 75}, 1175--89.

\bibitem[\protect\citeauthoryear{Hansen and Racine}{Hansen and
  Racine}{2012}]{HansenRacine_12}
Hansen, B. and J.~Racine (2012).
\newblock Jackknife model averaging.
\newblock {\em Journal of Econometrics\/}~{\em 157}, 38--46.

\bibitem[\protect\citeauthoryear{Hansen and Sargent}{Hansen and
  Sargent}{2014}]{HansenSargent_14}
Hansen, L. and T.~Sargent (2014).
\newblock {\em Uncertainty Within Economic Models}, Volume~6 of {\em World
  Scientific Series in Economic Theory}.
\newblock Singapore: World Scientific.

\bibitem[\protect\citeauthoryear{Hansen and Yu}{Hansen and
  Yu}{2001}]{HansenYu_01}
Hansen, M.~H. and B.~Yu (2001).
\newblock Model selection and the principle of minimum description length.
\newblock {\em Journal of the American Statistical Association\/}~{\em 96},
  746--74.

\bibitem[\protect\citeauthoryear{Hansen, Lunde, and Nason}{Hansen
  et~al.}{2011}]{Hansen_etal_11}
Hansen, P., A.~Lunde, and J.~Nason (2011).
\newblock The model confidence set.
\newblock {\em Econometrica\/}~{\em 79}, 453--97.

\bibitem[\protect\citeauthoryear{Hartwell}{Hartwell}{2016}]{Hartwell_16}
Hartwell, C. (2016).
\newblock The institutional basis of efficiency in resource-rich countries.
\newblock {\em Economic Systems\/}~{\em 40}, 519--38.

\bibitem[\protect\citeauthoryear{Hastie, Tibshirani, and Friedman}{Hastie
  et~al.}{2009}]{Hastie_etal_09}
Hastie, T., R.~Tibshirani, and J.~Friedman (2009).
\newblock {\em The Elements of Statistical Learning}.
\newblock New York: Springer.

\bibitem[\protect\citeauthoryear{Hauser, P{\"o}tscher, and Reschenhofer}{Hauser
  et~al.}{1999}]{Hauser_etal_99}
Hauser, M., B.~P{\"o}tscher, and E.~Reschenhofer (1999).
\newblock Measuring persistence in aggregate output: {ARMA} models,
  fractionally integrated {ARMA} models and nonparametric procedures.
\newblock {\em Empirical Economics\/}~{\em 24}, 243--69.

\bibitem[\protect\citeauthoryear{Havranek, Horvath, Irsova, and
  Rusnak}{Havranek et~al.}{2015}]{Havranek_et_al_15}
Havranek, T., R.~Horvath, Z.~Irsova, and M.~Rusnak (2015).
\newblock Cross-country heterogeneity in intertemporal substitution.
\newblock {\em Journal of International Economics\/}~{\em 96}, 100--18.

\bibitem[\protect\citeauthoryear{Havranek, Irsova, and Zeynalova}{Havranek
  et~al.}{2018}]{Havranek_etal_18}
Havranek, T., Z.~Irsova, and O.~Zeynalova (2018).
\newblock Tuition fees and university enrolment: A meta-regression analysis.
\newblock {\em Oxford Bulletin of Economics and Statistics\/}~{\em 80},
  1145--84.

\bibitem[\protect\citeauthoryear{Havranek, Rusnak, and Sokolova}{Havranek
  et~al.}{2017}]{Havranek_etal_17}
Havranek, T., M.~Rusnak, and A.~Sokolova (2017).
\newblock Habit formation in consumption: A meta-analysis.
\newblock {\em European Economic Review\/}~{\em 95}, 142--67.

\bibitem[\protect\citeauthoryear{Havranek and Sokolova}{Havranek and
  Sokolova}{2016}]{HavranekSokolova_16}
Havranek, T. and A.~Sokolova (2016).
\newblock Do consumers really follow a rule of thumb? three thousand estimates
  from 130 studies say ``probably not''.
\newblock Working Paper 8/2016, Czech National Bank.

\bibitem[\protect\citeauthoryear{Henderson and Parmeter}{Henderson and
  Parmeter}{2016}]{HendersonParmeter_16}
Henderson, D. and C.~Parmeter (2016).
\newblock Model averaging over nonparametric estimators.
\newblock In G.~Gonz\'alez-Rivera, R.~Carter~Hill, and T.-H. Lee (Eds.), {\em
  Essays in Honor of Aman Ullah}, Advances in Econometrics, Vol.~36, Emerald
  Group Publishing Limited, pp.\  539--60.

\bibitem[\protect\citeauthoryear{Hendry}{Hendry}{2001}]{Hendry_01}
Hendry, D. (2001).
\newblock Modelling {UK} inflation, 1875-1991.
\newblock {\em Journal of Applied Econometrics\/}~{\em 16}, 255--75.

\bibitem[\protect\citeauthoryear{Hendry and Krolzig}{Hendry and
  Krolzig}{2005}]{HendryKrolzig_05}
Hendry, D. and H.-M. Krolzig (2005).
\newblock he properties of automatic gets modelling.
\newblock {\em Economic Journal\/}~{\em 115}, C32--61.

\bibitem[\protect\citeauthoryear{Hendry and Krolzig}{Hendry and
  Krolzig}{2004}]{HendryKrolzig_04}
Hendry, D.~H. and H.-M. Krolzig (2004).
\newblock We ran one regression.
\newblock {\em Oxford Bulletin of Economics and Statistics\/}~{\em 66},
  799--810.

\bibitem[\protect\citeauthoryear{Hern\'andez, Raftery, Pennington, and
  Parnell}{Hern\'andez et~al.}{2018}]{Hernandez_etal_15}
Hern\'andez, B., A.~Raftery, S.~Pennington, and A.~Parnell (2018).
\newblock Bayesian additive regression trees using {B}ayesian model averaging.
\newblock {\em Statistics and Computing\/}~{\em 28}, 869--90.

\bibitem[\protect\citeauthoryear{Hjort and Claeskens}{Hjort and
  Claeskens}{2003}]{HjortClaeskens_03}
Hjort, N. and G.~Claeskens (2003).
\newblock Frequentist model average estimators.
\newblock {\em Journal of the American Statistical Association\/}~{\em 98},
  879--99.

\bibitem[\protect\citeauthoryear{Ho}{Ho}{2015}]{Ho_15}
Ho, T. (2015).
\newblock Looking for a needle in a haystack: Revisiting the cross-country
  causes of the 2008-9 crisis by {B}ayesian model averaging.
\newblock {\em Economica\/}~{\em 82}, 813--40.

\bibitem[\protect\citeauthoryear{Hoeting, Madigan, Raftery, and
  Volinksy}{Hoeting et~al.}{1999}]{Hoeting_etal_99}
Hoeting, J., D.~Madigan, A.~Raftery, and C.~Volinksy (1999).
\newblock Bayesian model averaging: A tutorial.
\newblock {\em Statistical Science\/}~{\em 14}, 382--417 (with discussion).

\bibitem[\protect\citeauthoryear{Hoeting, Raftery, and Madigan}{Hoeting
  et~al.}{1996}]{Hoeting_etal_96}
Hoeting, J., A.~Raftery, and D.~Madigan (1996).
\newblock A method for simultaneous variable selection and outlier
  identification in linear regression.
\newblock {\em Computational Statistics and Data Analysis\/}~{\em 22}, 251--70.

\bibitem[\protect\citeauthoryear{Hoeting, Raftery, and Madigan}{Hoeting
  et~al.}{2002}]{Hoeting_etal_02}
Hoeting, J., A.~Raftery, and D.~Madigan (2002).
\newblock A method for simultaneous variable and transformation selection in
  linear regression.
\newblock {\em Journal of Computational and Graphical Statistics\/}~{\em 11},
  485--507.

\bibitem[\protect\citeauthoryear{Hofmarcher, Crespo~Cuaresma, Gr{\"u}n, Humer,
  and Moser}{Hofmarcher et~al.}{2018}]{Crespo_etal_17_joint}
Hofmarcher, P., J.~Crespo~Cuaresma, B.~Gr{\"u}n, S.~Humer, and M.~Moser (2018).
\newblock Bivariate jointness measures in {B}ayesian model averaging: Solving
  the conundrum.
\newblock {\em Journal of Macroeconomics\/}~{\em 57}, 150--65.

\bibitem[\protect\citeauthoryear{Hoogerheide, Kleijn, Ravazzolo, {van}~Dijk,
  and Verbeek}{Hoogerheide et~al.}{2010}]{Hoogerheide_etal_10}
Hoogerheide, L., R.~Kleijn, F.~Ravazzolo, H.~{van}~Dijk, and M.~Verbeek (2010).
\newblock Forecast accuracy and economic gains from {B}ayesian model averaging
  using time-varying weights.
\newblock {\em Journal of Forecasting\/}~{\em 29}, 251--69.

\bibitem[\protect\citeauthoryear{Hoover and Perez}{Hoover and
  Perez}{1999}]{HooverPerez_99}
Hoover, K. and S.~Perez (1999).
\newblock Data mining reconsidered: encompassing and the general-to- specific
  approach to specification search.
\newblock {\em Econometrics Journal\/}~{\em 2}, 167--91.

\bibitem[\protect\citeauthoryear{Hortas-Rico and Rios}{Hortas-Rico and
  Rios}{2016}]{HortasRios_16}
Hortas-Rico, M. and V.~Rios (2016).
\newblock The drivers of income inequality in cities: A spatial {B}ayesian
  {M}odel {A}veraging approach.
\newblock Estudios sobre la Econom\'{\i}a Espa\~nola 2016/26, FEDEA.

\bibitem[\protect\citeauthoryear{Horvath, Horvatova, and Siranova}{Horvath
  et~al.}{2017}]{Horvath_etal_17}
Horvath, R., E.~Horvatova, and M.~Siranova (2017).
\newblock Financial development, rule of law and wealth inequality: {B}ayesian
  model averaging evidence.
\newblock Discussion Paper 12/2017, Bank of Finland Institute for Economies in
  Transition.

\bibitem[\protect\citeauthoryear{Ibrahim and Chen}{Ibrahim and
  Chen}{2000}]{IbrahimChen_00}
Ibrahim, J. and M.-H. Chen (2000).
\newblock Power prior distributions for regression models.
\newblock {\em Statistical Science\/}~{\em 15}, 46--60.

\bibitem[\protect\citeauthoryear{Iyke}{Iyke}{2018}]{Iyke_15}
Iyke, B. (2018).
\newblock Macro determinants of the real exchange rate in a small open small
  island economy: Evidence from {M}auritius via {BMA}.
\newblock {\em Bulletin of Monetary Economics and Banking\/}~{\em 21}, 1--24.

\bibitem[\protect\citeauthoryear{Jeffreys}{Jeffreys}{1961}]{jeffreys1961}
Jeffreys, H. (1961).
\newblock {\em Theory of Probability\/} (3rd ed.).
\newblock Oxford University Press.

\bibitem[\protect\citeauthoryear{Jetter and Parmeter}{Jetter and
  Parmeter}{2018}]{JetterParmeter_17}
Jetter, M. and C.~Parmeter (2018).
\newblock Sorting through global corruption determinants: Institutions and
  education matter - not culture.
\newblock {\em World Development\/}~{\em 109}, 279--94.

\bibitem[\protect\citeauthoryear{Jetter and Ram\'{\i}rex~Hassan}{Jetter and
  Ram\'{\i}rex~Hassan}{2015}]{JetterHassan_15}
Jetter, M. and A.~Ram\'{\i}rex~Hassan (2015).
\newblock Want export diversification? educate the kids first.
\newblock {\em Economic Inquiry\/}~{\em 53}, 1765--82.

\bibitem[\protect\citeauthoryear{Johnson and Rossell}{Johnson and
  Rossell}{2010}]{JohnsonRossell_10}
Johnson, V. and D.~Rossell (2010).
\newblock On the use of non-local prior densities in {B}ayesian hypothesis
  tests.
\newblock {\em Journal of the Royal Statistical Society, B\/}~{\em 72},
  143--70.

\bibitem[\protect\citeauthoryear{Jovanovic}{Jovanovic}{2017}]{Jovanovic_17}
Jovanovic, B. (2017).
\newblock Growth forecast errors and government investment and consumption
  multipliers.
\newblock {\em International Review of Applied Economics\/}~{\em 31}, 83--107.

\bibitem[\protect\citeauthoryear{Kapetanios, Mitchell, Price, and
  Fawcett}{Kapetanios et~al.}{2015}]{Kapetanios_etal_15}
Kapetanios, G., J.~Mitchell, S.~Price, and N.~Fawcett (2015).
\newblock Generalised density forecast combinations.
\newblock {\em Journal of Econometrics\/}~{\em 188}, 150--65.

\bibitem[\protect\citeauthoryear{Kapetanios and Papailias}{Kapetanios and
  Papailias}{2018}]{KapetaniosPap_18}
Kapetanios, G. and F.~Papailias (2018).
\newblock Big data \& macroeconomic nowcasting: Methodological review.
\newblock Discussion Paper 2018-12, Economic Statistics Centre of Excellence.

\bibitem[\protect\citeauthoryear{Karl and Lenkoski}{Karl and
  Lenkoski}{2012}]{KarlLenkoski_12}
Karl, A. and A.~Lenkoski (2012).
\newblock Instrumental variables {B}ayesian model averaging via conditional
  {B}ayes factors.
\newblock Technical Report arXiv:1202.5846v3, Heidelberg University.

\bibitem[\protect\citeauthoryear{Kass and Raftery}{Kass and
  Raftery}{1995}]{kassraftery1995}
Kass, R. and A.~Raftery (1995).
\newblock Bayes factors.
\newblock {\em Journal of the American Statistical Association\/}~{\em 90},
  773--795.

\bibitem[\protect\citeauthoryear{Kass and Wasserman}{Kass and
  Wasserman}{1995}]{KassWasserman_95}
Kass, R. and L.~Wasserman (1995).
\newblock A reference {B}ayesian test for nested hypotheses and its
  relationship to the {S}chwarz criterion.
\newblock {\em Journal of the American Statistical Association\/}~{\em 90},
  928--34.

\bibitem[\protect\citeauthoryear{Kiviet}{Kiviet}{1995}]{Kiviet_95}
Kiviet, J. (1995).
\newblock On bias, inconsistency, and efficiency of various estimators in
  dynamic panel data models.
\newblock {\em Journal of Econometrics\/}~{\em 68}, 53--78.

\bibitem[\protect\citeauthoryear{Koop}{Koop}{2003}]{Koop_03}
Koop, G. (2003).
\newblock {\em Bayesian Econometrics}.
\newblock Chichester: Wiley.

\bibitem[\protect\citeauthoryear{Koop}{Koop}{2017}]{Koop_17}
Koop, G. (2017).
\newblock Bayesian methods for empirical macroeconomics with big data.
\newblock {\em Review of Economic Analysis\/}~{\em 9}, 33--56.

\bibitem[\protect\citeauthoryear{Koop and Korobilis}{Koop and
  Korobilis}{2012}]{KoopKorobilis12}
Koop, G. and D.~Korobilis (2012).
\newblock Forecasting inflation using dynamic model averaging.
\newblock {\em International Economic Review\/}~{\em 53}, 867--86.

\bibitem[\protect\citeauthoryear{Koop and Korobilis}{Koop and
  Korobilis}{2016}]{KoopKorobilis_16}
Koop, G. and D.~Korobilis (2016).
\newblock Model uncertainty in panel vector autoregressive models.
\newblock {\em European Economic Review\/}~{\em 81}, 115--131.
\newblock Special issue on ``Model Uncertainty in Economics''.

\bibitem[\protect\citeauthoryear{Koop, Leon-Gonzalez, and Strachan}{Koop
  et~al.}{2012}]{Koop_et_al_12}
Koop, G., R.~Leon-Gonzalez, and R.~Strachan (2012).
\newblock Bayesian model averaging in the instrumental variable regression
  model.
\newblock {\em Journal of Econometrics\/}~{\em 171}, 237--50.

\bibitem[\protect\citeauthoryear{Koop, Ley, Osiewalski, and Steel}{Koop
  et~al.}{1997}]{Koop_etal_97}
Koop, G., E.~Ley, J.~Osiewalski, and M.~Steel (1997).
\newblock Bayesian analysis of long memory and persistence using {ARFIMA}
  models.
\newblock {\em Journal of Econometrics\/}~{\em 76}, 149--69.

\bibitem[\protect\citeauthoryear{Korobilis}{Korobilis}{2018}]{Korobilis_18}
Korobilis, D. (2018).
\newblock Machine learning macroeconometrics: A primer.
\newblock Working Paper 36: 07-2018, University of Essex.

\bibitem[\protect\citeauthoryear{Kourtellos, Marr, and Tan}{Kourtellos
  et~al.}{2016}]{Kourtellos_etal_16}
Kourtellos, A., C.~Marr, and C.~Tan (2016).
\newblock Robust determinants of intergenerational mobility in the land of
  opportunity.
\newblock {\em European Economic Review\/}~{\em 81}, 132--47.
\newblock Special issue on ``Model Uncertainty in Economics''.

\bibitem[\protect\citeauthoryear{Kourtellos and Tsangarides}{Kourtellos and
  Tsangarides}{2015}]{KourtellosTsangarides_16}
Kourtellos, A. and C.~Tsangarides (2015).
\newblock Robust correlates of growth spells: {D}o inequality and
  redistribution matter?
\newblock Working Paper 15-20, The Rimini Centre for Economic Analysis.

\bibitem[\protect\citeauthoryear{Lamnisos, Griffin, and Steel}{Lamnisos
  et~al.}{2009}]{Lamnisos_etal_09}
Lamnisos, D., J.~E. Griffin, and M.~F.~J. Steel (2009).
\newblock Transdimensional sampling algorithms for {B}ayesian variable
  selection in classification problems with many more variables than
  observations.
\newblock {\em Journal of Computational and Graphical Statistics\/}~{\em 18},
  592--612.

\bibitem[\protect\citeauthoryear{Lamnisos, Griffin, and Steel}{Lamnisos
  et~al.}{2013}]{Lamnisos_etal_13}
Lamnisos, D., J.~E. Griffin, and M.~F.~J. Steel (2013).
\newblock {Adaptive Monte Carlo for Bayesian Variable Selection in Regression
  Models}.
\newblock {\em Journal of Computational and Graphical Statistics\/}~{\em 22},
  729--48.

\bibitem[\protect\citeauthoryear{Lanzafame}{Lanzafame}{2016}]{Lanzafame_16}
Lanzafame, M. (2016).
\newblock Potential growth in {A}sia and its determinants: An empirical
  investigation.
\newblock {\em Asian Development Review\/}~{\em 33}, 1--27.

\bibitem[\protect\citeauthoryear{Leamer}{Leamer}{1978}]{leamer78}
Leamer, E. (1978).
\newblock {\em Specification Searches: Ad Hoc Inference with Nonexperimental
  Data}.
\newblock New York: Wiley.

\bibitem[\protect\citeauthoryear{Leamer}{Leamer}{1983}]{Leamer83}
Leamer, E. (1983).
\newblock Let's take the con out of econometrics.
\newblock {\em American Economic Review\/}~{\em 73}, 31--43.

\bibitem[\protect\citeauthoryear{Leamer}{Leamer}{1985}]{Leamer85}
Leamer, E. (1985).
\newblock Sensitivity analyses would help.
\newblock {\em American Economic Review\/}~{\em 75}, 308--13.

\bibitem[\protect\citeauthoryear{Leamer}{Leamer}{2016a}]{Leamer_16_EER}
Leamer, E. (2016a).
\newblock S-values and {B}ayesian weighted all-subsets regressions.
\newblock {\em European Economic Review\/}~{\em 81}, 15--31.
\newblock Special issue on ``Model Uncertainty in Economics''.

\bibitem[\protect\citeauthoryear{Leamer}{Leamer}{2016b}]{Leamer_16_JE}
Leamer, E. (2016b).
\newblock S-values: Conventional context-minimal measures of the sturdiness of
  regression coefficients.
\newblock {\em Journal of Econometrics\/}~{\em 193}, 147--61.

\bibitem[\protect\citeauthoryear{Lee and Chen}{Lee and Chen}{2018}]{LeeChen_18}
Lee, K.-J. and Y.-C. Chen (2018).
\newblock Of needles and haystacks: revisiting growth determinants by robust
  {B}ayesian variable selection.
\newblock {\em Empirical Economics\/}~{\em 54}, 1517--47.

\bibitem[\protect\citeauthoryear{Leeper, Sims, and Zha}{Leeper
  et~al.}{1996}]{Leeper_etal_96}
Leeper, E., C.~Sims, and T.~Zha (1996).
\newblock What does monetary policy do?
\newblock {\em Brookings Papers on Economic Activity\/}~{\em 2}, 1--78 (with
  discussion).

\bibitem[\protect\citeauthoryear{Lenkoski, Eicher, and Raftery}{Lenkoski
  et~al.}{2014}]{Lenkoski_etal_14}
Lenkoski, A., T.~Eicher, and A.~Raftery (2014).
\newblock Two-stage {B}ayesian model averaging in endogenous variable models.
\newblock {\em Econometric Reviews\/}~{\em 33}, 122--51.

\bibitem[\protect\citeauthoryear{Lenkoski, Karl, and Neudecker}{Lenkoski
  et~al.}{2014}]{ivbma}
Lenkoski, A., A.~Karl, and A.~Neudecker (2014).
\newblock ivbma: {B}ayesian instrumental variable estimation and model
  determination via conditional bayes factors, {R} package.
\newblock \url{https://cran.r-project.org/web/packages/ivbma}.

\bibitem[\protect\citeauthoryear{Le\'on-Gonz\'alez and
  Montolio}{Le\'on-Gonz\'alez and Montolio}{2015}]{Leon-GonzalezMontolio_15}
Le\'on-Gonz\'alez, R. and D.~Montolio (2015).
\newblock Endogeneity and panel data in growth regressions: A {B}ayesian model
  averaging approach.
\newblock {\em Journal of Macroeconomics\/}~{\em 46}, 23--39.

\bibitem[\protect\citeauthoryear{LeSage}{LeSage}{2014}]{LeSage_14}
LeSage, J. (2014).
\newblock Spatial econometric panel data model specification: a bayesian
  approach.
\newblock {\em Spatial Statistics\/}~{\em 9}, 122--45.

\bibitem[\protect\citeauthoryear{LeSage}{LeSage}{2015}]{LeSage_15}
LeSage, J. (2015).
\newblock Software for bayesian cross section and panel spatial model
  comparison.
\newblock {\em Journal of Geographical Systems\/}~{\em 17}, 297--310.

\bibitem[\protect\citeauthoryear{LeSage and Parent}{LeSage and
  Parent}{2007}]{LeSageParent_07}
LeSage, J.~P. and O.~Parent (2007).
\newblock Bayesian model averaging for spatial econometric models.
\newblock {\em Geographical Analysis\/}~{\em 39}, 241--67.

\bibitem[\protect\citeauthoryear{Levine and Renelt}{Levine and
  Renelt}{1992}]{LevineRenelt_92}
Levine, R. and D.~Renelt (1992).
\newblock A sensitivity analysis of cross-country growth regressions.
\newblock {\em American Economic Review\/}~{\em 82}, 942--63.

\bibitem[\protect\citeauthoryear{Ley and Steel}{Ley and
  Steel}{2007}]{LeySteel_07}
Ley, E. and M.~Steel (2007).
\newblock Jointness in {B}ayesian variable selection with applications to
  growth regression.
\newblock {\em Journal of Macroeconomics\/}~{\em 29}, 476--93.

\bibitem[\protect\citeauthoryear{Ley and Steel}{Ley and
  Steel}{2009a}]{LeySteel_09b}
Ley, E. and M.~Steel (2009a).
\newblock Comments on '{J}ointness of growth determinants'.
\newblock {\em Journal of Applied Econometrics\/}~{\em 24}, 248--51.

\bibitem[\protect\citeauthoryear{Ley and Steel}{Ley and Steel}{2009b}]{LS6}
Ley, E. and M.~Steel (2009b).
\newblock On the effect of prior assumptions in {B}ayesian model averaging with
  applications to growth regression.
\newblock {\em Journal of Applied Econometrics\/}~{\em 24}, 651--74.

\bibitem[\protect\citeauthoryear{Ley and Steel}{Ley and
  Steel}{2012}]{leysteel2012}
Ley, E. and M.~Steel (2012).
\newblock Mixtures of $g$-priors for {B}ayesian model averaging with economic
  applications.
\newblock {\em Journal of Econometrics\/}~{\em 171}, 251--66.

\bibitem[\protect\citeauthoryear{Li, Li, Racine, and Zhang}{Li
  et~al.}{2018}]{Li_etal_18}
Li, C., Q.~Li, J.~Racine, and D.~Zhang (2018).
\newblock Optimal model averaging of varying coefficient models.
\newblock {\em Statistica Sinica\/}.

\bibitem[\protect\citeauthoryear{Li}{Li}{1987}]{Li_87}
Li, K.-C. (1987).
\newblock Asymptotic optimality for $c_p$, $c_l$, cross-validation and
  generalized cross-validation: Discrete index set.
\newblock {\em Annals of Statistics\/}~{\em 15}, 958--75.

\bibitem[\protect\citeauthoryear{Li and Clyde}{Li and Clyde}{2018}]{LiClyde_17}
Li, Y. and M.~Clyde (2018).
\newblock Mixtures of $g$-priors in generalized linear models.
\newblock {\em Journal of the American Statistical Association\/}~{\em 113},
  1828--45.

\bibitem[\protect\citeauthoryear{Liang, Paulo, Molina, Clyde, and Berger}{Liang
  et~al.}{2008}]{Liang_etal_08}
Liang, F., R.~Paulo, G.~Molina, M.~Clyde, and J.~Berger (2008).
\newblock Mixtures of $g$ priors for {B}ayesian variable selection.
\newblock {\em Journal of the American Statistical Association\/}~{\em 103},
  410--23.

\bibitem[\protect\citeauthoryear{Liang and Wong}{Liang and
  Wong}{2000}]{LiangWong_00}
Liang, F. and W.~Wong (2000).
\newblock Evolutionary {M}onte {C}arlo: Applications to $c_p$ model sampling
  and change point problem.
\newblock {\em Statistica Sinica\/}~{\em 10}, 317--42.

\bibitem[\protect\citeauthoryear{Lindley}{Lindley}{1968}]{Lindley_68}
Lindley, D. (1968).
\newblock The choice of variables in multiple regression (with discussion).
\newblock {\em Journal of the Royal Statistical Society\/}~{\em Ser.~B, 30},
  31--66.

\bibitem[\protect\citeauthoryear{Liu}{Liu}{2015}]{Liu}
Liu, C.-A. (2015).
\newblock Distribution theory of the least squares averaging estimator.
\newblock {\em Journal of Econometrics\/}~{\em 186}, 142--59.

\bibitem[\protect\citeauthoryear{Liu, Okui, and Yoshimura}{Liu
  et~al.}{2016}]{Liu_etal_16}
Liu, Q., R.~Okui, and A.~Yoshimura (2016).
\newblock Generalized least squares model averaging.
\newblock {\em Econometric Reviews\/}~{\em 35}, 1692--752.

\bibitem[\protect\citeauthoryear{Ly\'ocsa, Moln\'ar, and Todorova}{Ly\'ocsa
  et~al.}{2017}]{Lyocsa_etal_17}
Ly\'ocsa, v., P.~Moln\'ar, and N.~Todorova (2017).
\newblock Volatility forecasting of non-ferrous metal futures: {C}ovariances,
  covariates or combinations?
\newblock {\em Journal of International Financial Markets, Institutions \&
  Money\/}~{\em 51}, 228--47.

\bibitem[\protect\citeauthoryear{Madigan, Gavrin, and Raftery}{Madigan
  et~al.}{1995}]{Madigan_etal_95}
Madigan, D., J.~Gavrin, and A.~Raftery (1995).
\newblock Eliciting prior information to enhance the predictive perfromance of
  {B}ayesian graphical models.
\newblock {\em Communications in Statistics, Theory and Methods\/}~{\em 24},
  2271--92.

\bibitem[\protect\citeauthoryear{Madigan and Raftery}{Madigan and
  Raftery}{1994}]{MadiganRaftery_94}
Madigan, D. and A.~Raftery (1994).
\newblock Model selection and accounting for model uncertainty in graphical
  models using {O}ccam's window.
\newblock {\em Journal of the American Statistical Association\/}~{\em 89},
  1535--46.

\bibitem[\protect\citeauthoryear{Madigan and York}{Madigan and
  York}{1995}]{MadiganYork}
Madigan, D. and J.~York (1995).
\newblock Bayesian graphical models for discrete data.
\newblock {\em International Statistical Review\/}~{\em 63}, 215--32.

\bibitem[\protect\citeauthoryear{Magnus and De~Luca}{Magnus and
  De~Luca}{2011}]{DeLucaMagnus_11}
Magnus, J. and G.~De~Luca (2011).
\newblock Bayesian model averaging and weighted-average least squares:
  Equivariance, stability, and numerical issues.
\newblock {\em Stata Journal\/}~{\em 11}, 518--44.

\bibitem[\protect\citeauthoryear{Magnus and De~Luca}{Magnus and
  De~Luca}{2016}]{MagnusSurvey}
Magnus, J. and G.~De~Luca (2016).
\newblock Weighted-average least squares ({WALS}): A survey.
\newblock {\em Journal of Economic Surveys\/}~{\em 30}, 117--48.

\bibitem[\protect\citeauthoryear{Magnus and Wang}{Magnus and
  Wang}{2014}]{MagnusWang_14}
Magnus, J. and W.~Wang (2014).
\newblock Concept-based bayesian model averaging and growth empirics.
\newblock {\em Oxford Bulletin Of Economics And Statistics\/}~{\em 76},
  874--97.

\bibitem[\protect\citeauthoryear{Magnus, Powell, and Pr\"ufer}{Magnus
  et~al.}{2010}]{Magnus_etal_10}
Magnus, J.~R., O.~Powell, and P.~Pr\"ufer (2010).
\newblock A comparison of two model averaging techniques with an application to
  growth empirics.
\newblock {\em Journal of Econometrics\/}~{\em 154}, 139--53.

\bibitem[\protect\citeauthoryear{Magnus, Wan, and Zhang}{Magnus
  et~al.}{2011}]{Magnus_etal_11}
Magnus, J.~R., A.~T. Wan, and X.~Zhang (2011).
\newblock Weighted average least squares estimation with nonspherical
  disturbances and an application to the {H}ong {K}ong housing market.
\newblock {\em Computational Statistics and Data Analysis\/}~{\em 55},
  1331--41.

\bibitem[\protect\citeauthoryear{Makie{\l}a and Osiewalski}{Makie{\l}a and
  Osiewalski}{2018}]{MakielaOsiewalski_18}
Makie{\l}a, K. and J.~Osiewalski (2018).
\newblock Cost efficiency analysis of electricity distribution sector under
  model uncertainty.
\newblock {\em The Energy Journal\/}~{\em 39}, 31--56.

\bibitem[\protect\citeauthoryear{Man}{Man}{2015}]{Man_15}
Man, G. (2015).
\newblock Competition and the growth of nations: International evidence from
  {B}ayesian model averaging.
\newblock {\em Economic Modelling\/}~{\em 51}, 491--501.

\bibitem[\protect\citeauthoryear{Man}{Man}{2018}]{Man_17}
Man, G. (2018).
\newblock Critical appraisal of jointness concepts in {B}ayesian model
  averaging: Evidence from life sciences, sociology, and other scientific
  fields.
\newblock {\em Journal of Applied Statistics\/}~{\em 45}, 845--67.

\bibitem[\protect\citeauthoryear{Marinacci}{Marinacci}{2015}]{Marinacci_15}
Marinacci, M. (2015).
\newblock Model uncertainty.
\newblock {\em Journal of the European Economic Association\/}~{\em 13},
  1022--1100.

\bibitem[\protect\citeauthoryear{Maruyama and George}{Maruyama and
  George}{2011}]{MaruyamaGeorge_11}
Maruyama, Y. and E.~George (2011).
\newblock Fully {B}ayes factors with a generalized $g$-prior.
\newblock {\em Annals of Statistics\/}~{\em 39}, 2740--2765.

\bibitem[\protect\citeauthoryear{Masanjala and Papageorgiou}{Masanjala and
  Papageorgiou}{2008}]{MasanPapa_08}
Masanjala, W. and C.~Papageorgiou (2008).
\newblock Rough and lonely road to prosperity: a reexamination of the sources
  of growth in {A}frica using {B}ayesian model averaging.
\newblock {\em Journal of Applied Econometrics\/}~{\em 23}, 671--82.

\bibitem[\protect\citeauthoryear{Mazerolle}{Mazerolle}{2017}]{AICcmodavg}
Mazerolle, M. (2017).
\newblock {AICcmodavg} - model selection and multimodel inference based on
  (q)aic(c), {R} package.
\newblock \url{https://cran.r-project.org/web/packages/AICcmodavg}.

\bibitem[\protect\citeauthoryear{McCullagh and Nelder}{McCullagh and
  Nelder}{1989}]{McCullaghNelder_89}
McCullagh, P. and J.~A. Nelder (1989).
\newblock {\em Generalized Linear Models}.
\newblock Chapman and Hall.

\bibitem[\protect\citeauthoryear{McKenzie}{McKenzie}{2016}]{McKenzie_16}
McKenzie, T. (2016).
\newblock Technological change and productivity in the rail industry: A
  {B}ayesian approach.
\newblock Technical report, University of Oregon.

\bibitem[\protect\citeauthoryear{Meng and Wong}{Meng and
  Wong}{1996}]{mengwong1996}
Meng, X. and W.~Wong (1996).
\newblock Simulating ratios of normalizing constants via a simple identity: A
  theoretical exploration.
\newblock {\em Statistica Sinica\/}~{\em 6}, 831--860.

\bibitem[\protect\citeauthoryear{Miloschewski}{Miloschewski}{2016}]{Miloschewski_16}
Miloschewski, A. (2016).
\newblock Model uncertainty and the endogeneity problem.
\newblock 9-month {PhD} progress report, University of Warwick.

\bibitem[\protect\citeauthoryear{Min and Zellner}{Min and
  Zellner}{1993}]{MinZellner}
Min, C.-K. and A.~Zellner (1993).
\newblock Bayesian and non-{B}ayesian methods for combining models and
  forecasts with applications to forecasting international growth rates.
\newblock {\em Journal of Econometrics\/}~{\em 56}, 89--118.

\bibitem[\protect\citeauthoryear{Min and Sun}{Min and Sun}{2016}]{MinSun_16}
Min, X. and D.~Sun (2016).
\newblock Bayesian model selection for a linear model with grouped covariates.
\newblock {\em Annals of the Institute of Statistical Mathematics\/}~{\em 68},
  877--903.

\bibitem[\protect\citeauthoryear{Mirestean and Tsangarides}{Mirestean and
  Tsangarides}{2016}]{MiresteanTsangarides_15}
Mirestean, A. and C.~G. Tsangarides (2016).
\newblock Growth determinants revisited using limited-information {B}ayesian
  model averaging.
\newblock {\em Journal of Applied Econometrics\/}~{\em 31}, 106--32.

\bibitem[\protect\citeauthoryear{Moral-Benito}{Moral-Benito}{2012}]{Moral-Benito12}
Moral-Benito, E. (2012).
\newblock Determinants of economic growth: a {B}ayesian panel data approach.
\newblock {\em Review of Economics and Statistics\/}~{\em 94}, 566--79.

\bibitem[\protect\citeauthoryear{Moral-Benito}{Moral-Benito}{2013}]{Moral-Benito_13}
Moral-Benito, E. (2013).
\newblock Likelihood-based estimation of dynamic panels with predetermined
  regressors.
\newblock {\em Journal of Business and Economic Statistics\/}~{\em 31},
  451--72.

\bibitem[\protect\citeauthoryear{Moral-Benito}{Moral-Benito}{2015}]{Moral-Benito15}
Moral-Benito, E. (2015).
\newblock Model averaging in economics: {A}n overview.
\newblock {\em Journal of Economic Surveys\/}~{\em 29}, 46--75.

\bibitem[\protect\citeauthoryear{Moral-Benito}{Moral-Benito}{2016}]{Moral-Benito_16}
Moral-Benito, E. (2016).
\newblock Growth empirics in panel data under model uncertainty and weak
  exogeneity.
\newblock {\em Journal of Applied Econometrics\/}~{\em 31}, 584--602.

\bibitem[\protect\citeauthoryear{Moral-Benito and Roehn}{Moral-Benito and
  Roehn}{2016}]{Moral-BenitoRoehn_16}
Moral-Benito, E. and O.~Roehn (2016).
\newblock The impact of financial regulation on current account balances.
\newblock {\em European Economic Review\/}~{\em 81}, 148--66.
\newblock Special issue on ``Model Uncertainty in Economics''.

\bibitem[\protect\citeauthoryear{Moreno, Gir\'on, and Casella}{Moreno
  et~al.}{2015}]{Moreno_etal_15}
Moreno, E., J.~Gir\'on, and G.~Casella (2015).
\newblock Posterior model consistency in variable selection as the model
  dimension grows.
\newblock {\em Statistical Science\/}~{\em 30}, 228--41.

\bibitem[\protect\citeauthoryear{Morey, Rouder, and Jamil}{Morey
  et~al.}{2015}]{MorRou15}
Morey, R.~D., J.~N. Rouder, and T.~Jamil (2015).
\newblock {BayesFactor}: Computation of {B}ayes factors for common designs, {R}
  package version 0.9.11-1.
\newblock \url{http://CRAN.R-project.org/package=BayesFactor}.

\bibitem[\protect\citeauthoryear{Moser and Hofmarcher}{Moser and
  Hofmarcher}{2014}]{MoserHofmarcher_14}
Moser, M. and P.~Hofmarcher (2014).
\newblock Model priors revisited: Interaction terms in {BMA} growth
  applications.
\newblock {\em Journal of Applied Econometrics\/}~{\em 29}, 344--47.

\bibitem[\protect\citeauthoryear{Mukhopadhyay and Samanta}{Mukhopadhyay and
  Samanta}{2017}]{Mukh_etal_17}
Mukhopadhyay, M. and T.~Samanta (2017).
\newblock A mixture of $g$-priors for variable selection when the number of
  regressors grows with the sample size.
\newblock {\em Test\/}~{\em 26}, 377--404.

\bibitem[\protect\citeauthoryear{Mukhopadhyay, Samanta, and
  Chakrabarti}{Mukhopadhyay et~al.}{2015}]{Mukh_etal_15}
Mukhopadhyay, M., T.~Samanta, and A.~Chakrabarti (2015).
\newblock On consistency and optimality of {B}ayesian variable selection based
  on $g$-prior in normal linear regression models.
\newblock {\em Annals of the Institute of Statistical Mathematics\/}~{\em 67},
  963--997.

\bibitem[\protect\citeauthoryear{Ng, Ibrahim, and Mirakhor}{Ng
  et~al.}{2016}]{Ng_etal_16}
Ng, A., M.~Ibrahim, and A.~Mirakhor (2016).
\newblock Does trust contribute to stock market development?
\newblock {\em Economic Modelling\/}~{\em 52}, 239--50.

\bibitem[\protect\citeauthoryear{Nott and Kohn}{Nott and
  Kohn}{2005}]{NottKohn_05}
Nott, D. and R.~Kohn (2005).
\newblock Adaptive sampling for {B}ayesian variable selection.
\newblock {\em Biometrika\/}~{\em 92}, 747--63.

\bibitem[\protect\citeauthoryear{Oberdabernig, Humer, and
  Crespo~Cuaresma}{Oberdabernig et~al.}{2018}]{Oberdabernig_etal_16}
Oberdabernig, D., S.~Humer, and J.~Crespo~Cuaresma (2018).
\newblock Democracy, geography and model uncertainty.
\newblock {\em Scottish Journal of Political Economy\/}~{\em 65}, 154--85.

\bibitem[\protect\citeauthoryear{Onorante and Raftery}{Onorante and
  Raftery}{2016}]{Onorante16}
Onorante, L. and A.~E. Raftery (2016).
\newblock Dynamic model averaging in large model spaces using dynamic {O}ccam's
  window.
\newblock {\em European Economic Review\/}~{\em 81}, 2--14.
\newblock Special issue on ``Model Uncertainty in Economics''.

\bibitem[\protect\citeauthoryear{Osiewalski and Steel}{Osiewalski and
  Steel}{1993}]{OsiewalskiSteel_93}
Osiewalski, J. and M.~Steel (1993).
\newblock Una perspectiva bayesiana en selecci\'on de modelos.
\newblock {\em Cuadernos Economicos\/}~{\em 55}, 327--51.
\newblock Original English version ``A Bayesian perspective on model
  selection'' at \url{http://www.cyfronet.krakow.pl/~eeosiewa/pubo.htm}.

\bibitem[\protect\citeauthoryear{Ouysse}{Ouysse}{2016}]{Ouysse_16}
Ouysse, R. (2016).
\newblock Bayesian model averaging and principal component regression forecasts
  in a data rich environment.
\newblock {\em International Journal of Forecasting\/}~{\em 32}, 763--87.

\bibitem[\protect\citeauthoryear{Papageorgiou}{Papageorgiou}{2011}]{Papa_11}
Papageorgiou, C. (2011).
\newblock How to use interaction terms in {BMA}: reply to {C}respo {C}uaresma's
  comment on {M}asanjala and {P}apageorgiou (2008).
\newblock {\em Journal of Applied Econometrics\/}~{\em 26}, 1048--50.

\bibitem[\protect\citeauthoryear{Parmeter, Wan, and Zhang}{Parmeter
  et~al.}{2016}]{Parmeter_etal_16}
Parmeter, C., A.~Wan, and X.~Zhang (2016).
\newblock Model averaging estimators for the stochastic frontier model.
\newblock Economics Working Paper 2016-09, University of Miami.

\bibitem[\protect\citeauthoryear{Pelster and Vilsmeier}{Pelster and
  Vilsmeier}{2018}]{PelsterVilsmeier_16}
Pelster, M. and J.~Vilsmeier (2018).
\newblock The determinants of {CDS} spreads: evidence from the model space.
\newblock {\em Review of Derivatives Research\/}~{\em 21}, 63--118.

\bibitem[\protect\citeauthoryear{P\'erez and Berger}{P\'erez and
  Berger}{2002}]{PerezBerger_02}
P\'erez, J. and J.~Berger (2002).
\newblock Expected-posterior prior distributions for model selection.
\newblock {\em Biometrika\/}~{\em 89}, 491--511.

\bibitem[\protect\citeauthoryear{Perrakis and Ntzoufras}{Perrakis and
  Ntzoufras}{2018}]{PerrakisNtzoufras_18}
Perrakis, K. and I.~Ntzoufras (2018).
\newblock Bayesian variable selection using the hyper-g prior in {WinBUGS}.
\newblock {\em WIREs Computational Statistics\/}, e1442.
\newblock https://doi.org/10.1002/wics.1442.

\bibitem[\protect\citeauthoryear{Pham}{Pham}{2017}]{Pham_17}
Pham, T. (2017).
\newblock Impacts of globalization on the informal sector: Empirical evidence
  from developing countries.
\newblock {\em Economic Modelling\/}~{\em 62}, 207--18.

\bibitem[\protect\citeauthoryear{Philips}{Philips}{2016}]{Philips_16}
Philips, A. (2016).
\newblock Seeing the forest through the trees: a meta-analysis of political
  budget cycles.
\newblock {\em Public Choice\/}~{\em 168}, 313--41.

\bibitem[\protect\citeauthoryear{Piironen and Vehtari}{Piironen and
  Vehtari}{2017}]{PiirVeht_17}
Piironen, J. and A.~Vehtari (2017).
\newblock Comparison of {B}ayesian predictive methods for model selection.
\newblock {\em Statistics and Computing\/}~{\em 27}, 711--35.

\bibitem[\protect\citeauthoryear{Piribauer}{Piribauer}{2016}]{Piribauer_16}
Piribauer, P. (2016).
\newblock Heterogeneity in spatial growth clusters.
\newblock {\em Empirical Economics\/}~{\em 51}, 659--80.

\bibitem[\protect\citeauthoryear{Piribauer and Crespo~Cuaresma}{Piribauer and
  Crespo~Cuaresma}{2016}]{PiribauerCuaresma_17}
Piribauer, P. and J.~Crespo~Cuaresma (2016).
\newblock Bayesian variable selection in spatial autoregressive models.
\newblock {\em Spatial Economic Analysis\/}~{\em 11}, 457--79.

\bibitem[\protect\citeauthoryear{Puy}{Puy}{2016}]{Puy_16}
Puy, D. (2016).
\newblock Mutual funds flows and the geography of contagion.
\newblock {\em Journal of International Money and Finance\/}~{\em 60}, 73--93.

\bibitem[\protect\citeauthoryear{Raftery}{Raftery}{1995}]{Raftery1995}
Raftery, A. (1995).
\newblock Bayesian model selection in social research.
\newblock {\em Sociological Methodology\/}~{\em 25}, 111--63.

\bibitem[\protect\citeauthoryear{Raftery}{Raftery}{1996}]{Raftery_96}
Raftery, A. (1996).
\newblock Approximate {B}ayes factors and accounting for model uncertainty in
  generalised linear models.
\newblock {\em Biometrika\/}~{\em 83}, 251--66.

\bibitem[\protect\citeauthoryear{Raftery, Gneiting, Balabdaoui, and
  Polakowski}{Raftery et~al.}{2005}]{Raftery_etal_05}
Raftery, A., T.~Gneiting, F.~Balabdaoui, and M.~Polakowski (2005).
\newblock Using {B}ayesian model averaging to calibrate forecast ensembles.
\newblock {\em Monthly Weather Review\/}~{\em 133}, 1155--74.

\bibitem[\protect\citeauthoryear{Raftery, Hoeting, Volinsky, Painter, and
  Yeung}{Raftery et~al.}{2010}]{Raftery_etal_soft}
Raftery, A., J.~Hoeting, C.~Volinsky, I.~Painter, and K.~Yeung (2010).
\newblock {B}ayesian model averaging. {R} package vs. 3.13.
\newblock {\url{http://CRAN.R-project.org/package=BMA}}.

\bibitem[\protect\citeauthoryear{Raftery, Madigan, and Hoeting}{Raftery
  et~al.}{1997}]{Raftery_etal_97}
Raftery, A., D.~Madigan, and J.~Hoeting (1997).
\newblock Bayesian model averaging for linear regression models.
\newblock {\em Journal of the American Statistical Association\/}~{\em 92},
  179--91.

\bibitem[\protect\citeauthoryear{Raftery and Zheng}{Raftery and
  Zheng}{2003}]{RafteryZheng_03}
Raftery, A. and Y.~Zheng (2003).
\newblock Discussion: Performance of {B}ayesian model averaging.
\newblock {\em Journal of the American Statistical Association\/}~{\em 98},
  931--8.

\bibitem[\protect\citeauthoryear{Raftery, K\'arn\'y, and Ettler}{Raftery
  et~al.}{2010}]{Raftery_etal_10}
Raftery, A.~E., M.~K\'arn\'y, and P.~Ettler (2010).
\newblock Online prediction under model uncertainty via dynamic model
  averaging: Application to a cold rolling mill.
\newblock {\em Technometrics\/}~{\em 52}, 52--66.

\bibitem[\protect\citeauthoryear{Ranciati, Galimberti, and Soffritti}{Ranciati
  et~al.}{2019}]{Ranciati_etal_19}
Ranciati, S., G.~Galimberti, and G.~Soffritti (2019).
\newblock Bayesian variable selection in linear regression models with
  non-normal errors.
\newblock {\em Statistical Methods and Applications\/}, forthcoming.

\bibitem[\protect\citeauthoryear{Ribeiro}{Ribeiro}{2017}]{Ribeiro_17}
Ribeiro, P. (2017).
\newblock Selecting exchange rate fundamentals by bootstrap.
\newblock {\em International Journal of Forecasting\/}~{\em 33}, 894--914.

\bibitem[\protect\citeauthoryear{Robert}{Robert}{2007}]{Robert2007}
Robert, C. (2007).
\newblock {\em The {B}ayesian choice: from decision-theoretic foundations to
  computational implementation\/} (2nd ed.).
\newblock Springer.

\bibitem[\protect\citeauthoryear{Robert and Casella}{Robert and
  Casella}{2004}]{RobertCasella_04}
Robert, C. and G.~Casella (2004).
\newblock {\em Monte Carlo Statistical Methods}.
\newblock New York: Springer.

\bibitem[\protect\citeauthoryear{Rockey and Temple}{Rockey and
  Temple}{2016}]{RockeyTemple_16}
Rockey, J. and J.~Temple (2016).
\newblock Growth econometrics for agnostics and true believers.
\newblock {\em European Economic Review\/}~{\em 81}, 86--102.
\newblock Special issue on ``Model Uncertainty in Economics''.

\bibitem[\protect\citeauthoryear{Rodrik, Subramanian, and Trebbi}{Rodrik
  et~al.}{2004}]{Rodrik_etal_04}
Rodrik, D., A.~Subramanian, and F.~Trebbi (2004).
\newblock Institutions rule: the primacy of insitutions over geography and
  integration in economic development.
\newblock {\em Journal of Economic Growth\/}~{\em 9}, 131--165.

\bibitem[\protect\citeauthoryear{Rose and Spiegel}{Rose and
  Spiegel}{2011}]{RoseSpiegel_11}
Rose, A. and M.~Spiegel (2011).
\newblock Cross-country causes and consequences of the crisis: {A}n update.
\newblock {\em European Economic Review\/}~{\em 55}, 309--24.

\bibitem[\protect\citeauthoryear{Rossell, Cook, Telesca, and Roebuck}{Rossell
  et~al.}{2014}]{Rossell14}
Rossell, D., J.~D. Cook, D.~Telesca, and P.~Roebuck (2014).
\newblock mombf: Moment and inverse moment {B}ayes factors, {R} package version
  1.5.9.
\newblock \url{http://CRAN.R-project.org/package=mombf}.

\bibitem[\protect\citeauthoryear{Rossell and Telesca}{Rossell and
  Telesca}{2017}]{RossellTelesca_17}
Rossell, D. and D.~Telesca (2017).
\newblock Nonlocal priors for high-dimensional estimation.
\newblock {\em Journal of the American Statistical Association\/}~{\em 112},
  254--65.

\bibitem[\protect\citeauthoryear{Rossi, Allenby, and McCulloch}{Rossi
  et~al.}{2006}]{Rossi_etal_06}
Rossi, P.~E., G.~Allenby, and R.~McCulloch (2006).
\newblock {\em Bayesian Statistics and Marketing}.
\newblock New York: Wiley.

\bibitem[\protect\citeauthoryear{Russell, Murphy, and Raftery}{Russell
  et~al.}{2015}]{Russell_etal_15}
Russell, N., T.~Murphy, and A.~Raftery (2015).
\newblock Bayesian model averaging in model-based clustering and density
  estimation.
\newblock technical report arXiv:1506.09035, University of Washington.

\bibitem[\protect\citeauthoryear{Saban\'es~Bov\'e}{Saban\'es~Bov\'e}{2011}]{hypergsplines}
Saban\'es~Bov\'e, D. (2011).
\newblock hyper-$g$ priors for {GAM} selection, {R} package.
\newblock \url{https://r-forge.r-project.org/projects/hypergsplines/}.

\bibitem[\protect\citeauthoryear{Saban\'es~Bov\'e and Held}{Saban\'es~Bov\'e
  and Held}{2011a}]{SabanesHeld_11}
Saban\'es~Bov\'e, D. and L.~Held (2011a).
\newblock Bayesian fractional polynomials.
\newblock {\em Statistics and Computing\/}~{\em 29}, 309--24.

\bibitem[\protect\citeauthoryear{Saban\'es~Bov\'e and Held}{Saban\'es~Bov\'e
  and Held}{2011b}]{BoveHeld_11}
Saban\'es~Bov\'e, D. and L.~Held (2011b).
\newblock Hyper-$g$ priors for generalized linear models.
\newblock {\em Bayesian Analysis\/}~{\em 6}, 387--410.

\bibitem[\protect\citeauthoryear{Saban\'es~Bov\'e, Held, and
  Kauermann}{Saban\'es~Bov\'e et~al.}{2015}]{Sabanes_etal_15}
Saban\'es~Bov\'e, D., L.~Held, and G.~Kauermann (2015).
\newblock Objective {B}ayesian model selection in generalised additive models
  with penalised splines.
\newblock {\em Journal of Computational and Graphical Statistics\/}~{\em 24},
  394--415.

\bibitem[\protect\citeauthoryear{Sachs and Warner}{Sachs and
  Warner}{2001}]{SachsWarner_01}
Sachs, J. and A.~Warner (2001).
\newblock The curse of natural resources.
\newblock {\em European Economic Review\/}~{\em 45}, 827--38.

\bibitem[\protect\citeauthoryear{Sala-{i-Martin}}{Sala-{i-Martin}}{1997}]{SiM_97}
Sala-{i-Martin}, X. (1997).
\newblock I just ran two million regressions.
\newblock {\em American Economic Review\/}~{\em 87\/}(2), 178--83.

\bibitem[\protect\citeauthoryear{Sala-{i-Martin}, Doppelhofer, and
  Miller}{Sala-{i-Martin} et~al.}{2004}]{SDM}
Sala-{i-Martin}, X., G.~Doppelhofer, and R.~Miller (2004).
\newblock Determinants of long-term growth: A {B}ayesian averaging of classical
  estimates ({BACE}) approach.
\newblock {\em American Economic Review\/}~{\em 94}, 813--35.

\bibitem[\protect\citeauthoryear{Schwarz}{Schwarz}{1978}]{Schwarz78}
Schwarz, G. (1978).
\newblock Estimating the dimension of a model.
\newblock {\em Annals of Statistics\/}~{\em 6}, 461--4.

\bibitem[\protect\citeauthoryear{Scott and Berger}{Scott and
  Berger}{2010}]{ScottBerger}
Scott, J. and J.~Berger (2010).
\newblock Bayes and empirical {B}ayes multiplicity adjustment in the
  variable-selection problem.
\newblock {\em Annals of Statistics\/}~{\em 38}, 2587--619.

\bibitem[\protect\citeauthoryear{Shao}{Shao}{1997}]{Shao_97}
Shao, J. (1997).
\newblock An asymptotic theory for linear model selection.
\newblock {\em Statistica Sinica\/}~{\em 7}, 221--64.

\bibitem[\protect\citeauthoryear{Shi}{Shi}{2016}]{Shi_16}
Shi, J. (2016).
\newblock Bayesian model averaging under regime switching with application to
  cyclical macro variable forecasting.
\newblock {\em Journal of Forecasting\/}~{\em 35}, 250--62.

\bibitem[\protect\citeauthoryear{Sickles}{Sickles}{2005}]{Sickles_05}
Sickles, R. (2005).
\newblock Panel estimators and the identification of firm-specific efficiency
  levels in parametric, semiparametric and nonparametric settings.
\newblock {\em Journal of Econometrics\/}~{\em 126}, 305--34.

\bibitem[\protect\citeauthoryear{Smith and Wallis}{Smith and
  Wallis}{2009}]{SmithWallis_09}
Smith, J. and K.~Wallis (2009).
\newblock A simple explanation of the forecast combination puzzle.
\newblock {\em Oxford Bulletin of Economics and Statistics\/}~{\em 71},
  331--55.

\bibitem[\protect\citeauthoryear{Som, Hans, and MacEachern}{Som
  et~al.}{2015}]{Som_etal_15}
Som, A., C.~Hans, and S.~MacEachern (2015).
\newblock Bayesian modeling with mixtures of block $g$ priors.
\newblock Technical report, {D}ept.~of {S}tatistics, Ohio State University.

\bibitem[\protect\citeauthoryear{Sparks, Khare, and Ghosh}{Sparks
  et~al.}{2015}]{Sparks_etal_15}
Sparks, D., K.~Khare, and M.~Ghosh (2015).
\newblock Necessary and sufficient conditions for high-dimensional posterior
  consistency under $g$-priors.
\newblock {\em Bayesian Analysis\/}~{\em 10}, 627--664.

\bibitem[\protect\citeauthoryear{Spiegelhalter, Best, Carlin, and van~der
  Linde}{Spiegelhalter et~al.}{2002}]{Spiegelhalter_etal02}
Spiegelhalter, D., N.~Best, B.~Carlin, and A.~van~der Linde (2002).
\newblock Bayesian measures of model complexity and fit (with discussion).
\newblock {\em Journal of the Royal Statistical Society, Series B\/}~{\em 64},
  583--640.

\bibitem[\protect\citeauthoryear{Stock and Watson}{Stock and
  Watson}{2004}]{StockWatson_04}
Stock, J. and M.~Watson (2004).
\newblock Combination forecasts of output growth in a seven-country data set.
\newblock {\em Journal of Forecasting\/}~{\em 23}, 405--30.

\bibitem[\protect\citeauthoryear{Stock and Watson}{Stock and
  Watson}{2006}]{StockWatson_06}
Stock, J. and M.~Watson (2006).
\newblock Forecasting with many predictors.
\newblock In C.~G. G.~Elliott and A.~Timmermann (Eds.), {\em Handbook of
  Economic Forecasting}, pp.\  515--54. Elsevier.

\bibitem[\protect\citeauthoryear{Strachan and van Dijk}{Strachan and van
  Dijk}{2013}]{StrachanvanDijk_13}
Strachan, R. and H.~van Dijk (2013).
\newblock Evidence on features of a {DSGE} business cycle model from {B}ayesian
  model averaging.
\newblock {\em International Economic Review\/}~{\em 54}, 385--402.

\bibitem[\protect\citeauthoryear{Strachan}{Strachan}{2009}]{Strachan_09}
Strachan, R.~W. (2009).
\newblock Comment on 'jointness of growth determinants' by {G}ernot
  {D}oppelhofer and {M}elvyn {W}eeks.
\newblock {\em Journal of Applied Econometrics\/}~{\em 24}, 245--47.

\bibitem[\protect\citeauthoryear{Strawderman}{Strawderman}{1971}]{Strawderman}
Strawderman, W. (1971).
\newblock Proper {B}ayes minimax estimators of the multivariate normal mean.
\newblock {\em Annals of Mathematical Statistics\/}~{\em 42}, 385--8.

\bibitem[\protect\citeauthoryear{Temple}{Temple}{2000}]{Temple_00}
Temple, J. (2000).
\newblock Growth regressions and what the textbooks don't tell you.
\newblock {\em Bulletin of Economic Research\/}~{\em 52}, 181--205.

\bibitem[\protect\citeauthoryear{Tibshirani}{Tibshirani}{1996}]{Tibshirani}
Tibshirani, R. (1996).
\newblock Regression shrinkage and selection via the lasso.
\newblock {\em Journal of the Royal Statistical Society, Series B\/}~{\em 58},
  267--88.

\bibitem[\protect\citeauthoryear{Tobias and Li}{Tobias and Li}{2004}]{TobiasLi}
Tobias, J. and M.~Li (2004).
\newblock Returns to schooling and {B}ayesian model averaging; a union of two
  literatures.
\newblock {\em Journal of Economic Surveys\/}~{\em 18}, 153--80.

\bibitem[\protect\citeauthoryear{Traczynski}{Traczynski}{2017}]{Traczynski_17}
Traczynski, J. (2017).
\newblock Firm default prediction: A {B}ayesian model-averaging approach.
\newblock {\em Journal Of Financial And Quantitative Analysis\/}~{\em 52},
  1211--45.

\bibitem[\protect\citeauthoryear{Tsangarides}{Tsangarides}{2004}]{Tsangarides04}
Tsangarides, C.~G. (2004).
\newblock A {B}ayesian approach to model uncertainty.
\newblock Working Paper 04/68, IMF.

\bibitem[\protect\citeauthoryear{Tu}{Tu}{2018}]{Tu_18}
Tu, Y. (2018).
\newblock Entropy-based model averaging estimation of nonparametric models.
\newblock Guanghua {S}chool of {M}anagement and {C}enter for {S}tatistical
  {S}cience, Peking University.

\bibitem[\protect\citeauthoryear{Ullah and Wang}{Ullah and
  Wang}{2013}]{UllahWang_13}
Ullah, A. and H.~Wang (2013).
\newblock Parametric and nonparametric frequentist model selection and model
  averaging.
\newblock {\em Econometrics\/}~{\em 1}, 157--79.

\bibitem[\protect\citeauthoryear{Vallejos and Steel}{Vallejos and
  Steel}{2017}]{VallejosSteel_17}
Vallejos, C. and M.~Steel (2017).
\newblock Bayesian survival modelling of university outcomes.
\newblock {\em Journal of the Royal Statistical Society, A\/}~{\em 180},
  613--31.

\bibitem[\protect\citeauthoryear{van~den Broeck, Koop, Osiewalski, and
  Steel}{van~den Broeck et~al.}{1994}]{Broeck_etal_94}
van~den Broeck, J., G.~Koop, J.~Osiewalski, and M.~Steel (1994).
\newblock Stochastic frontier models: a {B}ayesian perspective.
\newblock {\em Journal of Econometrics\/}~{\em 61}, 273--303.

\bibitem[\protect\citeauthoryear{van~der Maas}{van~der
  Maas}{2014}]{vanderMaas_14}
van~der Maas, J. (2014).
\newblock Forecasting inflation using time-varying {B}ayesian model averaging.
\newblock {\em Statistica Neerlandica\/}~{\em 68}, 149--82.

\bibitem[\protect\citeauthoryear{Varian}{Varian}{2014}]{Varian_14}
Varian, H. (2014).
\newblock Big data: {N}ew tricks for econometrics.
\newblock {\em Journal of Economic Perspectives\/}~{\em 28}, 3--28.

\bibitem[\protect\citeauthoryear{Va\v{s}\'{\i}\v{c}ek, \v{Z}igraiov\'a,
  Hoeberichts, Vermeulen, \v{S}m\'{i}dkov\'a, and de~Haan}{Va\v{s}\'{\i}\v{c}ek
  et~al.}{2017}]{Vasicek_etal_17}
Va\v{s}\'{\i}\v{c}ek, B., D.~\v{Z}igraiov\'a, M.~Hoeberichts, R.~Vermeulen,
  K.~\v{S}m\'{i}dkov\'a, and J.~de~Haan (2017).
\newblock Leading indicators of financial stress: {N}ew evidence.
\newblock {\em Journal of Financial Stability\/}~{\em 28}, 240--57.

\bibitem[\protect\citeauthoryear{Verdinelli and Wasserman}{Verdinelli and
  Wasserman}{1995}]{verdinelliwasserman1995}
Verdinelli, I. and L.~Wasserman (1995).
\newblock Computing {B}ayes factors by using a generalization of the
  {S}avage-{D}ickey density ratio.
\newblock {\em Journal of the American Statistical Association\/}~{\em 90},
  614--618.

\bibitem[\protect\citeauthoryear{Villa and Lee}{Villa and
  Lee}{2016}]{VillaLee_16}
Villa, C. and J.~Lee (2016).
\newblock Model prior distribution for variable selection in linear regression
  models.
\newblock technical report arXiv:1512.08077, University of Kent.

\bibitem[\protect\citeauthoryear{Villa and Walker}{Villa and
  Walker}{2015}]{VillaWalker_15}
Villa, C. and S.~Walker (2015).
\newblock An objective {B}ayesian criterion to determine model prior
  probabilities.
\newblock {\em Scandinavian Journal of Statistics\/}~{\em 42}, 947--66.

\bibitem[\protect\citeauthoryear{Volinsky, Madigan, Raftery, and
  Kronmal}{Volinsky et~al.}{1997}]{Volinsky_etal_97}
Volinsky, C., D.~Madigan, A.~Raftery, and R.~Kronmal (1997).
\newblock Bayesian model averaging in proportional hazards model: Predicting
  the risk of a stroke.
\newblock {\em Applied Statistics\/}~{\em 46}, 443--8.

\bibitem[\protect\citeauthoryear{Wagner and Hlouskova}{Wagner and
  Hlouskova}{2015}]{WagnerHlouskova_15}
Wagner, M. and J.~Hlouskova (2015).
\newblock Growth regressions, principal components augmented regressions and
  frequentist model averaging.
\newblock {\em Jahrb\"ucher fur National\"okonomie und Statistik\/}~{\em 235},
  642--62.

\bibitem[\protect\citeauthoryear{Wallis}{Wallis}{2005}]{Wallis_05}
Wallis, K. (2005).
\newblock Combining density and interval forecasts: {A} modest proposal.
\newblock {\em Oxford Bulletin of Economics and Statistics\/}~{\em 67},
  983--94.

\bibitem[\protect\citeauthoryear{Wang, Zhang, and Zou}{Wang
  et~al.}{2009}]{Wangetal09}
Wang, H., X.~Zhang, and G.~Zou (2009).
\newblock Frequentist model averaging estimation: A review.
\newblock {\em Journal of Systems Science and Complexity\/}~{\em 22}, 732--48.

\bibitem[\protect\citeauthoryear{Wang and Zhou}{Wang and
  Zhou}{2013}]{WangZhou_13}
Wang, H. and S.~Zhou (2013).
\newblock Interval estimation by frequentist model averaging.
\newblock {\em Communications in Statistics: Theory and Methods\/}~{\em 42},
  4342--56.

\bibitem[\protect\citeauthoryear{Wang}{Wang}{2017}]{Wang_17}
Wang, M. (2017).
\newblock Mixtures of $g$-priors for analysis of variance models with a
  diverging number of parameters.
\newblock {\em Bayesian Analysis\/}~{\em 12}, 511--32.

\bibitem[\protect\citeauthoryear{Wang and Maruyama}{Wang and
  Maruyama}{2016}]{WangMaruyama_16}
Wang, M. and Y.~Maruyama (2016).
\newblock Consistency of {B}ayes factor for nonnested model selection when the
  model dimension grows.
\newblock {\em Bernoulli\/}~{\em 22}, 2080--100.

\bibitem[\protect\citeauthoryear{Watson and Deller}{Watson and
  Deller}{2017}]{WatsonDeller_17}
Watson, P. and S.~Deller (2017).
\newblock Economic diversity, unemployment and the {G}reat {R}ecession.
\newblock {\em The Quarterly Review of Economics and Finance\/}~{\em 64},
  1--11.

\bibitem[\protect\citeauthoryear{Wei and Cao}{Wei and Cao}{2017}]{WeiCao_17}
Wei, Y. and Y.~Cao (2017).
\newblock Forecasting house prices using dynamic model averaging approach:
  {E}vidence from {C}hina.
\newblock {\em Economic Modelling\/}~{\em 61}, 147--55.

\bibitem[\protect\citeauthoryear{W\"olfel and Weber}{W\"olfel and
  Weber}{2017}]{WolfelWeber_17}
W\"olfel, K. and C.~Weber (2017).
\newblock Searching for the {F}ed's reaction function.
\newblock {\em Empirical Economics\/}~{\em 52}, 191--227.

\bibitem[\protect\citeauthoryear{Womack, Fuentes, and Taylor-Rodriguez}{Womack
  et~al.}{2015}]{Womack_etal_15}
Womack, A., C.~Fuentes, and D.~Taylor-Rodriguez (2015).
\newblock Model space priors for objective sparse {B}ayesian regression.
\newblock technical report arXiv:1511.04745v1.

\bibitem[\protect\citeauthoryear{Wright}{Wright}{2008}]{Wright_08}
Wright, J. (2008).
\newblock Bayesian model averaging and exchange rate forecasts.
\newblock {\em Journal of Econometrics\/}~{\em 146}, 329--41.

\bibitem[\protect\citeauthoryear{Wu, Ferreira, Elkhouly, and Ji}{Wu
  et~al.}{2019}]{Wu_etal_18}
Wu, H., M.~Ferreira, M.~Elkhouly, and T.~Ji (2019).
\newblock Hyper nonlocal priors for variable selection in generalized linear
  models.
\newblock {\em Sankhy\={a} A\/}, forthcoming.

\bibitem[\protect\citeauthoryear{Wu, Ferreira, and Gompper}{Wu
  et~al.}{2016}]{Wu_etal_15}
Wu, H., M.~Ferreira, and M.~Gompper (2016).
\newblock Consistency of hyper-$g$-prior-based {B}ayesian variable selection
  for generalized linear models.
\newblock {\em Brazilian Journal of Probability and Statistics\/}~{\em 30},
  691--709.

\bibitem[\protect\citeauthoryear{Xiang, Ghosh, and Khare}{Xiang
  et~al.}{2016}]{Xiang_etal_16}
Xiang, R., M.~Ghosh, and K.~Khare (2016).
\newblock Consistency of {B}ayes factors under hyper $g$-priors with growing
  model size.
\newblock {\em Journal of Statistical Planning and Inference\/}~{\em 173}, 64
  -- 86.

\bibitem[\protect\citeauthoryear{Zellner}{Zellner}{1971}]{zellner1971}
Zellner, A. (1971).
\newblock {\em An Introduction to Bayesian Inference in Econometrics}.
\newblock New York: Wiley.

\bibitem[\protect\citeauthoryear{Zellner}{Zellner}{1986}]{Zellner86}
Zellner, A. (1986).
\newblock On assessing prior distributions and {B}ayesian regression analysis
  with g-prior distributions.
\newblock In P.~K. Goel and A.~Zellner (Eds.), {\em Bayesian Inference and
  Decision Techniques: Essays in Honor of Bruno de Finetti}, Amsterdam:
  North-Holland, pp.\  233--43.

\bibitem[\protect\citeauthoryear{Zellner and Siow}{Zellner and
  Siow}{1980}]{ZellnerSiow80}
Zellner, A. and A.~Siow (1980).
\newblock Posterior odds ratios for selected regression hypotheses (with
  discussion).
\newblock In J.~Bernardo, M.~DeGroot, D.~Lindley, and A.~Smith (Eds.), {\em
  Bayesian Statistics}, Valencia: University Press, pp.\  585--603.

\bibitem[\protect\citeauthoryear{Zeugner and Feldkircher}{Zeugner and
  Feldkircher}{2015}]{FZ_15}
Zeugner, S. and M.~Feldkircher (2015).
\newblock Bayesian model averaging employing fixed and flexible priors: {T}he
  {BMS} package for {R}.
\newblock {\em Journal of Statistical Software\/}~{\em 68\/}(4).

\bibitem[\protect\citeauthoryear{Zhang, Huang, Gan, Karmaus, and
  Sabo-Attwood}{Zhang et~al.}{2016}]{Zhang_etal_16BA}
Zhang, H., X.~Huang, J.~Gan, W.~Karmaus, and T.~Sabo-Attwood (2016).
\newblock A two-component $g$-prior for variable selection.
\newblock {\em Bayesian Analysis\/}~{\em 11}, 353--80.

\bibitem[\protect\citeauthoryear{Zhang, Lu, and Zou}{Zhang
  et~al.}{2013}]{Zhang_etal_13}
Zhang, X., Z.~Lu, and G.~Zou (2013).
\newblock Adaptively combined forecasting for discrete response time series.
\newblock {\em Journal of Econometrics\/}~{\em 176}, 80--91.

\bibitem[\protect\citeauthoryear{Zhang and Wang}{Zhang and
  Wang}{2019}]{ZhangWang_18}
Zhang, X. and W.~Wang (2019).
\newblock Optimal model averaging estimation for partially linear models.
\newblock {\em Statistica Sinica\/}, forthcoming.

\bibitem[\protect\citeauthoryear{Zhang, Yu, Zou, and Liang}{Zhang
  et~al.}{2016}]{Zhang_etal_17}
Zhang, X., D.~Yu, G.~Zou, and H.~Liang (2016).
\newblock Optimal model averaging estimation for generalized linear models and
  generalized linear mixed- effects models.
\newblock {\em Journal of the American Statistical Association\/}~{\em 111},
  1775--90.

\bibitem[\protect\citeauthoryear{Zhu, Wan, Zhang, and Zou}{Zhu
  et~al.}{2019}]{Zhu_etal_18}
Zhu, R., A.~Wan, X.~Zhang, and G.~Zou (2019).
\newblock A {M}allows- type model averaging estimator for the
  varying-coefficient partially linear model.
\newblock {\em Journal of the American Statistical Association\/}, forthcoming.

\bibitem[\protect\citeauthoryear{Zigraiova and Havranek}{Zigraiova and
  Havranek}{2016}]{ZigraiovaHavranek_16}
Zigraiova, D. and T.~Havranek (2016).
\newblock Bank competition and financial stability: Much ado about nothing?
\newblock {\em Journal of Economic Surveys\/}~{\em 30}, 944--81.

\bibitem[\protect\citeauthoryear{Zou}{Zou}{2006}]{Zou06}
Zou, H. (2006).
\newblock The adaptive lasso and its oracle properties.
\newblock {\em Journal of the American Statistical Association\/}~{\em 101},
  1418--29.

\end{thebibliography}

\end{document}